\pdfoutput=1
\documentclass[%
 reprint,
superscriptaddress,
 amsmath,amssymb,
 aps,
floatfix,
]{revtex4-2}

\usepackage{graphicx}% Include figure files
\usepackage{dcolumn}% Align table columns on decimal point
\usepackage{bm}% bold math
\usepackage{amsthm}
\usepackage[colorlinks=true, hyperindex, breaklinks, linkcolor=blue, urlcolor=blue, citecolor=blue]{hyperref} % Physical Review style
\usepackage{amsmath,amsfonts,amssymb,color,graphicx,xcolor,physics}
\usepackage{tikz}
\usepackage{float}
\makeatletter
\let\newfloat\newfloat@ltx
\makeatother
\usepackage{algorithm} 
\usepackage{algorithmicx}
\usepackage[noend]{algpseudocode} 
\newcommand\nonpfrate[1]{\gamma_{X, Y}}
\makeatletter
\newcommand*{\rom}[1]{\expandafter\@slowromancap\romannumeral #1@}
\makeatother
\def\algbackskip{\hskip-\ALG@thistlm}

\begin{document}

\preprint{APS/123-QED}

\title{Autonomous quantum error correction and fault-tolerant quantum computation with squeezed cat qubits}

\author{Qian Xu}
\thanks{These authors contributed equally.}
\affiliation{Pritzker School of Molecular Engineering, The University of Chicago, Chicago 60637, USA}

\author{Guo Zheng}
\thanks{These authors contributed equally.}
\affiliation{Pritzker School of Molecular Engineering, The University of Chicago, Chicago 60637, USA}

\author{Yu-Xin Wang}
\affiliation{Pritzker School of Molecular Engineering, The University of Chicago, Chicago 60637, USA}

\author{Peter Zoller}
\affiliation{Institute for Theoretical Physics, University of Innsbruck, Innsbruck A-6020, Austria}
\affiliation{Institute for Quantum Optics and Quantum Information of the Austrian Academy of Sciences, Innsbruck A-6020, Austria}

\author{Aashish A. Clerk}
\affiliation{Pritzker School of Molecular Engineering, The University of Chicago, Chicago 60637, USA}

\author{Liang Jiang}
\email{liang.jiang@uchicago.edu}
\affiliation{Pritzker School of Molecular Engineering, The University of Chicago, Chicago 60637, USA}
%\affiliation{AWS Center for Quantum Computing, Pasadena, CA 91125, USA}

\date{\today}% It is always \today, today,

\begin{abstract}
We propose an autonomous quantum error correction scheme using squeezed cat (SC) code against the dominant error source, excitation loss, in continuous-variable systems. Through reservoir engineering, we show that a structured dissipation can stabilize a two-component SC while autonomously correcting the errors. The implementation of such dissipation only requires low-order nonlinear couplings among three bosonic modes or between a bosonic mode and a qutrit. While our proposed scheme is device independent, it is readily implementable with current experimental platforms such as superconducting circuits and trapped-ion systems. Compared to the stabilized cat, the stabilized SC has a much lower dominant error rate and a significantly enhanced noise bias. Furthermore, the bias-preserving operations for the SC have much lower error rates. In combination, the stabilized SC leads to substantially better logical performance when concatenating with an outer discrete-variable code. The surface-SC scheme achieves  more than one order of magnitude increase in the threshold ratio between the loss rate $\kappa_1$ and the engineered dissipation rate $\kappa_2$. Under a practical noise ratio $\kappa_1/\kappa_2 = 10^{-3}$, the repetition-SC scheme can reach a $10^{-15}$ logical error rate even with a small mean excitation number of 4, which already suffices for practically useful quantum algorithms.
\end{abstract}

\maketitle

%\tableofcontents

\section{Introduction}
Quantum information is fragile to errors introduced by the environment. Quantum error correction (QEC) protects quantum systems by correcting the errors and removing the entropy~\cite{nielsen_chuang_2010, lidar_brun_2013, aharonov1996limitations}. Based upon QEC, fault-tolerant quantum computation (FTQC) can be performed, provided that the physical noise strength is below an accuracy threshold~\cite{aharonov1997fault, kitaev1997quantum, knill1998resilient, aliferis2005quantum}. However, realizing FTQC is yet challenging due to the demanding threshold requirement and the significant resource overhead~\cite{fowler2012surface, litinski2019game, chao2020optimization, beverland2021cost}. Unlike discrete-variable (DV) systems, continuous-variable (CV) systems possess an infinite-dimensional Hilbert space. Encoding the quantum information in CV systems, therefore, provides a hardware-efficient approach to QEC~\cite{gottesman2001encoding, chuang1997bosonic, michael2016new, cochrane1999macroscopically, albert2018performance}. Various bosonic codes have been experimentally demonstrated to suppress errors in CV systems~\cite{ofek2016extending, HuL19, campagne2020quantum, lescanne2020exponential, Fluhmann19, grimm2020stabilization}. 

The standard QEC procedure relies on  actively measuring the error syndromes and performing feedback control~\cite{nielsen_chuang_2010}. However, such adaptive protocols demand fast, high-fidelity coherent operations and measurements, which poses significant experimental challenges. At this stage, the error rates in the encoded level are still higher than the physical error rates in current devices due to the errors during the QEC operations~\cite{egan2021fault, zhao2022realization, ryan2022implementing, acharya2022suppressing}. To address these challenges, we may implement QEC non-adaptively via engineered dissipation
– an approach called autonomous QEC (AutoQEC)~\cite{lebreuilly2021autonomous}. Such an approach avoids the measurement imperfection and overhead associated with the classical feedback loops. Autonomous QEC in bosonic systems that can magnificently suppress the dephasing noise has been demonstrated using the two-component cat code~\cite{lescanne2020exponential, grimm2020stabilization, berdou2022one}. However, AutoQEC against excitation loss, which is usually the dominant error source in a bosonic mode, remains challenging. It requires either large nonlinearities that are challenging to engineer (e.g., the multiphoton processes needed for the multi-component cat codes~\cite{mirrahimi2014dynamically}) or couplings to an intrinsically nonlinear DV system~\cite{royer2020stabilization, gertler2021protecting} that is much noisier than the bosonic mode. 

In this work, we propose an AutoQEC scheme against excitation loss with low-order nonlinearities and accessible experimental resources. Our scheme is, in principle, device-independent and readily implementable in superconducting circuits and trapped-ion systems. The scheme is based on the squeezed cat (SC) encoding, which involves the superposition of squeezed coherent state. 
% The same encoding has been considered in Ref.~\cite{schlegel2022quantum}. Although the capability of correcting loss errors has been analyzed using the QEC conditions, and a mathematically optimized recovery channel has been calculated in Ref.~\cite{schlegel2022quantum}, it is not clear how the loss errors can be corrected in practice. 
We introduce an explicit AutoQEC scheme for the SC against loss errors by engineering a nontrivial dissipation, which simultaneously stabilizes the SC states and corrects the loss errors. We show that the engineered dissipation is close to the optimal recovery obtained using a semidefinite programming~\cite{reimpell2005iterative, fletcher2007optimum, noh2018quantum}. Notably, our proposed dissipation can be implemented with the same order of nonlinearity as that required by the two-component cat, which has been experimentally demonstrated in superconducting circuits~\cite{lescanne2020exponential} and shown to be feasible in trapped-ion systems~\cite{poyatos1996quantum}.
% As such, our scheme requires an the lowest-order nonlinearity among all AutoQEC schemes against loss errors.
% As such, our scheme is a practical AutoQEC scheme against loss errors that requires only experimental resources that have been previously demonstrated. 

Furthermore, we show that similar to the stabilized cat qubits, the stabilized SC qubits also possess a biased noise channel (with one type of error dominant over others), with an even larger bias (defined to be the ratio between the dominant error rate and the others) $\sim e^{\bar{n}^2}$ (compared to $\sim e^{\bar{n}}$ for the cat), where $\bar{n}$ denotes the mean excitation number of the codewords. Consequently, we can concatenate the stabilized SC qubits with a DV code tailored towards the biased noise to realize low-overhead fault tolerant QEC and quantum computation~\cite{tuckett2018ultrahigh, tuckett2019tailoring, tuckett2020fault, ataides2021xzzx, roffe2022bias, xu2022tailored}. We develop a set of operations for the SC that are compatible with the engineered dissipation and can preserve the noise bias needed for the concatenation. Compared to those for the cat~\cite{guillaud2019repetition}, these operations suffer less from the loss errors because of the AutoQEC. Moreover, they can be implemented faster due to a larger dissipation gap and a cancellation of the leading-order non-adiabatic errors. In combination, the access to higher-quality operations leads to much better logical performance in the concatenated level using the SC qubits. For instance, we can achieve one-to-two orders of magnitude improvement in the $\kappa_1/\kappa_2$ threshold, where $\kappa_1$ is the excitation loss rate and $\kappa_2$ is the engineered dissipation rate, for the surface-SC and repetition-SC scheme (compared to surface-cat and repetition-cat, respectively). Furthermore, the repetition-SC can easily achieve a logical error rate as low as $10^{-15}$, which already suffices for many useful quantum algorithms~\cite{fowler2012surface, o2017quantum}, even using a small SC with $\bar{n} = 4$ under a practical noise ratio $\kappa_1/\kappa_2 = 10^{-3}$. 
% We note that neither the biased-noise nature of the SC nor the concatenated schemes are considered in Ref.~\cite{schlegel2022quantum}. These aspects give the better use of the SC qubits, given that the SC qubits alone cannot achieve low enough logical error rates required for fault-tolerant QEC and computation.  

% Note that the same SC encoding was considered in Ref.~\cite{schlegel2022quantum}. In Ref.~\cite{schlegel2022quantum}, the SC's capability of correcting loss and dephasing errors was analyzed using the QEC conditions, and a mathematically optimized recovery channel was calculated. However, unlike our work, it is unclear how their recovery channel can be implemented in practice. Furthermore, they did not consider and explore the biased-noise structure of the SC, nor concatenation with outer DV codes using bias-preserving operations. We believe that concatenating the SC qubits while exploiting their biased-noise structure gives the better use of the SC qubits, given that the SC qubits alone cannot achieve low enough logical error rates required for fault-tolerant QEC and computation. 

We note that aspects of the SC encoding were also recently studied in Ref.~\cite{schlegel2022quantum}, with an emphasis on the enhanced protection against dephasing provided by squeezing (a point already noted in Refs.~\cite{teh2020overcoming, lo2015spin, le2018slowing}).  Unlike our work, Ref.~\cite{schlegel2022quantum} neither explored the enhanced noise bias provided by squeezing, nor exploited the ability to concatenate the SC code with outer DV codes using bias-preserving operations; as we have discussed, these are key advantages of the SC approach.  Our work also goes beyond Ref.~\cite{schlegel2022quantum} in providing an explicit, fully autonomous approach to SC QEC that exploits low-order nonlinearities, and it is compatible with several experimental platforms.  In contrast, Ref.~\cite{schlegel2022quantum} studied an approach requiring explicit syndrome measurements and a formal, numerically-optimized recovery operation. It was unclear how such an operation could be feasibly implemented in experiment. We also note that the SC has also been studied in the context of quantum transduction~\cite{lau2019high} (a very different setting than that considered here).

\section{Results}
\noindent
\textbf{Squeezed cat encoding} \\
The codewords of the SC are defined by applying a squeezing along the displacement axis (which is taken to be real) to the cat codewords:
\begin{equation}
    |SC_{r, \alpha^{\prime}}^{\pm}\rangle := \hat{S}(r)|C_{\alpha^{\prime}}^{\pm}\rangle
    \label{eq:squeezed_cat_codewords}
\end{equation}
where $|C_{\alpha^{\prime}}^{\pm}\rangle := \mathcal{N}_{\pm}(|\alpha^{\prime}\rangle + |-\alpha^{\prime}\rangle)$ with $\mathcal{N}_{\pm} = \frac{1}{\sqrt{2(1 \pm  e^{-2\alpha^{\prime 2}})}}$ being normalization factors, and $\hat{S}(r) := \exp [\frac{1}{2}r(\hat{a}^2 - \hat{a}^{\dagger 2})]$ is the squeezing operator. The above codewords with even ($|SC_{r,\alpha^{\prime}}^{+}\rangle$) and odd ($|SC_{r,\alpha^{\prime}}^{-}\rangle$) excitation number parity are defined to be the X-basis eigenstates. The performance of the code is related to the mean excitation number $\bar{n}$ of its codewords. For a code with fixed $\bar{n}$, the amplitude $\alpha^{\prime}$ of the underlying coherent states varies with the squeezing parameter $r$ as 
\begin{equation}
    \alpha^{\prime} \approx \sqrt{\bar{n} - \sinh^2 r} e^r,
    \label{eq:alpha_prime_relation}
\end{equation}
which holds for the regime of interest where $\alpha^{\prime} > 1$. Note that $\alpha^{\prime}$ is closely related to how separated in phase space the two computational-basis states are, which determines their resilience against local error processes. 
At fixed $\bar{n}$, $\alpha^{\prime 2}$ can be written as a concave quadratic function of $e^{2r}$, which has a maximum $\alpha^{\prime 2}_{\textrm{max}}=\bar{n}^{2} + \bar{n}$. 
% Here, we have taken $r$ and $\alpha^{'}$ to be real such that the squeezing and displacements are in orthogonal directions.

% \todo{$\alpha^{\prime 2} = \eta \bar{n} e^{2r} \approx 4 \eta (1 - \eta) \bar{n}^2$. Set $\eta = 1/4$. }

For the SC, it is convenient to consider the subsystem decomposition of the oscillator Hilbert space $\mathcal{H} = \mathcal{H}_L \otimes \mathcal{H}_g$, where $\mathcal{H}_L$ represents a logical sector of dimension 2 (which we refer to as a logical qubit) and $\mathcal{H}_g$ represents a gauge sector of infinite dimension (which we refer to as a gauge mode). Analogous to the modular subsystem decomposition of the GKP qubit~\cite{pantaleoni2020modular}, whose logical sector carries the modular value of the quadratures, the logical sector of the SC carries the parity information (excitation number modulo 2). We can choose a basis under the subsystem decomposition spanned by squeezed displaced Fock states $|\pm\rangle_L \otimes |\hat{\tilde{n}} = n\rangle_g \approx \mathcal{N}_{\pm, n} \hat{S}(r)[\hat{D}(\alpha^{\prime}) \pm (-1)^n \hat{D}(-\alpha^{\prime})]|n\rangle$ (we use $\approx$ since the right-hand side should be orthonormalized within each parity branch. See supplement.~\cite{SM} for details). By choosing this basis, the SC codewords in Eq.~\eqref{eq:squeezed_cat_codewords} coincide with $|\pm\rangle_L \otimes |\hat{\tilde{n}} = 0\rangle_g$, i.e., the codespace is the two-dimensional subspace obtained by projecting the gauge mode to the ground state. Furthermore, the bosonic annihilation operator $\hat{a}$ can be expressed as 
\begin{equation}
    \hat{a} = \hat{Z}_L \otimes (e^{-r}\alpha^{\prime} + \cosh r \hat{\tilde{a}} - \sinh r \hat{\tilde{a}}^{\dagger} ) + \mathcal{O}(e^{-2\alpha^{\prime 2}}),
    \label{eq:subsystem_decomp_a}
\end{equation}
where $\hat{Z}_L$ is the Pauli Z operator acting on the logical qubit, and $\hat{\tilde{a}} = \sum_{n = 0}^{\infty}\sqrt{n + 1}|\hat{\tilde{n}} = n\rangle_g \langle \hat{\tilde{n}} = n + 1|$ is the annihilation operator acting on the gauge mode.
% \jcmt{Should we use notation $\hat{a}_{g}$?} 

% Since the SC is equivalent to the cat up to a squeezing transformation, it is useful to move to the squeezed frame and transform any operator $\hat{O}$ as $\hat{O}_r = S^{\dagger}(r) \hat{O} \hat{S}(r)$. 

\begin{figure}[h!]
    \centering
    \includegraphics[width = 0.48 \textwidth]{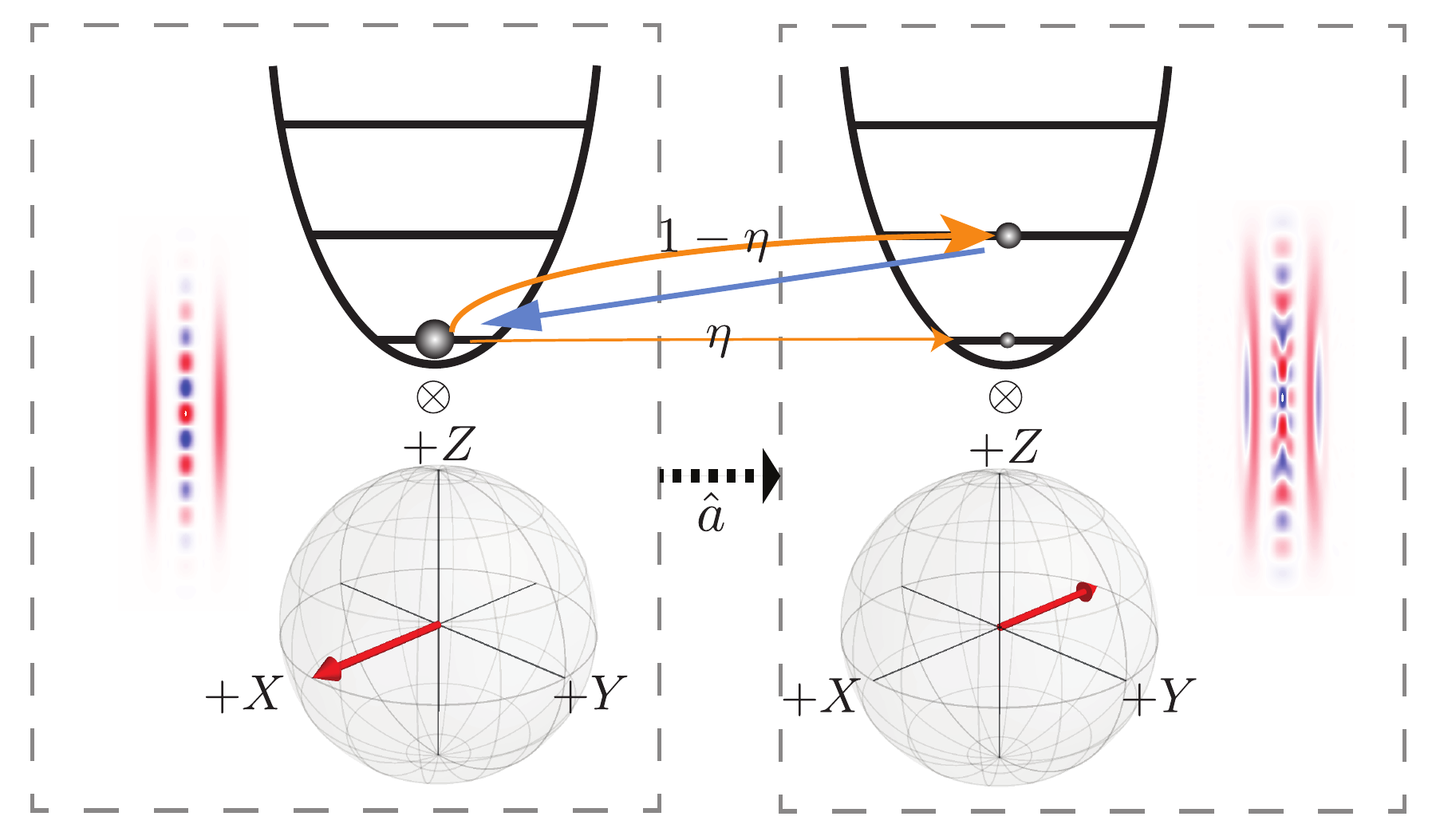}
     \caption{The illustration of a SC that suffers from a single excitation loss and then approximately corrects it. Each dashed box represents a state (visualized by the Wigner function) of the SC, which is decomposed as a product of a logical qubit and a gauge mode. A single excitation loss corrupts the codeword $|+\rangle_c$ (left) into the state $\hat{a}|+\rangle_c/\sqrt{\langle +|_c \hat{a}^{\dagger} \hat{a} |+\rangle_c}$ (right). During such a process, a phase flip happens on the logical qubit, and a fraction $1 - \eta$ of the gauge mode population gets excited (indicated by the thick orange arrow). The excited population can be detected and then corrected, as indicated by the blue arrow.}
    \label{fig:illus_diagram}
\end{figure}

Typtical bosonic systems suffer from excitation loss ($\hat{a}$), heating ($\hat{a}^{\dagger}$), and dephasing ($\hat{a}^{\dagger}\hat{a}$) errors, with loss being the prominent one. To understand why the SC code can correct excitation loss errors, we consider the Knill–Laflamme conditions~\cite{bennett1996mixed, knill1997theory} and evaluate the QEC matrices for loss errors~\cite{albert2018performance}. Consider a pure loss channel with a loss probability $\gamma$, the leading-order Kraus operators are $\{\hat{I}, \sqrt{\gamma} \hat{a}\}$. The detectability of a single excitation loss is quantified by the matrix:
% \begin{equation}
%     \hat{P}_{\textrm{code}} \hat{a} \hat{P}_{\textrm{code}} \approx \sqrt{\bar{n} - \sinh ^2 r} \hat{Z}_{c} + \mathcal{O}\left(e^{-2\bar{n}}\right)\hat{Y}_c,
%     \label{eq:QEC_matrix_detectability}
% \end{equation}
\begin{equation}
\begin{aligned}
    \hat{P}_{\textrm{code}} \hat{a} \hat{P}_{\textrm{code}} & = e^{-r} \alpha^{\prime} \frac{q + q^{-1}}{2}\hat{Z}_c + i e^r \alpha^{\prime} \frac{q - q^{-1}}{2}\hat{Y}_c \\
    & \approx \sqrt{\bar{n} - \sinh^2 r}\hat{Z}_c - i e^r \alpha^{\prime} e^{-2 \alpha^{\prime 2}} \hat{Y}_c,
\end{aligned}
    \label{eq:QEC_matrix_detectability}
\end{equation}
where $\hat{P}_{\textrm{code}}$ is the projection onto the code space, $\hat{Z}_c$ ($\hat{Y}_c$) is the Pauli $Z$ ($Y$) operator in the code space and $q := \sqrt{\frac{1 - e^{-2 \alpha^{\prime 2}}}{1 + e^{-2 \alpha^{\prime 2}}}}$. The approximation in the second line is made in the regime of interest where $\alpha^{\prime} \gg 1$. Eq.~\eqref{eq:QEC_matrix_detectability} indicates that a single excitation loss mostly leads to undetectable logical phase-flip errors, and the undetectability decreases with the squeezing parameter $r$. The increase of the detectability of single excitation loss events with the squeezing $r$ can be better understood by considering the action of the decomposed $\hat{a}$ operator (Eq.~\eqref{eq:subsystem_decomp_a}) on the codeword $\hat{a} (|+\rangle_L \otimes |0^{\prime}\rangle_g) = |-\rangle_L \otimes \sqrt{\bar{n}}(\sqrt{\eta}|0^{\prime}\rangle - \sqrt{1 - \eta}|1^{\prime}\rangle)$, where 
\begin{equation}
    \eta := (\bar{n} - \sinh^2 r)/\bar{n}.
    \label{eq:eta_definition}
\end{equation}
As shown in Fig.~\ref{fig:illus_diagram}, after a single excitation loss, the branch of the population (with ratio $\eta$) that stays in the ground state of the gauge mode leads to undetectable logical phase-flip errors. In contrast, the other branch (with ratio $1 - \eta$) that goes to the first excited gauge state is in principle detectable. The detectable branch is also approximately correctable since 
$\hat{P}_{\textrm{code}} \hat{a}^{\dagger} \hat{a} \hat{P}_{\textrm{code}} \approx \bar{n}\hat{I}_c + \mathcal{O}(e^{-2\alpha^{\prime 2}})\hat{X}_c$. Therefore, we expect that we can suppress the loss-induced phase flip errors by a factor $\eta$ that decreases with the squeezing $r$. 
% Note, however, that we can not perfectly distinguish between the loss errors and the dephasing errors $\kappa_{\phi}\mathcal{D}[\hat{a}^{\dagger} \hat{a}]$. (\todo{more details}) So the presence of both errors in the system leads to uncorrectable phase-flip errors. However, given that the loss rate $\kappa_1$ is much larger than the dephasing rate $\kappa_{\phi}$ in typical bosonic systems, we still expect a great reduction in the phase-flip rate due to the dominant loss errors. 
Moreover, the $\hat{X}_c$ and $\hat{Y_c}$ terms in the QEC matrices for both loss, heating, and dephasing are exponentially suppressed by $\alpha^{\prime 2}$. As shown in Eq.~\eqref{eq:alpha_prime_relation}, $\alpha^{\prime 2}$ can be greatly increased by adding squeezing (with $\alpha^{\prime 2}_{\textrm{max}} = \bar{n}^2 + \bar{n}$).
% ~\footnote{$\alpha^{\prime}$ first increases and then decreases with the squeezing $r$}. 
Consequently, we expect that the SC can also have significantly enhanced noise bias compared to the cat.  
% \jcmt{Move the Z part to next section?}

\noindent
\textbf{Autonomous quantum error correction} \\
% Fig. 2: Memory error rates
% (option: figure on realization)
While we have shown that the SC encoding can, in principle, detect and correct the loss errors, it remains a non-trivial task to find an explicit and practical recovery channel. In this section, we provide such a recovery channel, showing surprisingly that it requires only experimental resources that have been previously demonstrated.
% Here we propose an autonomous QEC scheme that corrects the detectable part of the loss errors. 
As shown by the blue arrow in Fig.~\ref{fig:illus_diagram}, we can, in principle, perform photon counting measurement on a probe field that is weakly coupled to the gauge mode, and apply a feedback parity flip $\hat{Z}_L$ on the logical qubit upon detecting an excitation in the probe field~\cite{gross2018qubit}. Such measurement and feedback process can be equivalently implemented by applying the dissipator:
\begin{equation}
    \hat{F} = (\hat{Z}_L\otimes \hat{\tilde{I}}) \hat{S}(r) (\hat{a}^2 - \alpha^{\prime 2})\hat{S}^{\dagger}(r).
    \label{eq:engineered_dissipator}
\end{equation}
When $\alpha^{\prime} \gg 1$, $\hat{F} \propto \hat{Z}_L \otimes \hat{\tilde{a}}$ represents a logical phase flip conditioned on the gauge mode losing an excitation. In the Fock basis, such an operator can be approximately given by
\begin{equation}
    \hat{F} \approx \frac{e^r}{\alpha^{\prime}} (c_1 \hat{a} + c_2 \hat{a}^{\dagger})\hat{S}(r)(\hat{a}^2 - \alpha^{\prime 2})\hat{S}^{\dagger}(r),
\end{equation}
with $c_1 + c_2 = 1$. In Methods, we propose two reservoir-engineering approaches to implement such a nontrivial dissipator 
using currently accessible experimental resources. We sketch the main ideas here. The first approach utilizes three bosonic modes that are nonlinearly coupled. As shown in Fig.~\ref{fig:physical_realization}(a), a high-quality mode $b$ and a lossy mode $c$, together, serve as a nonreciprocal bath that provides a directional interaction $e^{-i\theta \hat{Z}_L} \otimes \hat{\tilde{a}}$ from the gauge mode to the logical qubit in the storage mode $a$. Such a coupled system can be physically realized in, e.g., superconducting circuits~\cite{lescanne2020exponential, chamberland2022building}. 
% The dissipator can be implemented by engineering a nonreciprocal interaction $e^{-i\theta \hat{Z}_L} \otimes \hat{\tilde{a}}$ between the gauge mode and the logical qubit, which applies a qubit rotation conditioned on the gauge mode losing an excitation. Such an interaction can be engineered by introducing a high-quality ancilla mode b and a lossy mode c. By controlling the couplings between the b, c modes and the gauge ($a^{\prime}$) mode, an excitation on the gauge mode is swapped to the c mode through the b mode directionally and then dumped to the environment. While the excitation stays in the b mode, a dispersive coupling between the b mode and the logical qubit sector leads to a $Z_L$ logical rotation. In the regime where the coupling between the storage mode and the b, c modes are much weaker than the dissipation rate on the c mode, we can obtain the effective dissipator on the storage mode by simultaneously eliminating the b, c modes. 
% See Methods for details.
The second approach couples a bosonic mode nonlinearly to a qutrit $\{|g\rangle, |e\rangle, |f\rangle\}$. As shown in Fig.~\ref{fig:dissipation_realization_trapped_ion}, the bosonic mode is coupled to the $gf$ transition via $\hat{S}(r)(\hat{a}^2 - \alpha^{\prime 2})\hat{S}^{\dagger}(r)|f\rangle \langle g| + h.c.$ and to the $ef$ transition via $\hat{Z}_L |e\rangle \langle f| + h.c.$. By enhacing the decay from $|e\rangle$ to $|g\rangle$, we can obtain the effective dissipator $\hat{F}$ by adiabatically eliminating both $|e\rangle$ and $|f\rangle$. Such a system can be physically realized in, e.g., trapped-ion system~\cite{poyatos1996quantum}.   

With the engineered dissipator in Eq.~\eqref{eq:engineered_dissipator}, the SC can be autonomously protected from excitation loss, heating and dephasing. The dynamics of the system are described by the Lindblad master equation:
\begin{equation}
\begin{aligned}
    \frac{d \hat{\rho}}{dt} & = \kappa_2 \mathcal{D}[\hat{F}] \hat{\rho} + \kappa_1 (1 + n_{\textrm{th}}) \mathcal{D}[\hat{a}] \hat{\rho} \\
    & + \kappa_1 n_{\textrm{th}} \mathcal{D}[\hat{a}^{\dagger}] \hat{\rho} + \kappa_{\phi} \mathcal{D}[\hat{a}^{\dagger} \hat{a}] \hat{\rho},
    \label{eq:memory_dynamics}
\end{aligned}
\end{equation}
where $\mathcal{D}[\hat{A}]\hat{\rho} := \hat{A} \hat{\rho} \hat{A}^{\dagger} - \frac{1}{2}\{\hat{A}^{\dagger}\hat{A}, \hat{\rho}\}$. 
The logical phase-flip and bit-flip error rates of the SC under the dynamics described by Eq.~\eqref{eq:memory_dynamics} can be analytically obtained (see Methods for the derivations):
\begin{eqnarray} 
& \gamma_Z = [\kappa_1(1 + 2n_{\textrm{th}}) + \kappa_{\phi}e^{-2r}] (\bar{n} - \sinh^2 r), 
\label{eq:memory_error_rates_Z}\\
    & \nonpfrate{} = \kappa_{\phi} \frac{\left(\bar{n}-\sinh ^{2} r\right) e^{2 r}(\sinh ^{2} 2 r/4+\cosh 4 r)}{2\sinh \left(2\left(\bar{n}-\sinh ^{2} r\right) e^{2 r}\right)},
    \label{eq:memory_error_rates_X}
\end{eqnarray}
where $\gamma_{X,Y}$ denotes the sum of the logical $X$ and $Y$ error rates, which we refer to as the bitf-flip rate for simplicity~\footnote{Note that similar to the cat~\cite{puri2020bias}, the full error channel of the stabilized SC, which is analyzed in detail in supplement.~\cite{SM}, is not a Pauli error channel in general. For simplicity, we make the Pauli-twirling approximation only keeping the diagonal terms of the process matrix in the Pauli basis.}. We only consider the dephasing for $\nonpfrate{}$ since the loss-induced bit-flip rate has a more favorable scaling $\sim e^{-4\alpha^{\prime 2}} $with $\alpha^{\prime}$~\cite{guillaud2021error, chamberland2022building}. The loss and the heating contribute to $\gamma_Z$ in the same way (both suppressed by a factor $\eta$) since their undetectable portion ($\eta$) is the same (see Eq.~\eqref{eq:subsystem_decomp_a} and its hermitian conjugate). The dephasing also contributes to $\gamma_Z$, but with an extra $e^{-2r}$ suppression, when combined with the parity-flipping dissipator $\hat{F}$. See Methods for details. Setting $r = 0$ and removing the $\kappa_{\phi}$ term in $\gamma_Z$, we restore the error rates of the dissipative cat~\cite{guillaud2019repetition}. 

\begin{figure}[t]
    \centering
    \includegraphics[width = 0.48 \textwidth]{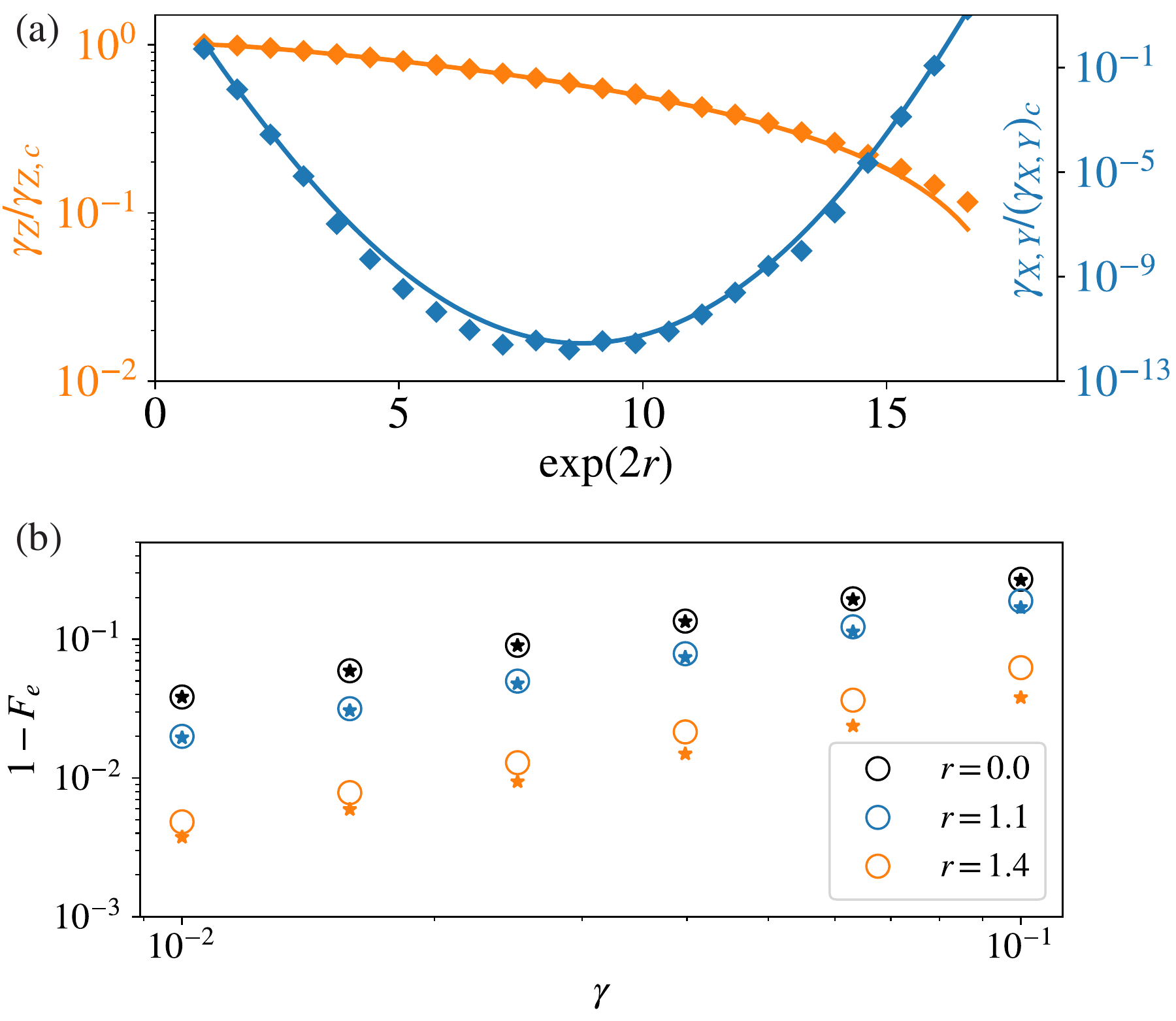}
     \caption{(a) The phase $\gamma_Z$ (orange) and bit $\nonpfrate{}$ (cyan) error rate of the dissipatively stabilized SC as a function of squeezing $r$ under the parameters $\bar{n} = 4, \kappa_1 = 100 \kappa_{\phi} = \kappa_2/100, n_{\textrm{th}} = 0.01$. The solid lines represent the analytical expressions Eqs.~\eqref{eq:memory_error_rates_Z} and \eqref{eq:memory_error_rates_X} while the diamonds represent the numerically extracted values. All the error rates are normalized by those of the dissipative cat $\gamma_{Z, c}, (\gamma_{X, Y})_c$, which are given by Eqs.~\eqref{eq:memory_error_rates_Z} and \eqref{eq:memory_error_rates_X} with $r = 0$. (b)
    %  \jcmt{color?} 
     The entanglement infidelity of a joint loss and recovery channel varying with the loss probability $\gamma$ for the SC encoding with $\bar{n} = 4$. The recovery channel is either the engineered dissipation (the circles) or the optimal recovery channel determined by an SDP program~\cite{reimpell2005iterative, fletcher2007optimum, noh2018quantum} (the stars).} 
    \label{fig:memory_error_rates}
\end{figure}

In the regime where $e^{-r} \ll 1$ and $\gamma_Z$ is mainly contributed by excitation loss, we can simplify Eqs.~\eqref{eq:memory_error_rates_Z} and \eqref{eq:memory_error_rates_X} as
\begin{equation}
    \gamma_Z \approx \eta \bar{n} \kappa_1, \nonpfrate{} \approx \frac{9}{16}\kappa_{\phi} \alpha^{\prime 2} e^{-2 \alpha^{\prime 2}} e^{4r},
\end{equation}
where
\begin{equation}
    \alpha^{\prime} \approx \sqrt{4 \eta (1 - \eta)}\bar{n}.
    \label{eq:alpha_approximation}
\end{equation} 
As plotted in Fig.~\ref{fig:memory_error_rates}(a), fixing $\bar{n}$, $\gamma_Z$ decreases monotonically with the squeezing $r$ (unless $r$ approaches the maximum squeezing $r_{\textrm{max}} \approx \sinh^{-1}(\sqrt{\bar{n}})$. See Methods for details) as the undetectable portion $\eta$ of the loss-induced errors decreases (see Eq.~\eqref{eq:eta_definition}). The change of $\nonpfrate{}$ with $r$ (or equivalently, $\eta$) is roughly captured by the change in the displacement amplitude $\alpha^{\prime}$ (see Eq.~\eqref{eq:alpha_approximation}), and $\nonpfrate{}$ takes the minima roughly when $\alpha^{\prime}$ reaches the maxima $\alpha^{\prime}_{\textrm{max}} = \sqrt{\bar{n}^2 + \bar{n}}$. Note that the minimal bit-flip rate of the SC enjoys a more favorable scaling $\nonpfrate{} \propto e^{-2\bar{n}^2}$ with $\bar{n}$, compared to $\nonpfrate{} \propto e^{-2\bar{n}}$ for the cat, so that the SC can have a much larger noise bias under the same excitation number constraint. 
% As shown in Fig.~\ref{fig:memory_error_rates}(a), as $r$ passes the optimal value for $\nonfprate{}$, $\nonpfrate{}$ starts increasing while $\gamma_Z$ keeps decreasing.
In principle, one needs to consider the tradeoff between $\gamma_Z$ and $\alpha^{\prime}$ and choose the optimal $\eta$ depending on the tasks of interest. 
% Certain tasks, e.g. idling, prefer smaller $\eta$ since the system suffers less from excitation losses. 
% While certain operations, e.g. the bias-preserving CX gate (introduced in the next section), prefer larger $\alpha^{\prime}$ since the dissipation gap ($\propto \alpha^{\prime 2}$) is larger, which supports faster operations. 
Smaller $\eta$ leads to better protection from excitation losses, which is preferred by, e.g., the idling operations. Larger $\alpha^{\prime}$, on the other hand, leads to a larger noise bias and a widened dissipation gap~\footnote{the nonzero eigenvalue of the Lindbladian $\mathcal{D}[\hat{F}]$ with the
smallest real part, which characterizes the rate of the population decaying into the steady-state manifold.} ($\propto \alpha^{\prime 2}$), which can support faster operations, e.g., the bias-preserving CX gate introduce in the next section. In the following, we fix $\bar{n} = 4$ and $\eta = 1/4$ if not specified otherwise, which corresponds to a squeezing of $r=1.32$ (11.5 dB). Such a parameter choice leads to $\gamma_Z \approx \kappa_1$, which removes the enhancement factor $\bar{n}$ present for the cat (for $\bar{n} = 4$). Meanwhile, $\alpha^{\prime 2} \approx \frac{3}{4}\bar{n}^{2}$ provides a sufficiently large noise bias and dissipation gap.

In Fig.~\ref{fig:memory_error_rates}(b), we benchmark the performance of our Auto-QEC scheme against loss errors by comparing it to the optimal recovery channel given by a semidefinite programming (SDP) method~\cite{reimpell2005iterative, fletcher2007optimum, noh2018quantum}. 
We consider the joint channel $\mathcal{N} = \mathcal{D} \cdot \mathcal{N}_{\gamma} \cdot \mathcal{E}$, where $\mathcal{E}$ denotes the encoding map from a qubit to the SC, $\mathcal{N}_{\gamma}$ denotes a Gaussian pure loss channel (corresponding to Eq.~\eqref{eq:memory_dynamics} with $\kappa_2 = \kappa_{\phi} = n_{\textrm{th}} = 0$) with loss probability $\gamma := \kappa_1 t$, and $\mathcal{D}$ denotes the recovery channel either using the autonomous QEC with the dissipator Eq.~\eqref{eq:engineered_dissipator} or the optimal recovery channel. We evaluate the entanglement infidelity (EI) $1 - F_e$, where $F_e$ denotes the entanglement fidelity, of the joint channel $\mathcal{N}$, as a function of the loss probability $\gamma$. Note that the EI is the objective function for the SDP. As shown in Fig.~\ref{fig:memory_error_rates}(b), the EI obtained using the Auto-QEC is close to the optimal EI, especially in the low-$\gamma$ regime, demonstrating that our proposed autonomous QEC scheme is close to optimal for correcting excitation loss errors. We note that it is crucial to have the phase-flip $\hat{Z}_L$ correction in the dissipator $\hat{F}$ in order to correct the loss-induced phase-flip errors. Otherwise, a simple dissipator $\hat{S}(r)(\hat{a}^2 - \alpha^{\prime 2})\hat{S}^{\dagger}(r)$ directly generalized from the dissipative cat would still give an unsuppressed phase-flip rate $\gamma_Z = \kappa_1 \bar{n}$.
% we evaluate the entanglement infidelity~\cite{} of the joint channel $\mathcal{R} = \mathcal{D} \cdot \mathcal{N}_{\gamma} \cdot \mathcal{E}$

We note that the SC encoding also emerges as the optimal or close-to-optimal single-mode bosonic code through a bi-convex optimization (alternating SDP) procedure for a loss and dephasing channel with dephasing being dominant, as shown in Ref.~\cite{leviant2022quantum}.

\noindent
\textbf{Bias-preserving operations} \\
To apply the autonomously protected SC for computational tasks, we need to develop a set of gate operations that are compatible with the engineered dissipation. Furthermore, the operations should preserve the biased noise channel of the SC, which can be utilized for resource-efficient concatenated QEC and fault-tolerant quantum computing~\cite{bonilla2021xzzx, xu2022tailored, roffe2022bias, chamberland2022building, darmawan2021practical}. Following the literature for the cat and the pair-cat~\cite{guillaud2019repetition, puri2020bias, xu2022engineering, yuan2022construction}, we develop a set of bias-preserving operations $\mathcal{B} = \{\mathcal{P}_{|\pm\rangle_{c}}, \mathcal{M}_{X}, X, Z(\theta), ZZ(\theta), \text { CNOT, Toffoli} \}$ for the SC, which suffice for many concatenated QEC schemes (e.g. concatenation with the repetition codes or the surface codes). The detailed design of each operation is presented in Methods.
\begin{figure}[t]
    \centering
    \includegraphics[width = 0.45 \textwidth]{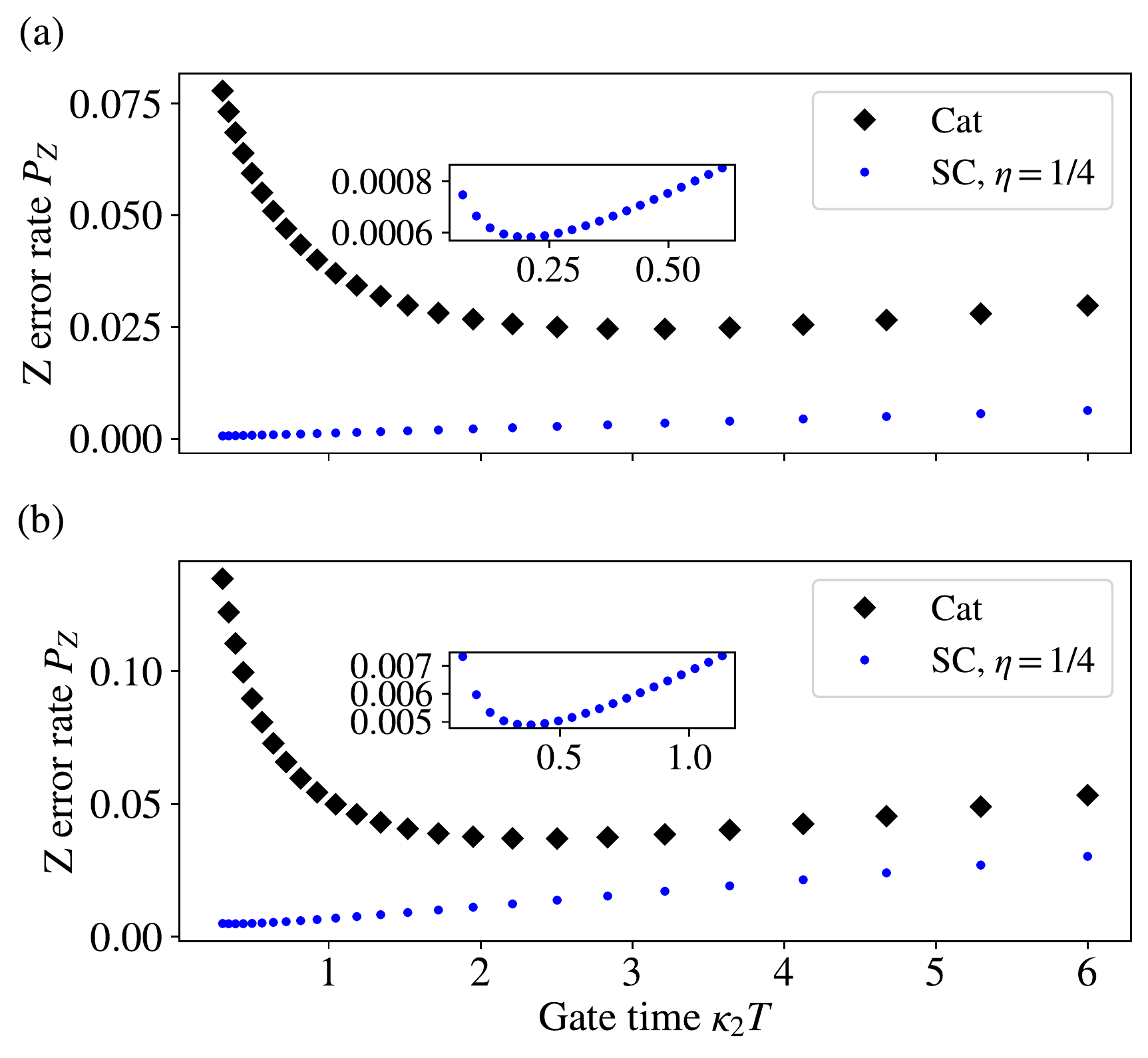}
    %  \caption{
    %  Comparison of the original cat gate scheme with the modified gate scheme with SC encoding of optimized $T_{cool}, T_{gate}, r$ while fixing $\bar{n} = 4$. The inset diagram is a zoomed-in plot of the SC gates' errors around their optimal points. (a) Z-error of a CX gate. (b) Z-error of a Z rotation of angle $\pi$.}
     \caption{The total $Z$ error probability of the $Z(\pi)$ gates (a) and the CX gates (b) as a function of the gate time. For the CX gate, $p_Z := p_{Z_c} + p_{Z_t} + p_{Z_c Z_t}$ is the sum of the control-mode, target-mode, and the correlated phase flip rates. $\kappa_{1}/\kappa_{2}$ is fixed at $10^{-3}$. The blue lines represent the gates on the cat qubits~\cite{guillaud2019repetition}, and the red lines represent our proposed gates on the SC qubits with $\eta = 1/4$. $\bar{n}$ is chosen as 4 for both cat and SC. The insets are the zoomed-in error rates of the SC gates around the optimal gate times.}
    \label{fig:gate_errors}
\end{figure}

Overall, the bias-preserving operations for the SC can achieve much higher fidelity (lower dominant $Z$-type error rates) than those for the cat for the following two reasons: (1). The operations suffer less from the excitation loss errors, which are (partially) autonomously corrected. (2). The non-adiabatic errors are significantly suppressed by the $\hat{Z}_L$ correction in the dissipator $\hat{F}$ (see Eq.~\eqref{eq:engineered_dissipator}) and the enlarged dissipation gap ($\propto \alpha^{\prime 2}$), so that the gate operations could be implemented faster. In Fig.~\ref{fig:gate_errors}, we show the total $Z$-type error rates for the Z-axis rotation $Z(\theta)$ and the CX gate as a function of the gate time. Compared to the cat gates in Ref.~\cite{guillaud2019repetition} with the same $\bar{n}$, the SC $Z(\theta)$ (CX) gate can achieve a 42.0 (7.56) times reduction in the lowest error rates. While we have fixed $\eta = 1/4$ as mentioned in last section, it is not necessarily the optimal choice of the squeezing. In fact, with $\eta$ approaching $1/2$, we obtain even lower errors at faster gate times.

We note that compared to the cat stabilized by $\hat{a}^2 - \alpha^2$ in the literature, a simple extension to a SC stabilized by $\hat{S}(r)(\hat{a}^2 - \alpha^{\prime 2}) \hat{S}^{\dagger}(r)$ can also lead to improvement in the gate speed and fidelities due to the enlarged dissipation gaps. However, adding the extra phase flip in the dissipator brings a much more significant improvement due to the suppression of the loss-induced errors and the leading-order non-adiabatic errors. See Discussion for more details.

% \todo{Complement the whole bias-preserving operation set.}

\begin{figure*}
    \centering
    \includegraphics[width = 1.0 \textwidth]{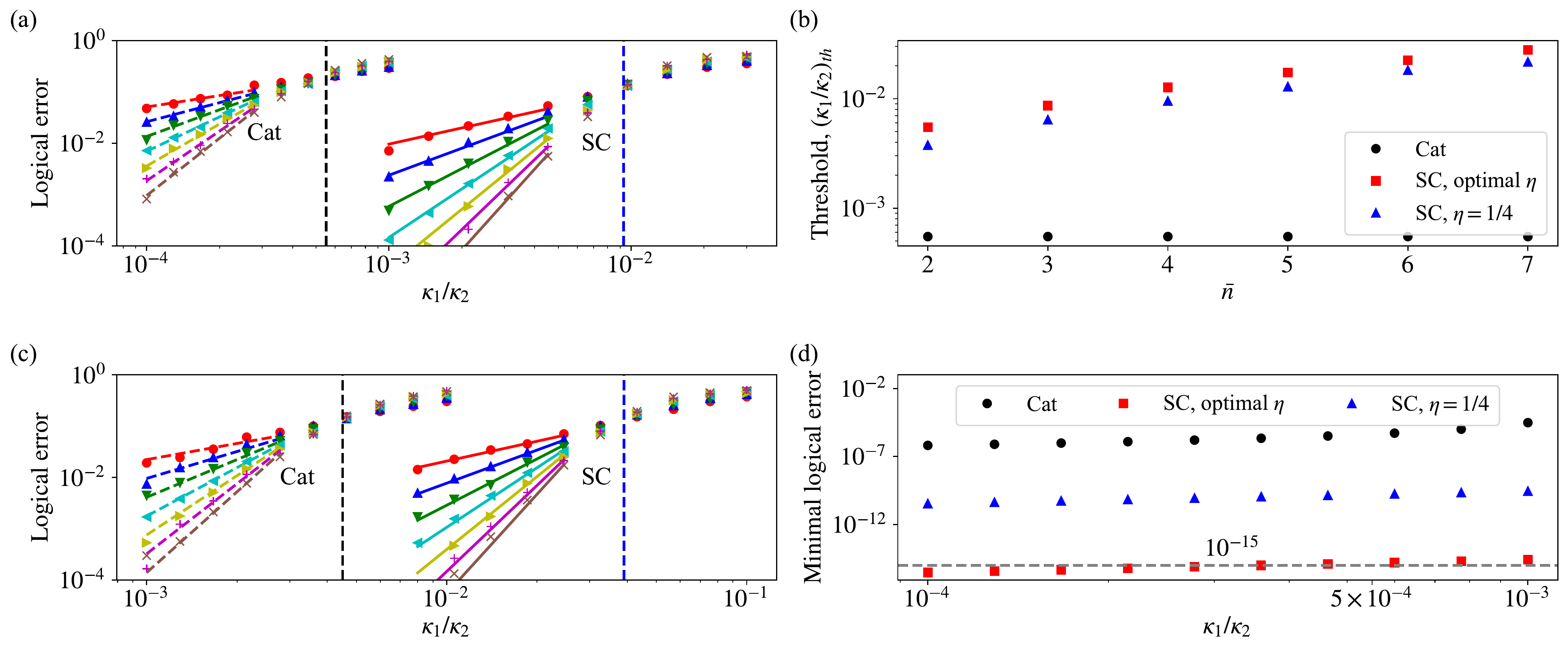}
     \caption{Logical errors of the SC and the cat concatenated with repetition codes or surface codes. (a) Surface code logical $Z$ error probabilities for a range of code distance $d_{Z} = 3,5,7,..., 15$ (from red to brown) with fixed $d_{X}=3$. The SC is fixed to $\bar{n} = 4, \eta = 1/4$. The dashed lines indicate the threshold values of $\kappa_1/\kappa_2$. (b) Surface code thresholds in $\kappa_{1}/\kappa_{2}$ varying with the average excitation number of the SC or the cat. (c) Repetition code logical $Z$ error probabilities for a range of code size $d_{Z}$. (d) Repetition code minimum total logical error probabilities at $\kappa_1/\kappa_2 = 10^{-4}$, under the long gate time constraint $T\geq 1/\kappa_2$. The cat we consider has an average excitation number $\bar{n} = 8$. The logical error probabilities for both the surface codes and the repetition codes are obtained from Monte Carlo simulations of $d_Z$ code cycles and one final round of perfect stabilizer measurement. We use the same minimum-weight-perfect-matching (MWPM) decoder as described in Ref.~\cite{chamberland2022building}}
    \label{fig: concatenated codes}
\end{figure*}

\noindent
\textbf{Concatenated quantum error correction} \\
With the bias-preserving operations, we can concatenate the SC with an outer discrete-variable code to suppress the logical error rates to the desired level. To compare the SC with the standard cat, we follow the literature and consider the concatenation with a repetition code~\cite{guillaud2019repetition} and a thin rotated surface code~\cite{chamberland2022building}. The surface-cat scheme can arbitrarily suppress the errors in a resource-efficient manner once the ratio between the loss rate $\kappa_2$ and the engineered dissipation rate $\kappa_2$ is below a certain threshold. The repetition-cat, on the other hand, cannot arbitrarily suppress the errors for a cat with constrained  $\bar{n}$. Below a $\kappa_1/\kappa_2$ threshold, as the repetition code size increases, the logical $Z$ error rate is exponentially suppressed while the logical $X$ error is linearly amplified. Thus, a minimal total logical error rate is present. The concatenated QEC schemes with the cat are yet challenging since the $\kappa_1/\kappa_2$ thresholds (e.g., $\sim 5\times 10^{-4}$ for the surface-cat, see Fig.~\ref{fig: concatenated codes}(a)) are very demanding because of the low-fidelity bias-preserving operations. Also, the minimal logical error probability of the repetition-cat (e.g., $\sim 10^{-5}$ for $\bar{n} = 8$, see Fig.~\ref{fig: concatenated codes}(d)) is still too high for fault-tolerant algorithms even using a relatively large cat because of the limited noise bias. In this section, we show that the $\kappa_1/\kappa_2$ thresholds for both the surface code and the repetition code can be significantly improved by concatenating with the dissipative SC. Moreover, the repetition-SC can reach sufficiently low logical error probability $\sim 10^{-15}$ even using a small SC $\bar{n} = 4$ (see Fig.~\ref{fig: concatenated codes}(d)).

% For the cat, it has been shown that once the ratio between the loss rate $\kappa_2$ and the engineered dissipation rate $\kappa_2$ is below a certain threshold, the concatenated codes can arbitrarily suppress the errors in a resource-efficient manner. However, this $\kappa_1/\kappa_2$ threshold (e.g., $\sim 4\times 10^{-4}$ for the surface-cat code~\cite{chamberland2022building}) is very demanding, which is one of the major challenges for the concatenated QEC schemes with cat. In this section, we show that the $\kappa_1/\kappa_2$ thresholds can be significantly improved by using the dissipative SC. The concatenated surface-SC scheme demonstrates a 17-to-51 times increase (depending on $\bar{n}$) in the $\kappa_1/\kappa_2$ threshold compared to the surface-cat. The repetition-SC scheme has about an order-of-magnitude increase of the $\kappa_1/\kappa_2$ threshold as well as a drastic reduction of the achievable minimal logical error rates. 

We first consider the concatenation of the SC with a $d_X$ by $d_Z$ thin surface code. Similar to Ref.~\cite{chamberland2022building}, we fix the X distance $d_X$ to 3, which suffices to suppress the logical $X$ error rate, and increase the Z distance $d_Z$ to suppress the logical $Z$ error rate. 
% We only consider the errors from the CX gates (\todo{which is justified by the fact that the CX gate is the most erroneous operation}) and extract the logical $Z$ error rate from a circuit-level Monte Carlo simulation:
% \begin{equation}
%     p_{L}^{Z} \approx 0.1\left(\frac{p_{Z}}{0.019}\right)^{\left(d_{Z}+1\right) / 2},
% \end{equation}
% where $p_Z$ denotes the total $Z$ error probability of each CX gate. The $1.9\%$ threshold for $p_Z$ also agrees with that obtained in Ref.~\cite{chamberland2022building}. Since $p_Z$ is a function of the dimensionless parameters $\kappa_1/\kappa_2, \kappa_2 T,  \kappa_2 T_{\textrm{cool}}, \bar{n}$ and $r$, the logical Z error rate is a function of  $\kappa_1/\kappa_2, \kappa_2 T, \kappa_2 T_{\textrm{cool}}, \bar{n}, r$ and $d_Z$. 
At fixed $\eta = 1/4$, we obtain the logical $Z$ error probability for $d_Z$ code cycles as a function of $\kappa_1/\kappa_2$ for different $d_Z$, as shown in Fig.~\ref{fig: concatenated codes}(a). The physical error rates of each physical operation involved in the surface-code QEC are presented in supplement~\cite{SM}. We obtain a $\kappa_1/\kappa_2$ threshold at $0.93\%$, which is around 20 times higher than that of the surface-cat~\cite{chamberland2022building}. Note that by optimizing over the choice of the squeezing, the maximum threshold we obtained for $\bar{n} = 4$ is around $1.2\%$.
% Given an intrinsic noise parameter $\kappa_1/\kappa_2$, a code $Z$ distance $d_Z$, and fixed mean excitation number $\bar{n}$, we can find the minimal logical $Z$ error rate $p_L^Z$ by numerically optimizing over the gate time $\kappa_2 T, \kappa_2 T_{\textrm{cool}}$ and the squeezing $r$.  In Fig.~\ref{fig: concatenated codes}(a) we plot the minimal $p_L^Z$ as a function of $\kappa_1/\kappa_2$ for different $d_Z$ under the constraint $\bar{n} = 4$ and observe a $\kappa_1/\kappa_2$ threshold higher than $1\%$.
% Compared to the surface-cat scheme in Ref.~\cite{chamberland2022building}, the $\kappa_1/\kappa_2$ threshold of the surface-SC is increased by around 30 times. 
Moreover, in Fig.~\ref{fig: concatenated codes}(b) we show that this threshold can be further increased to about $3\%$ by increasing $\bar{n}$ to $7$. 
% Here, we relaxed the constraint on $\eta$. 
% As a comparison, in the surface-cat scheme, the increase of excitation number $\bar{n}$ only increases the bias but has no effect on the $\kappa_1/\kappa_2$ threshold. 
Note that the $\kappa_1/\kappa_2$ threshold of the surface-cat remains the same when increasing $\bar{n}$. We attribute the increase of the $\kappa_1/\kappa_2$ threshold (for the concatenated SC schemes) to the reduced physical-operation error rates (see the previous section).
% \onecolumngrid

Next we consider the concatenation of the SC with a repetition code with size $d_Z$. 
% At large $\kappa_1/\kappa_2$, the total logical error is dominated by the logical Z error rate. 
As shown in Fig.~\ref{fig: concatenated codes}(c), we obtain a $3.9\%$ $\kappa_1/\kappa_2$ threshold for the logical $Z$ error rate (again, see supplement~\cite{SM} for the physical error rates used for the simulation), which is roughly 9 times higher than that of the repetition-cat. Below the $\kappa_1/\kappa_2$ threshold, as previously mentioned, a minimal total logical error rate is present.
To obtain the minimal total logical error rate (by optimizing over $d_Z$), we find approximate expressions for the logical $Z$ and $X$ error probabilies in the sub-threshold regime ($\kappa_1/\kappa_2 < 10^{-3}$):
% For small $\kappa_1/\kappa_2$, the total logical error rate is limited mostly by the logical X error rate, which is linearly amplified by increasing the code size. To obtain an expression that well describes the error rates of CX gate, we restrict the CX gates to be in the long gate time regime where $\kappa_2 T \geq 1$. In this regime, we numerically observe that the total X error rate of CX gate fits nicely to by $p_X \approx 5.57 \times \frac{e^{-2 \alpha^{\prime 2}}}{\alpha^{\prime 2}}\frac{1}{\kappa_2 T}$. The Z error rate of CX, $p_Z$, is dominated by loss-induced error in this regime. To obtain a better fit of the logical error rate, we throw away the control qubit error, which has negligible effect in our repetition code implementation. With CX gates subject to the long gate time constraint, the logical $Z$ and $X$ error rate of a repetition code with length $d_Z$ can be well described by:
\begin{equation}
    \begin{aligned}
p_{L}^{Z} &\approx 0.059d_Z\left(\frac{p_{Z}^{\prime}}{0.056}\right)^{0.48d_Z}, \\
p_{L}^{X} &\approx 2d_Z(d_Z-1)p_{X, Y},
\end{aligned}
\end{equation}
where $p_Z^{\prime} := p_{Z_t} + p_{Z_{c} Z_t}$ denotes the sum of the target-mode and the correlated phase-flip rate of the CX gate (phase flips on the control mode have negligible contribution to the logical error rate for the repetition code), $p_{X, Y}$ the total non-$Z$ error rates of the CX gate (the total rates of all the two-qubit Pauli errors that do not contain $Z$ terms). $p_Z^{\prime}$ and $p_{X, Y}$ are in general functions of the CX gate time $\kappa_2 T$. To obtain simple expressions for them, we restrict the CX gate time to be $\kappa_2 T \geq 1$. In this regime, we have $p_Z^{\prime} \approx \kappa_1 \bar{n} T$, $p_{X, Y} \approx 5.57 \times \frac{e^{-2 \alpha^{\prime 2}}}{\alpha^{\prime 2}}\frac{1}{\kappa_2 T}$. 
% The logistics for fitting to the logical Z error expression above is discussed in \cite{chamberland2022building}. \jcmt{discuss briefly here?}

In Fig.~\ref{fig: concatenated codes}(d), we plot the minimal total logical error probability $p_L = p_L^Z + p_L^X$ of the repetition-SC by optimizing $d_Z$ and $\kappa_2 T$ for $\bar{n} = 4$ and $\eta = 1/4$. As a comparison, we also included minimal logical error probabilities of the repetition-cat with $\bar{n} = 8$ using the physical error rates in Ref.~\cite{chamberland2022building}.
% \jedit{Since the error could be forbiddingly small to simulate numerically, we limit the gate time to be in the long gate time regime and apply the numerically obtained relation of $p_{X}$}.
% Moreover, the minimal logical error rates could be further decreased if the long gate time constraint is relaxed. See Methods for details. 
For a practical noise ratio $\kappa_1/\kappa_2 = 10^{-3}$, the minimal logical error probability of the repetition-SC can reach $\sim 10^{-15}$, which suffices for many useful quantum computational tasks~\cite{fowler2012surface, o2017quantum}. In contrast, the logical error probability of the repetition-cat can only reach $\sim 10^{-5}$, which is far from being useful. To reach a similar level of logical error probability as the repetition-SC, we need either a much larger cat with $\bar{n} \gg 10$ (with the repetition code), or a more sophisticated outer code, e.g., the surface code.  We attribute the drastic reduction in the minimal logical error rate of the repetition-SC, compared to the repetition-cat, to the significantly enhanced noise bias (or equivalently, the reduced physical bit-flip rates) of the SC. 

% We numerically observe the logical $Z$ and $X$ error rate of a repetition code with length $d_Z$ can be well described by \jcmt{update formula?}:
% \begin{equation}
%     \begin{aligned}
% p_{L}^{Z} &=0.067\left(\frac{p_{Z}}{0.059}\right)^{\frac{d_Z+1}{2}}, \\
% p_{L}^{X} &=2d_Z(d_Z-1)p_{X},
% \end{aligned}
% \end{equation}
% where, again, $p_Z$ and $p_X$ denotes the total physical $Z$ and $X$ probability of the CX gates, respectively. In the long gate time regime where $\kappa_2 T \geq 1$, we numerically observe that the physical bit flip rate is well described by $p_X = p_X^{\textrm{NA}} \approx 1.38 \times \frac{e^{-2 \alpha^{\prime 2}}}{\alpha^{\prime 2}}\frac{1}{\kappa_2 T}$. At large $\kappa_1/\kappa_2$, the total logical error is dominated by the logical Z error rate. We obtain a $3.9\%$ $\kappa_1/\kappa_2$ threshold, as shown in Fig.~\ref{fig: concatenated codes}(c). Note that this threshold value can be further improved in principle by optimizing all the parameters. For small $\kappa_1/\kappa_2$, the total logical error rate is limited mostly by the logical X error rate, which is linearly amplified by increasing the code size. 

\section{Discussion}
\noindent
\textbf{Additional comparisons with the cat}\\
Although in this work we benchmark the performance of the concatenated codes as a function of $\kappa_1/\kappa_2$ for both the cat and the SC, it might be of different difficulty level to engineer the same dissipation rate $\kappa_2$ for the cat and the SC, depending on the hardware implementation. Therefore, we can compare the performance of the concatenated codes as a function of $\kappa_1/M$, where $M$ is the physical rate that is most challenging to engineer in practice. Here we focus on the implementation in superconducting circuits.

For example, stabilizing a cat with $\bar{n}_c = \alpha^2$ requires the nonlinear coupling $g_2 (\hat{a}^2 - \alpha^2 )\hat{c}^{\dagger} + h.c.$ between the cat and a lossy mode $c$ with a loss rate $\kappa_c$. The adiabatic elimination condition requires that $2\alpha g_2 \ll \kappa_c$, which can be satisfied by setting $2\alpha g_2 = \epsilon \kappa_c$ where $\epsilon \ll 1$ is a constant. Then $\kappa_2$ is proportional to $\kappa_c$: $\kappa_2 = \frac{4 g_2^2}{\kappa_c} = \epsilon^2 \frac{\kappa_c}{\alpha^2}$. If the major hardware challenge is to engineer large $\kappa_c$~\footnote{For instance, $\kappa_{c}$ is limited by filter bandwidth in Ref.~\cite{chamberland2022building}.}, we can benchmark the performance of the concatenated codes as a function of $\kappa_1/\kappa_c$. For the SC, we have $\kappa_2 = \epsilon^2 \frac{\kappa_c}{\alpha^{\prime 2}}$.
% then the engineered dissipation rate $\kappa_2 = \frac{4 g_2^2}{\kappa_c} = \epsilon^2 \frac{\kappa_c}{\alpha^2}$ is limited by both $\kappa_c$ and $\alpha^2$. Similarly for the SC, $\kappa_2 = \epsilon^2 \frac{\kappa_c}{\alpha^{\prime 2}}$ is limited by $\kappa_c$ and $\alpha^{\prime 2}$. In other words, to engineer the same $\kappa_2$, the SC (cat) requires a bath loss rate $\kappa_c$ to be proportional to $\alpha^{\prime 2}$ ($\alpha^2 = \bar{n}_c$).
Using these relations between $\kappa_2$ and $\kappa_c$, we can compare the $\kappa_1/\kappa_c$ thresholds instead of $\kappa_1/\kappa_2$ in Fig,~\ref{fig: concatenated codes}(a). The $\kappa_1/\kappa_c$ threshold of the surface-SC is still 12 times that of the surface cat. 

Another potential hardware challenge is to engineer strong nonlinear couplings. In this case, we can compare the concatenated codes as a function of $\kappa_1/J_{\textrm{max}}$, where $J_{\textrm{max}}$ denotes the largest nonlinear coupling strength required. For the cat, $J_{\textrm{max}}$ is simply given by $g_2$ and $\kappa_2 = \frac{2\epsilon J_{\textrm{max}}}{\alpha}$. For the SC, since the challenging nonlinear coupling is in the form (see Eq.~\eqref{eq:full_three_modes_Hamiltonian} in Methods) $J \hat{S}(r) (\hat{a}^2 - \alpha^{\prime 2}) \hat{S}(r)^{\dagger} = J [\cosh^2 r \hat{a}^2 + \sinh^2 r \hat{a}^{\dagger} + \sinh r \cosh r \hat{a}^{\dagger} \hat{a} + \sinh r \cosh r - \alpha^{\prime 2}]\hat{c}^{\dagger} + h.c.$, the largest coupling strength is $J_m = J\cosh^2 r = \frac{\alpha^{\prime} \cosh^2 r \kappa_2}{2\epsilon}$. Using these relations, we can change the horizontal axis in Fig.~\ref{fig: concatenated codes}(a) to $\kappa_1/J_m$ and obtain about a 7 times increase in the $\kappa_1/J_m$ threshold for the surface-SC compared to the surface-cat.

% strength $g_2$, then $\kappa_2 = \frac{2\epsilon g_2}{\alpha}$ is limited by both $g_2$ and $\alpha$ for the cat. Then, to obtain a target $\kappa_2$, the required nonlinear coupling strength $J_m = g_2 = \frac{\alpha \kappa_2}{2\epsilon}$ is proportional to $\alpha$. For the SC, since the required nonlinear coupling is in the form (see Eq.~\eqref{eq:full_three_modes_Hamiltonian} in Methods) $J \hat{S}(r) (\hat{a}^2 - \alpha^{\prime 2}) \hat{S}(r)^{\dagger} = J [\cosh^2 r \hat{a}^2 + \sinh^2 r \hat{a}^{\dagger} + \sinh r \cosh r \hat{a}^{\dagger} \hat{a} + \sinh r \cosh r - \alpha^{\prime 2}]\hat{c}^{\dagger} + h.c.$, the largest coupling strength is $J_m = J\cosh^2 r = \frac{\alpha^{\prime} \cosh^2 r \kappa_2}{2\epsilon}$. Accordingly, we can change the horizontal axis in Fig.~\ref{fig: concatenated codes}(a) to $\kappa_1/J_m$. Under this consideration, the $\kappa_1/J_m$ threshold for the surface-SC is around 7 times that of the surface-cat.
% % we replotx Fig.~\ref{fig: concatenated codes}(a) with horizontal axis rescaled to $\kappa_{1}/J_{m}$ in Fig.~\ref{fig: concatenated codes alternative schemes}(b). Under this consideration, the threshold for SC is around 7 times that of cat.

% \noindent
% \textbf{Concatenated QEC with larger SC}\\
\noindent
\textbf{Comparison with the squeezed cat stabilized by a parity-preserving dissipator}\\
To better understand the novelty and necessity of the partity-flipping dissipator $\hat{F}$ we introduced in Eq.~\eqref{eq:engineered_dissipator}, we compare it with a parity-preserving dissipator
\begin{equation}
\hat{F}^{\prime}=\hat{S}(r) (\hat{a}^2 - \alpha^{\prime 2})\hat{S}^{\dagger}(r) \approx 2\alpha^{\prime}\hat{I}_{L}\otimes \hat{\tilde{a}},
\label{eq:engineered_dissipator_parity_preserving}
\end{equation}
which is a straightforward extension from $\hat{a}^2 - \alpha^2$ that stabilizes the cat. We show that the extra phase-flip correction in $\hat{F}$ is essential for reducing SC's error rate in both the memory level and gate operations, which then leads to better logical performance in the concatenated level. 

In the memory level, the change of a parity flip on the dissipator does not affect the bit-flip error rate we derived in Eq.~\eqref{eq:memory_error_rates_X}. Nevertheless, $\hat{F}^{\prime}$ lacks the parity flip $Z_{L}$ that corrects the detectable portion of the loss-induced errors, as shown clearly from Fig.~\ref{fig:illus_diagram} (the missing of the blue arrow). Therefore, a SC stabilized by $\hat{F}^{\prime}$ is not capable of correcting the loss errors. As such, it suffers from the same phase-flip error rate as a cat, $\gamma_{Z} = \kappa_1 \bar{n}$.

Regarding the gate operations, we take the Z rotation and the CNOT gate as examples. For the Z rotation, a SC stabilized by $\hat{F}^\prime$ only enjoys a suppression in the non-adiabatic errors by the the increased adiabatic gap, $4\kappa_2 \alpha^{\prime 2}$, compared to conventional cat of the same $\bar{n}$. In contrast, a SC stabilized by $\hat{F}$ corrects the leading-order non-adiabatic error in $1/\alpha^{\prime 2}$, since the the extra $\hat{Z}_L$ in $\hat{F}$ compensates the parity-flip associated with the non-adiabatic transition (to the leading order). The residual errors are proportion to the correction factor, $\xi\propto 1/\alpha^{\prime 2}$, as discussed in Methods (see Eq.~\eqref{eq:non_adiabatic_error_Z_rotation}). Therefore, while the minimal $Z(\theta)$ gate error for the SC with $\hat{F}^{\prime}$ is roughly suppressed by a factor $1/\bar{n}$ compared to the cat, that for the SC with $\hat{F}$ is suppressed by an $1/\bar{n}^2$ factor (see Table~\ref{tab:gate_error_comp}).
% The ratios of the gate errors are presented in Table.~\ref{tab:gate_error_comp}, where all errors are normalized by the respective gate error of cat. 

\begin{table}[h]
    \centering
    \begin{tabular}{|p{2cm}<{\centering}|p{2.5cm}<{\centering}|p{2.5cm}<{\centering}|}
         \hline
         Normalized gate error & SC with $\hat{F}^{\prime}$ & SC with $\hat{F}$\\
         \hline 
         $Z(\theta)$ & $1/(\bar{n}+1)$ & $\sim \bar{n}^{-2}$\\
         \hline 
         CNOT & $ 2/\sqrt{\bar{n}+1}$ & $\sim \bar{n}^{-3/2}$\\
         \hline 
    \end{tabular}
    \caption{Optimal gate error rate of the SC gates compared to the cat. All errors are normalized by the optimal gate errors of the cat, which are given by $p_{Z(\theta)} = \frac{\theta}{2}\sqrt{\frac{1}{\bar{n}}\frac{\kappa_1}{\kappa_2}}$ and $p_{\textrm{CNOT}} = \frac{\pi}{2\sqrt{2}}\sqrt{\frac{\kappa_1}{\kappa_2}}$~\cite{chamberland2022building}. The definitions of $\hat{F}$ and $\hat{F}^{\prime}$ are given in Eq.~\eqref{eq:engineered_dissipator} and Eq.~\eqref{eq:engineered_dissipator_parity_preserving} respectively. The gate errors are optimized over both gate time and the squeezing for SC. Since the cooling time is mostly assumed to be constant in our gate scheme, it is neglected for simplicity. We only provide the scaling of the gate errors with $\bar{n}$ for the SC since the exact expressions are complicated, as shown in Methods.}
    \label{tab:gate_error_comp}
\end{table}

The errors of CNOT operation can be analyzed in a similar fashion. 
% Nevertheless, we are adopting a different gate design for CNOT gates. The new gate design has no dissipator on the target qubit to enable faster gates and lower error rates. \jcmt{Moreover, the conventional CNOT scheme of cat cannot be naively extended to SC since the rotating squeezing direction would require a dissipator of high order nonlinearity.}
% The new gate scheme and the widened adiabatic gap altogether contributed to the gate error of SC with $\hat{F}^{\prime}$, whose minimal value is a factor of $\frac{2}{\sqrt{\bar{n}+1}}$ smaller than that of cat.
Due to the enlarged adiabatic gap, the minimal $Z$ error rate of our SC gate with $\hat{F}^{\prime}$ is a factor of $\frac{2}{\sqrt{\bar{n}+1}}$ smaller than that of the cat gate~\cite{guillaud2019repetition}.
For the mean excitation number we consider, $\bar{n}=4$, this factor is only slightly less than 1. However, with the parity-flipping dissipator $\hat{F}$, the gate error enjoys a $\eta$ suppression in the loss errors and an additional $\propto 1/\alpha^{'2}$ suppression in the non-adiabatic error. Combining these advantages, the CNOT gate error ratio with that of the cat roughly scales as $\bar{n}^{-3/2}$ (see Table~\ref{tab:gate_error_comp}).

% Since the fault-tolerant threshold is mostly limited by errors of the CNOT and the idling operation, the threshold of the concatenated SC schemes using $\hat{F}^{\prime}$ is comparable to that of the concatenated cat scheme even at optimal squeezing for small mean excitation number. As such, having the extra phase-flip correction in the dissipator $\hat{F}$ is crucial for concatenated QEC and fault-tolerant quantum computing. 

Since the fault-tolerant threshold is mostly limited by errors of the CNOT and the idling operation, the thresholds of the concatenated SC schemes using $\hat{F}^{\prime}$ is comparable to that of the concatenated cat scheme even at optimal squeezing for small mean excitation number. As such, having the extra phase-flip correction in the dissipator $\hat{F}$ is crucial for concatenated QEC and fault-tolerant quantum computing. 

\noindent
\textbf{Applications in superconducting circuits and trapped-ion systems}\\
The stabilized cat qubits have been considered as a candidate for hardware-efficient, fault-tolerant, and scalable computation tasks in superconducting circuits~\cite{chamberland2022building, darmawan2021practical}. The dissipative SC, which we show has an overall advantage over the cat, could play an important role along this direction. 

The dissipative SC could also find its application in trapped-ion systems. On the one hand, encoding into the motional states of the ions provides an alternative approach for storing and protecting the quantum information. How to process the information (e.g., implementing quantum gates) remains to be explored. On the other hand, if the information is stored in the internal states of the ions (the conventional approach), the bosonic codes like the SC could lead to more robust information processing. One could utilize multi-species ions~\cite{inlek2017multispecies, bruzewicz2019dual} with multiple levels~\cite{ringbauer2022universal} and dissipatively protect the motional modes while leaving a subset of the ions' internal states that carry the information intact. The protected motional modes can, for instance, be used for scalable, parallel, and high-quality entangling gates mediated by localized phonon modes~\cite{olsacher2020scalable}.

% \section{Conclusion and outlook}
% \todo{change to discussion}
% In this work we propose an AutoQEC scheme against the major error sources in bosonic systems using the SC code. Compared to the dissipative cat, the dissipatively stabilized SC has a suppressed dominant error rate and an enhanced noise bias. Furthermore, it supports faster operations with lower error rates. In combination, the SC can lead to concatenated QEC and fault-tolerant quantum computing with higher noise thresholds and less overhead. 

% In addition to its application in superconducting circuits, our scheme could also contribute to trapped-ion systems. \todo{more about superconducting circuits} On the one hand, encoding into the motional states of the ions provides an alternative approach for storing and protecting the quantum information. How to process the information (e.g., implementing quantum gates) remains to be explored. On the other hand, if the information is stored in the internal states of the ions (the conventional approach), the bosonic codes like the SC could lead to more robust information processing. One could utilize the multi-level structure \todo{with potentially multi-species with references} of the ions~\cite{ringbauer2022universal} and dissipatively protect the motional mode while leaving a subset of the ions' internal states that carry the information intact. The protected motional modes can, for instance, be used for scalable, parallel, and high-quality entangling gates mediated by localized phonon modes~\cite{olsacher2020scalable}. 

\section{Methods}
\noindent
\textbf{Physical realization of the dissipator} \\
In this section, we present the details of the two approaches implementing the dissipator in Eq.~\eqref{eq:engineered_dissipator}. 
% The first approach utilizes three bosonic modes that are nonlinearly coupled. One of the mode is made lossy to serve as the reservoir. Such a system can be physically realized in, e.g., superconducting circuits~\cite{lescanne2020exponential, chamberland2022building}. The second approach couples a bosonic mode nonlinearly to a qutrit $\{|g\rangle, |e\rangle, |f\rangle\}$, with one of the transition $|e\rangle \rightarrow |g\rangle$ being lossy. Such a system can be physically realized in, e.g., trapped-ion system~\cite{poyatos1996quantum}.   
Before describing our recipes, it is worth discussing the challenges involved here. The most straightforward method to realizing a generic Lindblad dissipator 
$ \mathcal{D}[\hat{F}]$ is to couple the system to an auxiliary reservoir mode $c$ 
(with decay rate $\kappa_c $) via a coupling Hamiltonian
$ g (\hat{F} \hat{c} ^\dag 
+ h.c. ) $. In the limit where mode $c$ acts as a Markovian environment for the system, i.e.~$\kappa_c \gg g$, we realize the target dissipator $\hat{F}$ with an effective dissipation rate $4g^2 / \kappa_c $. For the dissipator in Eq.~\eqref{eq:engineered_dissipator}, this simple route requires a strong fourth-order nonlinear coupling, which has not been demonstrated yet due to the experimental challenges. 

Here we present two approaches for realizing the desired nonlinear dissipator using accessible experimental resources: The first approach utilizes three nonlinearly-coupled bosonic modes, which can be physically realized in, e.g., superconducting circuits~\cite{lescanne2020exponential, chamberland2022building}; The second approach couples a bosonic mode nonlinearly to a qutrit, which can be physically realized in, e.g., trapped-ion system~\cite{poyatos1996quantum}.   
% the first route straightforwardly implements the reservoir engineering protocol but requires a higher-order nonlinearity; the second route, in contrast, requires only third-order nonlinear couplings but involves a second auxiliary mode. 

\begin{figure}[t]
    \centering
    \includegraphics[width = 0.45 \textwidth]{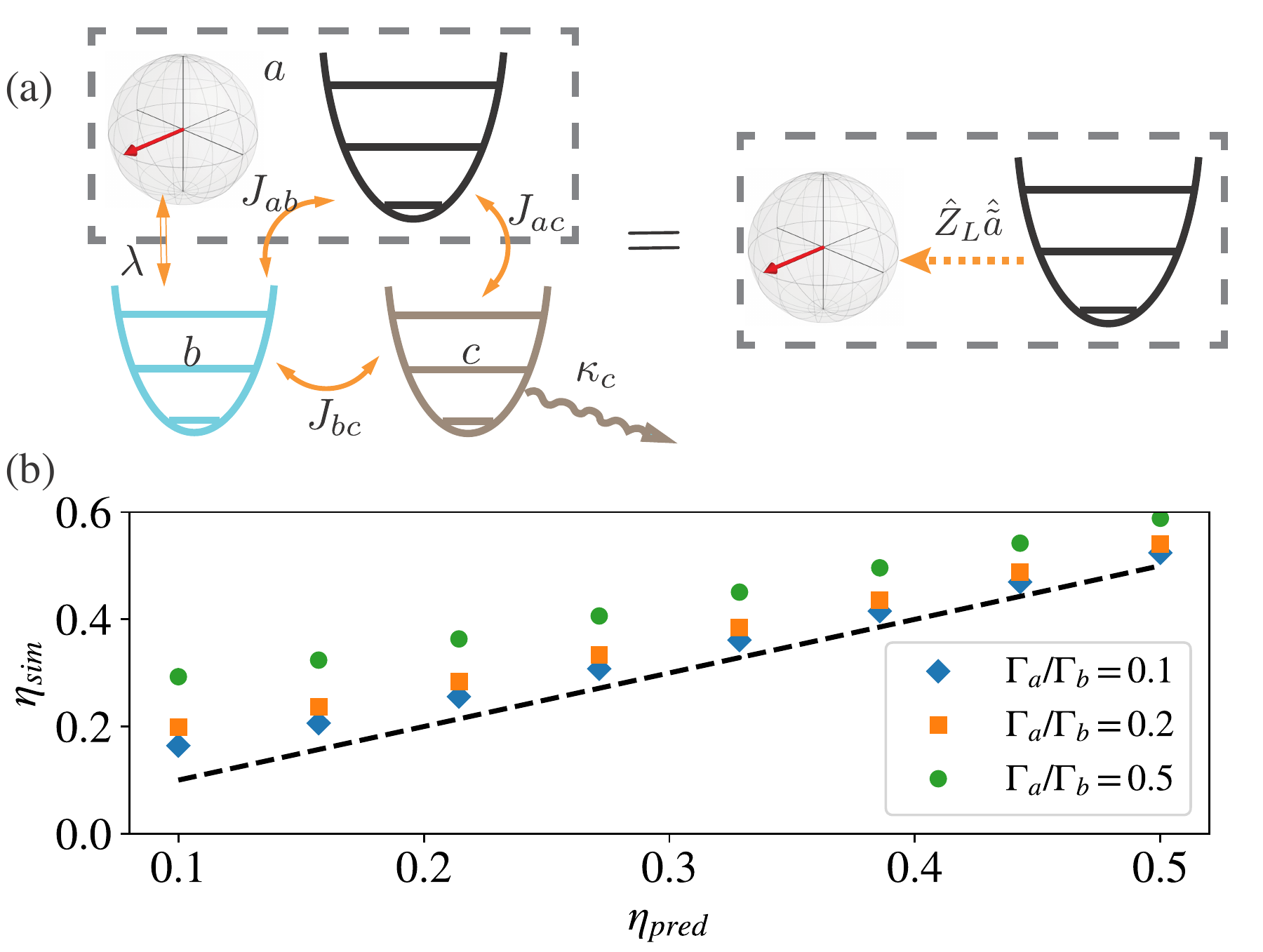}
     \caption{(a) Realization of the parity-flipping dissipator $\hat{Z}_L\otimes \hat{\tilde{a}}$ using three nonlinearly coupled bosonic modes. (b) Comparison between the numerically extracted $\eta$ ($\eta_{\textrm{sim}}$) and the theoretically predicted $\eta$ ($\eta_{\textrm{pred}}$ in Eq.~\eqref{eq:eta_definition}) for a range of finite $\Gamma_a/\Gamma_b$. The dashed line indicates the ideal case where $\eta_{\textrm{sim}} = \eta_{\textrm{pred}}$.}
    \label{fig:physical_realization}
\end{figure}

The first approach only requires third-order nonlinearities to implement our desired dissipator, making use of a more structured engineered dissipation proposed in Ref.~\cite{wang2022qnr}. Under the subsystem decomposition of the storage mode $a$ encoding the SC, one can realize a general nonlinear dissipator of the form 
$ \mathcal{D}[e^{- i \theta\hat{Z}_L} \otimes \hat{\tilde{A}}]$ (with an angle $\theta$), by coupling a gauge-mode operator $\hat{\tilde{A}}$ and an auxiliary mode $b$ to the input and output ports of a directional waveguide, respectively, and introducing a dispersive interaction between an auxiliary mode $b$ and the logical qubit: $\hat{H} _{\mathrm{disp.}}= \frac{\lambda}{2} 
\hat{Z}_L \hat{b}^{\dagger} \hat{b}$.
For the dissipator in Eq.~\eqref{eq:engineered_dissipator}, 
we choose $\hat{\tilde{A}} = \hat{\tilde{a}}$. The physical interactions (in the Fock basis) can be obtained from the mapping 
$\hat{\tilde{a}} \to \frac{1}{2 \alpha^{\prime}} \hat{S}(r)\left(\hat{a}^{2}-\alpha^{\prime 2}\right) \hat{S}^{\dagger}(r)$, and 
$\hat{Z}_L \to \frac{1}{2 \alpha^{\prime}} \hat{S}(r)\left(\hat{a}+\hat{a}^{\dagger}\right) \hat{S}^{\dagger}(r)$, which means that we need a nonlinear coupling between the storage mode $a$ and the waveguide port. While it is challenging to directly achieve this using e.g.~a physical circulator, the directional dynamics can be synthetically engineered by adding another reservoir mode $c$. The whole setup is illustrated in Fig.~\ref{fig:physical_realization}(a), whose dynamics is given by master equation 
\begin{equation}
    \frac{d}{dt}\hat{\rho} = 
    -i [\hat{H} _{\mathrm{disp.}} 
    + \hat{H} _{\mathrm{tun.}} , \hat{\rho}] + \kappa_c \mathcal{D}[\hat{c}],
\label{eq:full_three_modes_Lindbladian}
\end{equation}
where the tunnel coupling Hamiltonian $ \hat{H} _{\mathrm{tun.}} $ of the total system-reservoir is given by 
\begin{align} 
\label{eq:full_three_modes_Hamiltonian}
\hat{H} _{\mathrm{tun.}}&  =J_{ab} 
\hat{\tilde{a}}^{\dagger} \hat{b}
+ 
( J_{ac} \hat{\tilde{a}}
-i J_{bc} \hat{b} ) \hat{c}^{\dagger}+h . c .
, \\
J_{ab} = & {\sqrt{\Gamma_a \Gamma_b}} / {2}
, \, 
J_{ac} =  {\sqrt{\Gamma_a \kappa_c}} / {2}
, \, J_{bc} =  {\sqrt{\Gamma_b \kappa_c}} /{2}
. 
\end{align}
In the regime where the joint $b,c$ modes act as a Markovian reservoir for mode $a$, i.e.~$\kappa_c \gg \sqrt{\Gamma_a \Gamma_b}$ and $\Gamma_b \gg \Gamma_a$, we can adiabatically eliminate both $b$ and $c$ to obtain an effective dissipator (using the effective operator formalism~\cite{reiter2012effective}), as
\begin{equation}
\begin{aligned}
\frac{d}{dt}\hat{\rho}  &=\Gamma_{a} \mathcal{D}\left[\frac{i \lambda \hat{Z}_L -\Gamma_{b}}{i \lambda \hat{Z}_L +\Gamma_{b}} \hat{\tilde{a}}\right] \hat \rho. \\
\end{aligned}
\label{eq:effective_operator_three_modes}
\end{equation}
Setting $\lambda = \Gamma_b$, we obtain the desired dissipator $\hat{Z}_L \otimes \hat{\tilde{a}}$ to stabilize the SC (see supplement.~\cite{SM} for a detailed derivation).

Note that Eq.~\eqref{eq:full_three_modes_Hamiltonian} still involves nonlinear couplings between the $a,b$ and $a,c$ modes. Fortunately, all the nonlinear terms are now cubic, which can be realized by parametrically pumping nonlinear elements (e.g., ATS~\cite{chamberland2022building}, SNAIL~\cite{grimm2020stabilization}) that couples different modes with appropriate drive tones. We note that the nonlinearity required in Eq.~\eqref{eq:full_three_modes_Hamiltonian} has been demonstrated for stabilizing a cat~\cite{lescanne2020exponential, chamberland2022building}. 
% \todo{Add more remarks}

% \begin{figure}[h!]

When deriving Eq.~\eqref{eq:effective_operator_three_modes}, we require the physical setup Eq.~\eqref{eq:full_three_modes_Lindbladian} to operate in the regime where adiabatic elimination remains valid. It is thus natural to ask what are the imperfections given realistic physical parameters, i.e.~when the decay rates $ \kappa_c $, $\Gamma_b $ of auxiliary modes $b,c$ cannot be infinitely large. In that case, one can show the dominating error due to finite reservoir bandwidth is due to the finite decay rates $ \kappa_c $ and $ \Gamma_b $, and it is preferable to set  $ \kappa_c \sim \Gamma_b $ to optimize over hardware resources (see supplement.~\cite{SM} for details). In this regime, the extra error introduced by physical implementation is determined by the ratio $\Gamma_a/\Gamma_b$, which heuristically describes the branching ratio between the logical qubit population that does not undergo the parity flip (uncorrected error) and the population that does (corrected error) whenever a gauge mode excitation decays into the environment. More specifically, as shown in supplement~\cite{SM}, we can approximately derive the discrepancy between the desired suppression factor for the loss-induced phase flip rate $\eta_{\textrm{pred}}$ (using Eq.~\eqref{eq:eta_definition}) and the numerically extracted (achievable) value $\eta_{\textrm{sim}}$, as $\eta_{\textrm{sim}} - \eta_{\textrm{pred}} = (1 - \eta_{\textrm{pred}}) ({\Gamma_a}/{2\Gamma_b})
$. 
% + \eta_{\textrm{res}}  $.  
% (\todo{See supplementary material for derivations}). 
% As shown in Fig.~\ref{fig:physical_realization_error_rates}(b), our analytical argument is further verified by the full system simulation of Eq.~\eqref{eq:full_three_modes_Lindbladian}. 
As shown in Fig.~\ref{fig:physical_realization}(b), by setting $\Gamma_a/\Gamma_b = 0.1$, we can realize the desired $\eta$ within $50\%$ accuracy. 

% \begin{figure}[h]
%     \centering
%     \includegraphics[width = 0.5 \textwidth]{Physical_Implementation_error_rates.pdf}
%      \caption{(a) Comparison between the numerically extracted $\eta$ ($\eta_{\textrm{sim}}$) and the theoretically predicted $\eta$ ($\eta_{\textrm{pred}}$) for a range of finte $\Gamma_a/\Gamma_b$. The dashed line indicates the ideal case where $\eta_{\textrm{sim}} = \eta_{\textrm{pred}}$. (b) The difference between $\eta_{\textrm{sim}}$ and $\eta_{\textrm{pred}}$ as a function of $\Gamma_a/\Gamma_b$ when $\eta_{\textrm{pred}} = 0.2$. }
%     \label{fig:physical_realization_error_rates}
% \end{figure}

\begin{figure}[t]
    \centering
    \includegraphics[width = 0.45 \textwidth]{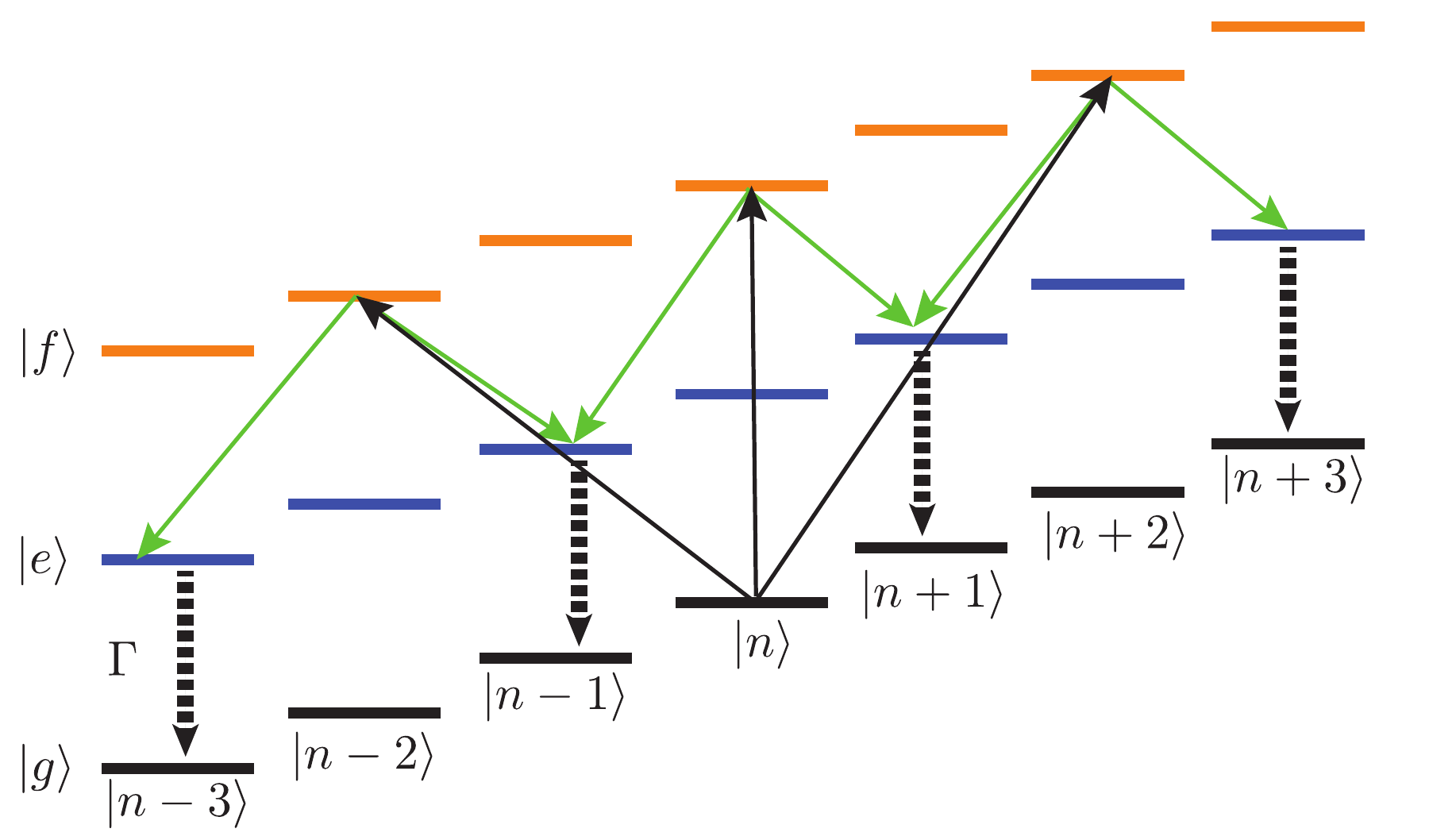}
     \caption{Laser configuration for the coupling Hamilotnian in Eq.~\eqref{eq:coupling_Hamiltonian} for implementing the SC in trapped-ion system. The motional mode of the ion is coupled to three internal states via the sideband transitions, represented by the black and the green arrows. Starting from $|g\rangle \otimes |\psi\rangle$ ($|\psi\rangle$ is an arbitrary motional state), the system goes through a two-step coherent transition $|g\rangle \otimes |\psi\rangle \rightarrow |f\rangle \otimes \hat{F}_1 |\psi\rangle \rightarrow |e\rangle \otimes \hat{F}_2 \hat{F}_1 |\psi\rangle$ (indicated by the black and the green solid arrows, respectively) and decays rapidly to $|g\rangle \otimes \hat{F}_2 \hat{F}_1 |\psi\rangle$ (indicated by the black dashed arrows). Here $\hat{F}_1 \propto \hat{S}(r) (\hat{a}^2 - \alpha^{\prime 2})\hat{S}^{\dagger}(r)$ and $\hat{F}_2 \propto \hat{S}(r)\hat{a}\hat{S}^{\dagger}(r)$. Adiabatically eliminating the $|e\rangle, |f\rangle$ states, we obtain the effective dissipator on the motional mode $\hat{F} = \hat{F}_2 \hat{F}_1$. }
    \label{fig:dissipation_realization_trapped_ion}
\end{figure}

Now we present the second approach for implementing the dissipator $\hat{F} = \frac{1}{\alpha^{\prime}}\hat{S}(r)  \hat{a} (\hat{a}^2 - \alpha^{\prime 2}) \hat{S}^{\dagger}(r)$ using a coupled boson-qutrit system. Note that a simpler dissipator stabilizing a cat $\hat{a}^2 - \alpha^2$ was obtained using a coupled boson-qubit system in trapped-ion platform in Ref.~\cite{poyatos1996quantum}. However, the dissipator $\hat{F}$ cannot be directly engineered using their approach since there are many frequency-degenerate terms, e.g., $\hat{a}$ and $\hat{a}^{\dagger} \hat{a}^2$, that cannot be independently controlled by a single sideband drive. To resolve this, we generalize their approach by introducing a third internal level of the ion, and implementing the dissipator $\hat{F}$ in two steps associated with different electronic transitions. Specifically, we use the motional mode of the ions in a 1D harmonic trap as the bosnonic mode, which is coupled to three internal levels $|g\rangle, |e\rangle$ and $|f\rangle$ via several laser beams:
\begin{equation}
    \frac{d}{dt}\hat{\rho} = -i[\hat{H}_{\textrm{eff}}, \hat{\rho}] + \mathcal{J}\hat{\rho},
\end{equation}
where $\hat{H}_{\textrm{eff}} = \nu \hat{a}^{\dagger}\hat{a} + \omega_e |e\rangle \langle e| + \omega_f |f\rangle \langle f| +  \frac{1}{2}\Omega_0 (|f\rangle \langle g|e^{-i\omega_f t} + h.c.) +  \hat{H}_{\textrm{coup}} - i\frac{\Gamma}{2}|e\rangle \langle e|$, with 
\begin{equation}
\begin{aligned}
    \hat{H}_{\textrm{coup}}
    & = \sum_{i=1}^3 \Omega_i \cos{[\eta_0 (\hat{a} + \hat{a}^{\dagger})]}(|f\rangle \langle g|e^{-i(\omega_f + \delta_i)t} + h.c.) \\
    & + \sum_{i = 4}^5\Omega_i \sin{[\eta_0 (\hat{a} + \hat{a}^{\dagger})]}(|e\rangle \langle f|e^{-i(\omega_e - \omega_f + \delta_i)t} + h.c.),
  \end{aligned}
\label{eq:coupling_Hamiltonian}
\end{equation}
and
\begin{equation}
\begin{aligned}
J \hat{\rho}=& \Gamma \int_{-1}^1 d u N(u) e^{-i \eta_0 u\left(\hat{a}+\hat{a}^{\dagger}\right)} \\
& \times |g\rangle \langle e| \hat{\rho} |e\rangle \langle g| e^{i \eta_0 u\left(\hat{a}+\hat{a}^{\dagger}\right)}.
\end{aligned}
\label{eq:momentum_kick}
\end{equation}
Here $\nu$ is the trap frequency, $\eta_0$ the Lamb-Dick parameter, $\Gamma$ the engineered decay rate from $|e\rangle$ to $|g\rangle$, and $N(u)$ the normalized dipole pattern. $\hat{H}_{\textrm{coup}}$ describes the coupling between the motional mode and the internal states, illustrated in Fig.~\ref{fig:dissipation_realization_trapped_ion}, and $\mathcal{J}\hat{\rho}$ describes the spontaneous emission of the ion from $|e\rangle$ to $|g\rangle$ and its associated momentum kicks. The drive with amplitude $\Omega_0$ in $\hat{H}_{\textrm{eff}}$ comes from a laser that is coupled to the ion along a constrained (transverse) direction, thereby only driving the internal transitions. By tuning the laser detunings $\delta_1 = - 2\nu, \delta_2 = 2\nu, \delta_3 = 0, \delta_4 = - \nu$, and $\delta_5 = \nu$, and choosing appropriate driving strength $\{\Omega_i\}$ (see supplement.~\cite{SM}), we can obtain a coupling Hamiltonian (neglecting the fast-rotating terms):
\begin{equation}
\begin{aligned}
    \hat{H}_{\textrm{coup}} & = \Omega^{\prime}_{gf} \hat{S}(r) (\hat{a}^2 - \alpha^{\prime 2})\hat{S}^{\dagger}(r) |f\rangle \langle g| \\ 
    & + \Omega^{\prime}_{ef} \frac{1}{\alpha^{\prime}} \hat{S}(r) \hat{a} \hat{S}^{\dagger}(r) |e\rangle \langle f| + h.c.
\end{aligned}
\end{equation}
In the regime where $2\alpha^{\prime} \Omega_{gf}^{\prime} \ll \Gamma$, $\Omega^{\prime}_{gf} \ll \Omega^{\prime}_{ef}$, we can obtain a reduced dynamics on the motional mode by adiabtically eliminating the $|e\rangle, |f\rangle$ states:
\begin{equation}
    \frac{d}{dt}\hat{\rho}_m = \kappa_2 \mathcal{D}[\hat{F}]\hat{\rho}_m,
\end{equation}
where $\hat{\rho}_m$ is the reduced density matrix on the motional mode. Through numerically simulations we find that we can obtain the dissipator $\hat{F}$ with the desired rate by setting $\Omega^{\prime}_{ef} = 0.5 \Gamma, \Omega^{\prime}_{gf}/\Omega^{\prime}_{ef} = 1/20$. A large $\kappa_2$, therefore, demands large $\Gamma$ and driving strength. Note that we have assumed that $\Gamma$ and $\{\Omega_i\}, i = 1,2,3,4,5$ are much smaller than $\nu$, so that the off-resonant terms can be safely neglected (secular approximation). In practice, however, one might be able to go beyond this weak-drive regime by carefully cancelling the effects from the off-resonant terms. We have also neglected the effects from the momentum kicks here, which only lead to a small increase in the phase-flip suppression factor $\eta \rightarrow \eta + \mathcal{O}(\eta_0^2)$. See supplement.~\cite{SM} for a more detailed analysis. We stress that our proposed approach requires the same order of nonlinearity as that required by a two-component cat, which has been considered to be feasible in trapped-ion system~\cite{poyatos1996quantum}.

\noindent
\textbf{The memory error rates of the squeezed cat} \\
% The memory error rates of the SC under the dynamics Eq.~\eqref{eq:memory_dynamics} can be analytically derived:
% \begin{eqnarray} 
% & \gamma_Z = [\kappa_1(1 + 2n_{\textrm{th}}) + \kappa_{\phi}e^{-2r}] (\bar{n} - \sinh^2 r), 
% \label{eq:memory_error_rates_Z}\\
%     & \gamma_{X} = \frac{1}{2} \kappa_{\phi} \frac{\left(\bar{n}-\sinh ^{2} r\right) e^{2 r}}{\sinh \left(2\left(\bar{n}-\sinh ^{2} r\right) e^{2 r}\right)}[\frac{\sinh ^{2} 2 r}{4}+\cosh 4 r].
%     \label{eq:memory_error_rates_X}
% \end{eqnarray}
% Note that we restore the error rates of the dissipative cat~\cite{guillaud2019repetition} when $r = 0$, and we obtain the approximate expression 
In this section, we provide the derivation of the memory error rates for the SC in Eqs.~\eqref{eq:memory_error_rates_Z} and \eqref{eq:memory_error_rates_X}. 

Since the bit-flip error rate is exponentially small in $\alpha^{\prime}$, the subsystem decomposition is insufficient to obtain an analytical expression of it. 
% Notice that our proposed dissipator $\hat{F} = \hat{Z}_{L}\otimes \hat{\tilde{a}}$ only modifies the Z-error rates compared to the original dissipation stabilized cat. 
Thus, we derive the bit-flip error rate  using the conserved quantities of the system~\cite{mirrahimi2014dynamically, albert2018lindbladians}. To facilitate the analysis, we first neglect the $\hat{Z}_L$ term in the dissipator in Eq.~\eqref{eq:engineered_dissipator} since it does not contribute to the bit-flip rate, and then analyze the system dynamics in the squeezed frame: 
\begin{equation}
\begin{aligned}
    \frac{d \hat{\rho}_s}{dt} = \kappa_2 \mathcal{D}[\hat{a}^2 - \alpha^{\prime 2}] \hat{\rho}_s + \kappa_{\phi} \mathcal{D}[\hat{a}_s^{\dagger} \hat{a}_s] \hat{\rho}_s,
    \label{eq:memory_dynamics_squeezed_frame}
\end{aligned}
\end{equation}
where $\hat{A}_s := \hat{S}^{\dagger}(r) \hat{A} \hat{S}(r)$ for any operator $\hat{A}$. Note that we we consider the dephasing here, which is the dominant source for the bit-flip errors.
The two conserved quantities associated with the dominant dissipator $\hat{a}^2 - \alpha^{\prime 2}$ are
\begin{equation}
\begin{aligned}
&\hat{J}_{++} = \sum_{n=0}^{\infty}\ket{2n}\bra{2n}, \\
&\hat{J}_{+-}=\sqrt{\frac{2\alpha^{\prime 2}}{\sinh{2\alpha^{\prime 2}}}}\sum_{q = -\infty}^{\infty}\frac{\left(-1\right)^{q}}{2q+1}I_{q}\left(\alpha^{\prime 2}\right)\hat{J}_{+-}^{\left(q\right)},
\end{aligned}
\end{equation}
 where $I_{q}\left(\cdot\right)$ is the modified Bessel function of the first kind, and $\hat{J}_{+-}^{\left(q\right)} = \frac{\left(\hat{a}^{\dagger}\hat{a} - 1\right)!!}{\left(\hat{a}^{\dagger}\hat{a} + 2q\right)!!}\hat{J}_{++}\hat{a}^{2q+1}$ for $q\geq 0$ and $\hat{J}_{+-}^{\left(q\right)} = \hat{J}_{++}\hat{a}^{\dagger 2q+1}\frac{\left(\hat{a}^{\dagger}\hat{a}\right)!!}{\left(\hat{a}^{\dagger}\hat{a} + 2\abs{q}-1\right)!!}$ for $q< 0$. The steady state coherence of the system initialized in $\hat{\rho}(0)$ can be computed through $c_{++}\left(\infty\right) = \text{tr}\left\{J_{++}^{\dagger}\rho\left(0\right)\right\}$ and $c_{+-}\left(\infty\right) = \text{tr}\left\{J_{+-}^{\dagger}\rho\left(0\right)\right\}$. Thus, we compute the bir-flip rate perturbatively by considering the dephasing in the squeezed frame,
 \begin{equation}
     \nonpfrate{} = -\kappa_{\phi}\text{tr}\left\{J_{+-}^{\dagger}\mathcal{D}\left[\hat{S}^{\dagger}\left(r\right)\hat{a}^\dagger \hat{a}\hat{S}\left(r\right)\right]\ket{C_{\alpha}^{+}}\bra{C_{\alpha}^{-}}\right\},
 \end{equation}
which is then simplified to Eq.~\eqref{eq:memory_error_rates_X}.

The phase-flip error rate Eq.~\eqref{eq:memory_error_rates_Z} can be easily derived by analyzing the errors under the subsystem decomposition. The loss and heating errors are in the form $\hat{a} \approx \hat{Z}_L \otimes (e^{-r}\alpha^{\prime} + \cosh r \hat{\tilde{a}} - \sinh r \hat{\tilde{a}}^{\dagger})$, $\hat{a}^{\dagger} \approx \hat{Z}_L \otimes (e^{-r}\alpha^{\prime} + \cosh r \hat{\tilde{a}}^{\dagger} - \sinh r \hat{\tilde{a}})$. They both contribute to the phase-flip rate via the undetectable term $e^{-r}\alpha^{\prime} \hat{Z}_L = \sqrt{\eta \bar{n}}\hat{Z}_L$ (the detectable part associated with the $\hat{Z}_L \otimes \hat{\tilde{a}}^{\dagger}$ term is approximately correctable by $\hat{F}$). The dephasing is in the form $\hat{a}^{\dagger}\hat{a} \approx \hat{I}_L \otimes [e^{-2r}\alpha^{\prime 2} + e^{-2r}\alpha^{\prime}(\hat{\tilde{a}} + \hat{\tilde{a}}^{\dagger}) + \cosh^2 r \hat{\tilde{a}}^{\dagger} \hat{\tilde{a}} + \sinh^2 r \hat{\tilde{a}} \hat{\tilde{a}}^{\dagger} - \cosh r \sinh r (\hat{\tilde{a}}^2 + \hat{\tilde{a}}^{\dagger 2})]$. It contributes to the phase-flip rate dominantly by the $e^{-2r}\alpha^{\prime} \hat{I}_L \otimes \hat{\tilde{a}}^{\dagger}$ term, which creates an excitation in the gauge mode that is subsequently destroyed by $\hat{F}$ with a residual phase flip. Therefore, the dephasing contributes to the phase-flip rate by $\kappa_{\phi}e^{-4r}\alpha^{\prime 2} = \kappa_{\phi}e^{-2r} \eta \bar{n}$. 
% calculating the undetectable errors induced by excitation loss and heating, and the leading-order uncorrectable errors induced by the dephasing. 

Eq.~\eqref{eq:memory_error_rates_Z} is valid in the regime where $\alpha^{\prime} \gg 1$, which is violated when $r$ approaches the maximum squeezing allowed by the energy constraint. We now provide a leading-order correction to the loss-induced phase flip rate in such a regime.
% In the analysis of the Z-error rate, we similarly could compute the rate through the conserved quantity by considering excitation loss in the squeezed frame. If the dissipator is kept the same as the cat code, $\hat{a}^{2} - \alpha^{2}$, then the Z-error rate is $-\kappa_{1}\text{tr}\left\{J_{++}^{\dagger}\mathcal{D}\left[\hat{S}^{\dagger}\left(r\right)\hat{a}\hat{S}\left(r\right)\right]\ket{C_{\alpha}^{+}}\bra{C_{\alpha}^{-}}\right\} = \kappa_{1}\abs{\alpha}^{2}$, independent of squeezing r. Note that the subsystem decomposition of the excitation loss operator in the squeezed frame is $\hat{Z}_{L}\otimes \left(\cosh{r}\hat{\tilde{a}} - \sinh{r}\hat{a}^{' \dagger} + \sqrt{\bar{n} - \sinh^{2}{r}}\hat{I}\right)$. Therefore, the uncorrectable error associated with excitation loss occurs at a rate of $\kappa_{1}\left(\bar{n} - \sinh^{2}{r}\right)$. 
% The modified dissipator $\hat{F}=\hat{Z}_{L}(\hat{a}^{2}-\alpha^{2}) \approx \hat{Z}_{L}\otimes (\hat{\tilde{a}}^{2}+2\alpha^{\prime} \hat{\tilde{a}})$ is introduced to correct the portion of errors that lead to both a logical phase flip and an excitation in the gauge basis. 
We have assumed that the dissipator $\hat{F}=\hat{Z}_{L}(\hat{a}^{2}-\alpha^{2}) \approx \hat{Z}_{L}\otimes (\hat{\tilde{a}}^{2}+2\alpha^{\prime} \hat{\tilde{a}})$ can perfectly correct the detectable part of the loss-(or heating-)induced errors by removing the excitation in the gauge mode while applying a phase-flip correction on the logical qubit. 
However, it is not a perfect correction because of the non-Hermitian part of the dynamics induced by $F^{\dagger}F \approx \hat{I}_{L}\otimes [\hat{\tilde{a}}^{\dagger 2} \hat{\tilde{a}}^{2} + 2\alpha^{\prime} (\hat{\tilde{a}}^{\dagger 2}\hat{\tilde{a}} + \hat{\tilde{a}}^{\dagger}\hat{\tilde{a}}^{2}) + 4\alpha^{\prime 2}\hat{\tilde{a}}^{\dagger} \hat{\tilde{a}}]$. The second term above further excites the gauge mode, which introduces additional non-negligible Z errors when $\alpha^{\prime}\gg 1$ does not hold. Through analysis of a simplified 3-level system, we obtain a correction factor for the phase-flip rate in the form of
% We analyzed the steady state of a system with an initial state of $\ket{+}\otimes \ket{1}$ subject to the dissipator $\hat{F}$. 
\begin{equation}
    \xi = \frac{1}{2(1+3\alpha^{\prime 2})},
\end{equation}
which works well for $\alpha^{\prime}\geq 1.5$. This factor represents that, if the qubit evolves from an initial state of $\ket{\pm}_{L}\otimes \ket{\tilde{n}=1}$ under the dissipator $\hat{F}$, a population of $1 - \xi$ would end up in $\ket{\mp}_{L}\otimes \ket{\tilde{n}=0}$ and $\xi$ would be in $\ket{\pm}_{L}\otimes \ket{\tilde{n}=0}$ in steady state. Therefore, the phase-flip rate in Eq.~\eqref{eq:memory_error_rates_Z} has an extra correction:
\begin{equation}
    \gamma_Z \rightarrow \gamma_Z + \kappa_1 (1 + n_{\textrm{th}}) \bar{n} (1 - \eta)\xi + \kappa_1 n_{\textrm{th}}(1 - \eta^{\prime})\xi - \kappa_{\phi}e^{-2r}\eta \xi,
    \label{eqn: corrected Z error}
\end{equation}
where $\eta^{\prime} =(\bar{n}-\cosh^2r)/\bar{n}$, which approaches $\eta$ in the large squeezing limit.

% \textcolor{red}{Insert plot?}
The correction factor's effect becomes significant as $\eta$ approaches 0. In the limit of large $\bar{n}$ and only considering the dominant loss error, the Z error rate has a minimum value $\gamma_{Z, \textrm{min}}\approx \frac{\sqrt{2}}{4}\kappa_1$. Worth noticing, this minimum rate is independent of $\bar{n}$. Therefore, the SC enjoys an exponential suppression of the bit-flip rate while maintaining a bounded phase-flip rate by increasing $\bar{n}$, which is drastically different from the cat code or its DV counterpart, the repetition code.

% Similar to the Z-errors induced by photon loss, Eq.~\eqref{eqn: corrected Z error} can be extended to heating errors. The only correction comes from the rate at which populations are transferred into the first excited state in the gauge basis. As a result, corrected form of heating-induced Z-error is $\kappa_1\bar{n}n_{\textrm{th}}\left[\eta + (1-\eta^{\prime})\xi\right]$. Here $\eta^{\prime} = \frac{\bar{n}-\cosh^2r}{\bar{n}}$, which approaches $\eta$ in the large squeezing limit.

\noindent
\textbf{Bias-preserving operations for the squeezed cat} \\
In this section, we present the detailed design and error analysis for the $Z$ rotation $Z(
\theta)$ and the CX gate for the SC, which are representatives of bias-preserving opearations $\mathcal{B}$. See supplement.~\cite{SM} for the rest of the operations in $\mathcal{B}$. 

Similarly to the cat, the $Z$-axis rotation $Z(\theta)$ can be generated by a resonant linear drive $\hat{H}_{Z} = \frac{\theta}{4 \alpha^{\prime} T} e^{r} (\hat{a} + \hat{a}^{\dagger})$ in the presence of the engineered dissipation in Eq.~\eqref{eq:engineered_dissipator} for a time $T$. In the subsystem basis, $H_{Z} \approx \frac{\theta}{4\alpha^{'}T}\hat{Z}_{L}\otimes(2\alpha^{\prime} + \hat{\tilde{a}} + \hat{\tilde{a}}^{\dagger})$. The total phase flip error probability of the $Z$ rotation is $p_Z = p_Z^{\textrm{NA}}(T) + \kappa_1 \eta \bar{n} T$, where the second term represents the loss-induced phase flips and the first term represents the non-adiabatic errors due to the non-adiabatic excitation $\hat{Z}_L \otimes \hat{\tilde{a}}^{\dagger}$ in $\hat{H}_Z$. We note that compared to the parity-preserving dissipator $\mathcal{D}[\hat{I}_L \otimes \hat{\tilde{a}}]$, which is used in the literature for the cat (by applying a driven two-photon dissipation), the parity-flipping dissipator $\hat{F}$ in Eq.~\eqref{eq:engineered_dissipator} can significantly reduce the non-adiabatic errors induced by $\hat{Z}_L \otimes \hat{\tilde{a}}^{\dagger}$. The reason is that the majority of the parity flips associated with the non-adiabatic transitions can be flipped back through the application of the dissipator.
% the linear drive (or any non-adiabatic drive that does not commute with the parity). The reason is that the majority of the parity flips associated with the non-adiabatic transitions can be flipped back through the application of the dissipator. Moreover, the non-adiabatic errors are further suppressed by applying the squeezing since the dissipation gap $\Delta \approx 4 \kappa_2 \alpha^{\prime 2}$ increases. 
% \jedit{However, not all non-adiabatic errors can be corrected because the parity flipping dissipator $\hat{F}$ is not ideal and thus can not perfectly correct all populations with excited gauge levels. For details, see Methods? Nevertheless, we are only left with a small portion of the non-adiabatic error, and the adiabatic gap $\Delta \approx 4 \kappa_2 \alpha^{\prime 2}$ increases as well. The optimized gate time is orders of magnitude smaller than the cat scheme. Therefore, we also apply an idling time as part of the gate scheme, where the system is cooled by the dissipator $\hat{F}$, to reduce leakage error.} 
% A residual non-adiabatic error $p_Z^{\textrm{NA}}(T)$ exists because that $\hat{F}$ only provides a partial correction with efficiency $1 - \xi$ as described previously in Methods. 
The remaining errors with a fraction $\xi$ leads to the residual non-adiabatic error $p_{Z}^{\textrm{NA}}$ proportional to $\xi$ (see the previous Methods section).
Under the adiabatic limit $\frac{\theta}{4\alpha^{'}T}\ll 4\kappa_{2}\alpha^{'2}$, the system's evolution under the dissipator $\hat{F}$ can be approximated by the dynamics of the density matrix $\hat{\rho}_{\textrm{trunc}}$ with a truncated 2-level gauge basis:
\begin{equation}
\begin{aligned}
\kappa_{2}\mathcal{D}[\hat{F}]\hat{\rho} \approx
    4\kappa_{2}\alpha^{'2}((1 - \xi)\mathcal{D}[\hat{Z}_{L}\otimes \hat{\tilde{a}}] +
    \xi\mathcal{D}[\hat{I}_{L}\otimes \hat{\tilde{a}}])\hat{\rho}_{\textrm{trunc}}.
\end{aligned}
\end{equation}
Performing first-order adiabatic elimination~\cite{reiter2012effective} on the gauge excited state results in an effective Z error rate $\frac{\xi \theta^{2}}{16\kappa_{2}\alpha^{'4}T}$. Notice that adiabatic elimination does not capture the higher-order errors and the result only holds under the adiabatic limit. A more accurate expression can be derived through solving the ordinary differential equations of the two level system. As a result, the modified non-adiabatic error has the form:
\begin{equation}
    p_Z^{\textrm{NA}}(T) = \frac{\xi \theta^{2}}{16\kappa_{2}\alpha^{'4}T^{2}}(c_1 T + c_2 \frac{e^{-2\kappa_{2}\alpha^{'2}T} - 1}{2\kappa_{2}\alpha^{'2}}).
    \label{eq:non_adiabatic_error_Z_rotation}
\end{equation}
Performing numerical fit, we obtain $c_1 = 1.5, c_2 = 1.8$.

% As shown in Fig.~\ref{fig:gate_errors}(a), compared to the $Z(\theta)$ gate on the cat in the literature, our proposed $Z(\theta)$ gate on the SC has a much lower $Z$ error rate for any gate time $T$ due to the simultaneous suppression of the loss-induced errors and the non-adiabatic errors. 
% % \textcolor{red}{Z gate also needs cooling}

The CX gate is implemented by applying the engineered dissipation only on the control mode and a Hamiltonian term that drives a phase rotation on the target mode conditioned on the states of the control mode:
\begin{equation}
\begin{array}{c}
    \frac{d}{dt} \hat{\rho} = \kappa_2 \mathcal{D}[\hat{F_c}]\hat{\rho} - i [\hat{H}_{\textrm{CX}}, \hat{\rho}], \\
     \hat{H}_{\textrm{CX}} = \frac{\pi}{4 \alpha^{\prime} T}\left[e^{r}(\hat{a}_{c}+\hat{a}_{c}^{\dagger})-2 \alpha^{\prime}\right](\hat{a}_{t}^{\dagger} \hat{a}_{t}-\alpha^{\prime 2}),\\
\end{array}
\end{equation}
where $\hat{F}_c$ denotes the engineered dissipator in Eq.~\eqref{eq:engineered_dissipator} on the control mode. The noise terms are not shown for simplicity. We note that compared to the standard CX gate on the cat~\cite{guillaud2019repetition, chamberland2022building}, we turn off the dissipation on the target mode during the gate to circumvent the need for high-order coupling terms between the two modes. Although the target mode temporarily loses the protection against the excitation loss, we can still implement a high-quality gate if the gate time is short enough and the leakage on the target mode can be subsequently returned to the code space without introducing too many errors. Similar strategy and insights have been made in Ref.~\cite{gautier2022combined}. 
% \jedit{While the non-adiabatic error introduces one logical parity flip per gauge level excitation on the control mode, it excites the gauge levels of the target mode without affecting the parity. Therefore, in the cooling time $T_{cool}$, we apply parity-flipping $\hat{F}$ on the control mode and parity-preserving dissipation $\kappa_2 \mathcal{D}[\hat{S}(r)(\hat{a}_t^2 - \alpha^{\prime 2}) \hat{S}^{\dagger}(r)]$ on the target mode.} 
To deal with the non-adiabatic transitions on the target mode, which preserve the parity, we apply a parity-preserving dissipation $\kappa_2 \mathcal{D}[\hat{S}(r)(\hat{a}_t^2 - \alpha^{\prime 2}) \hat{S}^{\dagger}(r)]$ on the target mode (while the control mode is, as always, protected by the parity-flipping dissipation) for a time $T_{\textrm{cool}}$. In our simulations, we fix the cooling time $T_{\textrm{cool}} = 8\times \frac{1}{4\kappa_2\alpha^{\prime 2}}$ to ensure that the leakage is suppressed to below $0.5\%$. Using the Pauli-twirling approximation, the Z-type errors of the CX gate are
\begin{equation}
    \begin{aligned}
p_{Z_{c}} &=\kappa_{1} \eta \bar{n}\left(T+T_{\mathrm{cool}}\right)+p_{Z}^{\textrm{NA}}(T), \\
p_{Z_{t}} &=\kappa_{1} \bar{n}\left(\frac{T}{2} +T_{\mathrm{cool}}\right), \\
p_{Z_{c} Z_{t}} &=\frac{1}{2} \kappa_{1} \bar{n} T,
\label{eq:CX_Z_error}
\end{aligned}
\end{equation}
where $p_{Z_c}, p_{Z_t}$ and $p_{Z_c Z_t}$ denote the $Z$ error on the control, target mode and the correlated $Z$ error, respectively. They sum to the total Z error probability $p_Z = \kappa_1 \bar{n} (1 + \eta)(T + T_{\textrm{cool}}) + p_Z^{\textrm{NA}}(T)$. Note that, unlike the $Z$ rotation, the CX gate does not enjoy a full suppression of the loss-induced errors (by a factor $\eta$) due to the lack of the engineered dissipation on the target mode during the gate. The non-adiabatic error $p_Z^{\textrm{NA}}(T)$ on the control mode has a similar form as Eq.~\eqref{eq:non_adiabatic_error_Z_rotation}:
\begin{equation}
    p_{Z}^{\textrm{NA}}(T) = \frac{\xi \pi^{2}}{16\kappa_{2}\alpha^{'2}T^{2}}\left( 1.5T + 0.6\frac{e^{-2\kappa_{2}\alpha^{'2}T} - 1}{2\kappa_{2}\alpha^{'2}}\right)
\end{equation}

% As shown in Fig.~\ref{fig:gate_errors}(b), since the non-adiabatic error $p_Z^{\textrm{NA}}$ is significantly suppressed, the CX gate on the SC can be implemented much faster ($T \ll 1/\kappa_2$) with much lower $Z$ error rates compared to the cat gate.

We also present the non-$Z$ error rate of the CX gate here. Note that the CX gate has a significantly larger non-$Z$ error rate than all other bias-preserving operations in $\mathcal{B}$.
As discussed numerically in Ref.~\cite{chamberland2022building}, the non-$Z$ error of a cat's CX gate scales approximately as $1.8\frac{e^{-2\alpha^{2}}}{\alpha^{2}}\frac{1}{\kappa_{2}T}$. For our CX gate on the SC, we find a similar expression 
\begin{equation}
    p_{X, Y} = 5.57\frac{e^{-2\alpha^{\prime 2}}}{\alpha^{\prime 2}}\frac{1}{\kappa_{2}T},
\end{equation}
in the regime where $\kappa_2 T > 1$. Note that for shorter gate time, we cannot find a simple expression for $p_{X, Y}$ and a numerical simulation of the gate has to be performed to determine $p_{X, Y}$.

\begin{acknowledgments}
We thank Ramesh Bhandari, Yvonne Gao, Kyungjoo Noh for helpful discussions. 
We acknowledge support from the ARO (W911NF-18-1-0020, W911NF-18-1-0212, W911NF-19-1-0380), ARO MURI (W911NF-16-1-0349, W911NF-21-1-0325), AFOSR MURI (FA9550-19-1-0399, FA9550-21-1-0209), AFRL (FA8649-21-P-0781), DoE Q-NEXT, NSF (PHY-1748958, OMA-1936118, ERC-1941583, OMA-2137642), NTT Research, and the Packard Foundation (2020-71479).
\end{acknowledgments}

% \bibliographystyle{apsrev4-2}
% \bibliography{ref}

\begin{thebibliography}{72}%
\makeatletter
\providecommand \@ifxundefined [1]{%
 \@ifx{#1\undefined}
}%
\providecommand \@ifnum [1]{%
 \ifnum #1\expandafter \@firstoftwo
 \else \expandafter \@secondoftwo
 \fi
}%
\providecommand \@ifx [1]{%
 \ifx #1\expandafter \@firstoftwo
 \else \expandafter \@secondoftwo
 \fi
}%
\providecommand \natexlab [1]{#1}%
\providecommand \enquote  [1]{``#1''}%
\providecommand \bibnamefont  [1]{#1}%
\providecommand \bibfnamefont [1]{#1}%
\providecommand \citenamefont [1]{#1}%
\providecommand \href@noop [0]{\@secondoftwo}%
\providecommand \href [0]{\begingroup \@sanitize@url \@href}%
\providecommand \@href[1]{\@@startlink{#1}\@@href}%
\providecommand \@@href[1]{\endgroup#1\@@endlink}%
\providecommand \@sanitize@url [0]{\catcode `\\12\catcode `\$12\catcode
  `\&12\catcode `\#12\catcode `\^12\catcode `\_12\catcode `\%12\relax}%
\providecommand \@@startlink[1]{}%
\providecommand \@@endlink[0]{}%
\providecommand \url  [0]{\begingroup\@sanitize@url \@url }%
\providecommand \@url [1]{\endgroup\@href {#1}{\urlprefix }}%
\providecommand \urlprefix  [0]{URL }%
\providecommand \Eprint [0]{\href }%
\providecommand \doibase [0]{https://doi.org/}%
\providecommand \selectlanguage [0]{\@gobble}%
\providecommand \bibinfo  [0]{\@secondoftwo}%
\providecommand \bibfield  [0]{\@secondoftwo}%
\providecommand \translation [1]{[#1]}%
\providecommand \BibitemOpen [0]{}%
\providecommand \bibitemStop [0]{}%
\providecommand \bibitemNoStop [0]{.\EOS\space}%
\providecommand \EOS [0]{\spacefactor3000\relax}%
\providecommand \BibitemShut  [1]{\csname bibitem#1\endcsname}%
\let\auto@bib@innerbib\@empty
%</preamble>
\bibitem [{\citenamefont {Nielsen}\ and\ \citenamefont
  {Chuang}(2010)}]{nielsen_chuang_2010}%
  \BibitemOpen
  \bibfield  {author} {\bibinfo {author} {\bibfnamefont {M.~A.}\ \bibnamefont
  {Nielsen}}\ and\ \bibinfo {author} {\bibfnamefont {I.~L.}\ \bibnamefont
  {Chuang}},\ }\href {https://doi.org/10.1017/CBO9780511976667} {\emph
  {\bibinfo {title} {Quantum Computation and Quantum Information: 10th
  Anniversary Edition}}}\ (\bibinfo  {publisher} {Cambridge University Press},\
  \bibinfo {year} {2010})\BibitemShut {NoStop}%
\bibitem [{lid(2013)}]{lidar_brun_2013}%
  \BibitemOpen
  \href {https://doi.org/10.1017/CBO9781139034807} {\emph {\bibinfo {title}
  {Quantum Error Correction}}}\ (\bibinfo  {publisher} {Cambridge University
  Press},\ \bibinfo {year} {2013})\BibitemShut {NoStop}%
\bibitem [{\citenamefont {Aharonov}\ \emph {et~al.}(1996)\citenamefont
  {Aharonov}, \citenamefont {Ben-Or}, \citenamefont {Impagliazzo},\ and\
  \citenamefont {Nisan}}]{aharonov1996limitations}%
  \BibitemOpen
  \bibfield  {author} {\bibinfo {author} {\bibfnamefont {D.}~\bibnamefont
  {Aharonov}}, \bibinfo {author} {\bibfnamefont {M.}~\bibnamefont {Ben-Or}},
  \bibinfo {author} {\bibfnamefont {R.}~\bibnamefont {Impagliazzo}},\ and\
  \bibinfo {author} {\bibfnamefont {N.}~\bibnamefont {Nisan}},\ }\href
  {https://arxiv.org/abs/quant-ph/9611028} {\bibfield  {journal} {\bibinfo
  {journal} {arXiv preprint quant-ph/9611028}\ } (\bibinfo {year}
  {1996})}\BibitemShut {NoStop}%
\bibitem [{\citenamefont {Aharonov}\ and\ \citenamefont
  {Ben-Or}(1997)}]{aharonov1997fault}%
  \BibitemOpen
  \bibfield  {author} {\bibinfo {author} {\bibfnamefont {D.}~\bibnamefont
  {Aharonov}}\ and\ \bibinfo {author} {\bibfnamefont {M.}~\bibnamefont
  {Ben-Or}},\ }in\ \href@noop {} {\emph {\bibinfo {booktitle} {Proceedings of
  the twenty-ninth annual ACM symposium on Theory of computing}}}\ (\bibinfo
  {year} {1997})\ pp.\ \bibinfo {pages} {176--188}\BibitemShut {NoStop}%
\bibitem [{\citenamefont {Kitaev}(1997)}]{kitaev1997quantum}%
  \BibitemOpen
  \bibfield  {author} {\bibinfo {author} {\bibfnamefont {A.~Y.}\ \bibnamefont
  {Kitaev}},\ }\href
  {https://iopscience.iop.org/article/10.1070/RM1997v052n06ABEH002155/meta?casa_token=ZP_ozstfbdEAAAAA:4RIW0rCdB7GYdXisct3ZYUzRYBmuRHvhfYo44Ivp1F_GmsrGIW04QYja74ePCkXVVkNMX00TIuj7WBfboQY}
  {\bibfield  {journal} {\bibinfo  {journal} {Russian Mathematical Surveys}\
  }\textbf {\bibinfo {volume} {52}},\ \bibinfo {pages} {1191} (\bibinfo {year}
  {1997})}\BibitemShut {NoStop}%
\bibitem [{\citenamefont {Knill}\ \emph {et~al.}(1998)\citenamefont {Knill},
  \citenamefont {Laflamme},\ and\ \citenamefont {Zurek}}]{knill1998resilient}%
  \BibitemOpen
  \bibfield  {author} {\bibinfo {author} {\bibfnamefont {E.}~\bibnamefont
  {Knill}}, \bibinfo {author} {\bibfnamefont {R.}~\bibnamefont {Laflamme}},\
  and\ \bibinfo {author} {\bibfnamefont {W.~H.}\ \bibnamefont {Zurek}},\ }\href
  {https://doi.org/10.1098/rspa.1998.0166} {\bibfield  {journal} {\bibinfo
  {journal} {Proceedings of the Royal Society of London. Series A:
  Mathematical, Physical and Engineering Sciences}\ }\textbf {\bibinfo {volume}
  {454}},\ \bibinfo {pages} {365} (\bibinfo {year} {1998})}\BibitemShut
  {NoStop}%
\bibitem [{\citenamefont {Aliferis}\ \emph {et~al.}(2005)\citenamefont
  {Aliferis}, \citenamefont {Gottesman},\ and\ \citenamefont
  {Preskill}}]{aliferis2005quantum}%
  \BibitemOpen
  \bibfield  {author} {\bibinfo {author} {\bibfnamefont {P.}~\bibnamefont
  {Aliferis}}, \bibinfo {author} {\bibfnamefont {D.}~\bibnamefont
  {Gottesman}},\ and\ \bibinfo {author} {\bibfnamefont {J.}~\bibnamefont
  {Preskill}},\ }\href {https://arxiv.org/abs/quant-ph/0504218} {\bibfield
  {journal} {\bibinfo  {journal} {arXiv preprint quant-ph/0504218}\ } (\bibinfo
  {year} {2005})}\BibitemShut {NoStop}%
\bibitem [{\citenamefont {Fowler}\ \emph {et~al.}(2012)\citenamefont {Fowler},
  \citenamefont {Mariantoni}, \citenamefont {Martinis},\ and\ \citenamefont
  {Cleland}}]{fowler2012surface}%
  \BibitemOpen
  \bibfield  {author} {\bibinfo {author} {\bibfnamefont {A.~G.}\ \bibnamefont
  {Fowler}}, \bibinfo {author} {\bibfnamefont {M.}~\bibnamefont {Mariantoni}},
  \bibinfo {author} {\bibfnamefont {J.~M.}\ \bibnamefont {Martinis}},\ and\
  \bibinfo {author} {\bibfnamefont {A.~N.}\ \bibnamefont {Cleland}},\ }\href
  {https://doi.org/10.1103/PhysRevA.86.032324} {\bibfield  {journal} {\bibinfo
  {journal} {Physical Review A}\ }\textbf {\bibinfo {volume} {86}},\ \bibinfo
  {pages} {032324} (\bibinfo {year} {2012})}\BibitemShut {NoStop}%
\bibitem [{\citenamefont {Litinski}(2019)}]{litinski2019game}%
  \BibitemOpen
  \bibfield  {author} {\bibinfo {author} {\bibfnamefont {D.}~\bibnamefont
  {Litinski}},\ }\href {https://doi.org/10.22331/q-2019-03-05-128} {\bibfield
  {journal} {\bibinfo  {journal} {Quantum}\ }\textbf {\bibinfo {volume} {3}},\
  \bibinfo {pages} {128} (\bibinfo {year} {2019})}\BibitemShut {NoStop}%
\bibitem [{\citenamefont {Chao}\ \emph {et~al.}(2020)\citenamefont {Chao},
  \citenamefont {Beverland}, \citenamefont {Delfosse},\ and\ \citenamefont
  {Haah}}]{chao2020optimization}%
  \BibitemOpen
  \bibfield  {author} {\bibinfo {author} {\bibfnamefont {R.}~\bibnamefont
  {Chao}}, \bibinfo {author} {\bibfnamefont {M.~E.}\ \bibnamefont {Beverland}},
  \bibinfo {author} {\bibfnamefont {N.}~\bibnamefont {Delfosse}},\ and\
  \bibinfo {author} {\bibfnamefont {J.}~\bibnamefont {Haah}},\ }\href
  {https://doi.org/10.22331/q-2020-10-28-352} {\bibfield  {journal} {\bibinfo
  {journal} {Quantum}\ }\textbf {\bibinfo {volume} {4}},\ \bibinfo {pages}
  {352} (\bibinfo {year} {2020})}\BibitemShut {NoStop}%
\bibitem [{\citenamefont {Beverland}\ \emph {et~al.}(2021)\citenamefont
  {Beverland}, \citenamefont {Kubica},\ and\ \citenamefont
  {Svore}}]{beverland2021cost}%
  \BibitemOpen
  \bibfield  {author} {\bibinfo {author} {\bibfnamefont {M.~E.}\ \bibnamefont
  {Beverland}}, \bibinfo {author} {\bibfnamefont {A.}~\bibnamefont {Kubica}},\
  and\ \bibinfo {author} {\bibfnamefont {K.~M.}\ \bibnamefont {Svore}},\ }\href
  {https://doi.org/10.1103/PRXQuantum.2.020341} {\bibfield  {journal} {\bibinfo
   {journal} {PRX Quantum}\ }\textbf {\bibinfo {volume} {2}},\ \bibinfo {pages}
  {020341} (\bibinfo {year} {2021})}\BibitemShut {NoStop}%
\bibitem [{\citenamefont {Gottesman}\ \emph {et~al.}(2001)\citenamefont
  {Gottesman}, \citenamefont {Kitaev},\ and\ \citenamefont
  {Preskill}}]{gottesman2001encoding}%
  \BibitemOpen
  \bibfield  {author} {\bibinfo {author} {\bibfnamefont {D.}~\bibnamefont
  {Gottesman}}, \bibinfo {author} {\bibfnamefont {A.}~\bibnamefont {Kitaev}},\
  and\ \bibinfo {author} {\bibfnamefont {J.}~\bibnamefont {Preskill}},\ }\href
  {https://doi.org/10.1103/PhysRevA.64.012310} {\bibfield  {journal} {\bibinfo
  {journal} {Physical Review A}\ }\textbf {\bibinfo {volume} {64}},\ \bibinfo
  {pages} {012310} (\bibinfo {year} {2001})}\BibitemShut {NoStop}%
\bibitem [{\citenamefont {Chuang}\ \emph {et~al.}(1997)\citenamefont {Chuang},
  \citenamefont {Leung},\ and\ \citenamefont {Yamamoto}}]{chuang1997bosonic}%
  \BibitemOpen
  \bibfield  {author} {\bibinfo {author} {\bibfnamefont {I.~L.}\ \bibnamefont
  {Chuang}}, \bibinfo {author} {\bibfnamefont {D.~W.}\ \bibnamefont {Leung}},\
  and\ \bibinfo {author} {\bibfnamefont {Y.}~\bibnamefont {Yamamoto}},\ }\href
  {https://doi.org/10.1103/PhysRevA.56.1114} {\bibfield  {journal} {\bibinfo
  {journal} {Physical Review A}\ }\textbf {\bibinfo {volume} {56}},\ \bibinfo
  {pages} {1114} (\bibinfo {year} {1997})}\BibitemShut {NoStop}%
\bibitem [{\citenamefont {Michael}\ \emph {et~al.}(2016)\citenamefont
  {Michael}, \citenamefont {Silveri}, \citenamefont {Brierley}, \citenamefont
  {Albert}, \citenamefont {Salmilehto}, \citenamefont {Jiang},\ and\
  \citenamefont {Girvin}}]{michael2016new}%
  \BibitemOpen
  \bibfield  {author} {\bibinfo {author} {\bibfnamefont {M.~H.}\ \bibnamefont
  {Michael}}, \bibinfo {author} {\bibfnamefont {M.}~\bibnamefont {Silveri}},
  \bibinfo {author} {\bibfnamefont {R.}~\bibnamefont {Brierley}}, \bibinfo
  {author} {\bibfnamefont {V.~V.}\ \bibnamefont {Albert}}, \bibinfo {author}
  {\bibfnamefont {J.}~\bibnamefont {Salmilehto}}, \bibinfo {author}
  {\bibfnamefont {L.}~\bibnamefont {Jiang}},\ and\ \bibinfo {author}
  {\bibfnamefont {S.~M.}\ \bibnamefont {Girvin}},\ }\href
  {https://doi.org/10.1103/PhysRevX.6.031006} {\bibfield  {journal} {\bibinfo
  {journal} {Physical Review X}\ }\textbf {\bibinfo {volume} {6}},\ \bibinfo
  {pages} {031006} (\bibinfo {year} {2016})}\BibitemShut {NoStop}%
\bibitem [{\citenamefont {Cochrane}\ \emph {et~al.}(1999)\citenamefont
  {Cochrane}, \citenamefont {Milburn},\ and\ \citenamefont
  {Munro}}]{cochrane1999macroscopically}%
  \BibitemOpen
  \bibfield  {author} {\bibinfo {author} {\bibfnamefont {P.~T.}\ \bibnamefont
  {Cochrane}}, \bibinfo {author} {\bibfnamefont {G.~J.}\ \bibnamefont
  {Milburn}},\ and\ \bibinfo {author} {\bibfnamefont {W.~J.}\ \bibnamefont
  {Munro}},\ }\href {https://doi.org/10.1103/PhysRevA.59.2631} {\bibfield
  {journal} {\bibinfo  {journal} {Physical Review A}\ }\textbf {\bibinfo
  {volume} {59}},\ \bibinfo {pages} {2631} (\bibinfo {year}
  {1999})}\BibitemShut {NoStop}%
\bibitem [{\citenamefont {Albert}\ \emph {et~al.}(2018)\citenamefont {Albert},
  \citenamefont {Noh}, \citenamefont {Duivenvoorden}, \citenamefont {Young},
  \citenamefont {Brierley}, \citenamefont {Reinhold}, \citenamefont {Vuillot},
  \citenamefont {Li}, \citenamefont {Shen}, \citenamefont {Girvin} \emph
  {et~al.}}]{albert2018performance}%
  \BibitemOpen
  \bibfield  {author} {\bibinfo {author} {\bibfnamefont {V.~V.}\ \bibnamefont
  {Albert}}, \bibinfo {author} {\bibfnamefont {K.}~\bibnamefont {Noh}},
  \bibinfo {author} {\bibfnamefont {K.}~\bibnamefont {Duivenvoorden}}, \bibinfo
  {author} {\bibfnamefont {D.~J.}\ \bibnamefont {Young}}, \bibinfo {author}
  {\bibfnamefont {R.}~\bibnamefont {Brierley}}, \bibinfo {author}
  {\bibfnamefont {P.}~\bibnamefont {Reinhold}}, \bibinfo {author}
  {\bibfnamefont {C.}~\bibnamefont {Vuillot}}, \bibinfo {author} {\bibfnamefont
  {L.}~\bibnamefont {Li}}, \bibinfo {author} {\bibfnamefont {C.}~\bibnamefont
  {Shen}}, \bibinfo {author} {\bibfnamefont {S.}~\bibnamefont {Girvin}}, \emph
  {et~al.},\ }\href {https://doi.org/10.1103/PhysRevA.97.032346} {\bibfield
  {journal} {\bibinfo  {journal} {Physical Review A}\ }\textbf {\bibinfo
  {volume} {97}},\ \bibinfo {pages} {032346} (\bibinfo {year}
  {2018})}\BibitemShut {NoStop}%
\bibitem [{\citenamefont {Ofek}\ \emph {et~al.}(2016)\citenamefont {Ofek},
  \citenamefont {Petrenko}, \citenamefont {Heeres}, \citenamefont {Reinhold},
  \citenamefont {Leghtas}, \citenamefont {Vlastakis}, \citenamefont {Liu},
  \citenamefont {Frunzio}, \citenamefont {Girvin}, \citenamefont {Jiang} \emph
  {et~al.}}]{ofek2016extending}%
  \BibitemOpen
  \bibfield  {author} {\bibinfo {author} {\bibfnamefont {N.}~\bibnamefont
  {Ofek}}, \bibinfo {author} {\bibfnamefont {A.}~\bibnamefont {Petrenko}},
  \bibinfo {author} {\bibfnamefont {R.}~\bibnamefont {Heeres}}, \bibinfo
  {author} {\bibfnamefont {P.}~\bibnamefont {Reinhold}}, \bibinfo {author}
  {\bibfnamefont {Z.}~\bibnamefont {Leghtas}}, \bibinfo {author} {\bibfnamefont
  {B.}~\bibnamefont {Vlastakis}}, \bibinfo {author} {\bibfnamefont
  {Y.}~\bibnamefont {Liu}}, \bibinfo {author} {\bibfnamefont {L.}~\bibnamefont
  {Frunzio}}, \bibinfo {author} {\bibfnamefont {S.}~\bibnamefont {Girvin}},
  \bibinfo {author} {\bibfnamefont {L.}~\bibnamefont {Jiang}}, \emph {et~al.},\
  }\href {https://doi.org/10.1038/nature18949} {\bibfield  {journal} {\bibinfo
  {journal} {Nature}\ }\textbf {\bibinfo {volume} {536}},\ \bibinfo {pages}
  {441} (\bibinfo {year} {2016})}\BibitemShut {NoStop}%
\bibitem [{\citenamefont {Hu}\ \emph {et~al.}(2019)\citenamefont {Hu},
  \citenamefont {Ma}, \citenamefont {Cai}, \citenamefont {Mu}, \citenamefont
  {Xu}, \citenamefont {Wang}, \citenamefont {Wu}, \citenamefont {Wang},
  \citenamefont {Song}, \citenamefont {Zou}, \citenamefont {Girvin},
  \citenamefont {Duan},\ and\ \citenamefont {Sun}}]{HuL19}%
  \BibitemOpen
  \bibfield  {author} {\bibinfo {author} {\bibfnamefont {L.}~\bibnamefont
  {Hu}}, \bibinfo {author} {\bibfnamefont {Y.}~\bibnamefont {Ma}}, \bibinfo
  {author} {\bibfnamefont {W.}~\bibnamefont {Cai}}, \bibinfo {author}
  {\bibfnamefont {X.}~\bibnamefont {Mu}}, \bibinfo {author} {\bibfnamefont
  {Y.}~\bibnamefont {Xu}}, \bibinfo {author} {\bibfnamefont {W.}~\bibnamefont
  {Wang}}, \bibinfo {author} {\bibfnamefont {Y.}~\bibnamefont {Wu}}, \bibinfo
  {author} {\bibfnamefont {H.}~\bibnamefont {Wang}}, \bibinfo {author}
  {\bibfnamefont {Y.~P.}\ \bibnamefont {Song}}, \bibinfo {author}
  {\bibfnamefont {C.~L.}\ \bibnamefont {Zou}}, \bibinfo {author} {\bibfnamefont
  {S.~M.}\ \bibnamefont {Girvin}}, \bibinfo {author} {\bibfnamefont {L.~M.}\
  \bibnamefont {Duan}},\ and\ \bibinfo {author} {\bibfnamefont
  {L.}~\bibnamefont {Sun}},\ }\href {https://doi.org/10.1038/s41567-018-0414-3}
  {\bibfield  {journal} {\bibinfo  {journal} {Nature Physics}\ }\textbf
  {\bibinfo {volume} {15}},\ \bibinfo {pages} {503} (\bibinfo {year}
  {2019})}\BibitemShut {NoStop}%
\bibitem [{\citenamefont {Campagne-Ibarcq}\ \emph {et~al.}(2020)\citenamefont
  {Campagne-Ibarcq}, \citenamefont {Eickbusch}, \citenamefont {Touzard},
  \citenamefont {Zalys-Geller}, \citenamefont {Frattini}, \citenamefont
  {Sivak}, \citenamefont {Reinhold}, \citenamefont {Puri}, \citenamefont
  {Shankar}, \citenamefont {Schoelkopf} \emph {et~al.}}]{campagne2020quantum}%
  \BibitemOpen
  \bibfield  {author} {\bibinfo {author} {\bibfnamefont {P.}~\bibnamefont
  {Campagne-Ibarcq}}, \bibinfo {author} {\bibfnamefont {A.}~\bibnamefont
  {Eickbusch}}, \bibinfo {author} {\bibfnamefont {S.}~\bibnamefont {Touzard}},
  \bibinfo {author} {\bibfnamefont {E.}~\bibnamefont {Zalys-Geller}}, \bibinfo
  {author} {\bibfnamefont {N.~E.}\ \bibnamefont {Frattini}}, \bibinfo {author}
  {\bibfnamefont {V.~V.}\ \bibnamefont {Sivak}}, \bibinfo {author}
  {\bibfnamefont {P.}~\bibnamefont {Reinhold}}, \bibinfo {author}
  {\bibfnamefont {S.}~\bibnamefont {Puri}}, \bibinfo {author} {\bibfnamefont
  {S.}~\bibnamefont {Shankar}}, \bibinfo {author} {\bibfnamefont {R.~J.}\
  \bibnamefont {Schoelkopf}}, \emph {et~al.},\ }\href
  {https://doi.org/10.1038/s41586-020-2603-3} {\bibfield  {journal} {\bibinfo
  {journal} {Nature}\ }\textbf {\bibinfo {volume} {584}},\ \bibinfo {pages}
  {368} (\bibinfo {year} {2020})}\BibitemShut {NoStop}%
\bibitem [{\citenamefont {Lescanne}\ \emph {et~al.}(2020)\citenamefont
  {Lescanne}, \citenamefont {Villiers}, \citenamefont {Peronnin}, \citenamefont
  {Sarlette}, \citenamefont {Delbecq}, \citenamefont {Huard}, \citenamefont
  {Kontos}, \citenamefont {Mirrahimi},\ and\ \citenamefont
  {Leghtas}}]{lescanne2020exponential}%
  \BibitemOpen
  \bibfield  {author} {\bibinfo {author} {\bibfnamefont {R.}~\bibnamefont
  {Lescanne}}, \bibinfo {author} {\bibfnamefont {M.}~\bibnamefont {Villiers}},
  \bibinfo {author} {\bibfnamefont {T.}~\bibnamefont {Peronnin}}, \bibinfo
  {author} {\bibfnamefont {A.}~\bibnamefont {Sarlette}}, \bibinfo {author}
  {\bibfnamefont {M.}~\bibnamefont {Delbecq}}, \bibinfo {author} {\bibfnamefont
  {B.}~\bibnamefont {Huard}}, \bibinfo {author} {\bibfnamefont
  {T.}~\bibnamefont {Kontos}}, \bibinfo {author} {\bibfnamefont
  {M.}~\bibnamefont {Mirrahimi}},\ and\ \bibinfo {author} {\bibfnamefont
  {Z.}~\bibnamefont {Leghtas}},\ }\href
  {https://doi.org/10.1038/s41567-020-0824-x} {\bibfield  {journal} {\bibinfo
  {journal} {Nature Physics}\ }\textbf {\bibinfo {volume} {16}},\ \bibinfo
  {pages} {509} (\bibinfo {year} {2020})}\BibitemShut {NoStop}%
\bibitem [{\citenamefont {Flühmann}\ \emph {et~al.}(2019)\citenamefont
  {Flühmann}, \citenamefont {Nguyen}, \citenamefont {Marinelli}, \citenamefont
  {Negnevitsky}, \citenamefont {Mehta},\ and\ \citenamefont
  {Home}}]{Fluhmann19}%
  \BibitemOpen
  \bibfield  {author} {\bibinfo {author} {\bibfnamefont {C.}~\bibnamefont
  {Flühmann}}, \bibinfo {author} {\bibfnamefont {T.~L.}\ \bibnamefont
  {Nguyen}}, \bibinfo {author} {\bibfnamefont {M.}~\bibnamefont {Marinelli}},
  \bibinfo {author} {\bibfnamefont {V.}~\bibnamefont {Negnevitsky}}, \bibinfo
  {author} {\bibfnamefont {K.}~\bibnamefont {Mehta}},\ and\ \bibinfo {author}
  {\bibfnamefont {J.~P.}\ \bibnamefont {Home}},\ }\href
  {https://doi.org/10.1038/s41586-019-0960-6} {\bibfield  {journal} {\bibinfo
  {journal} {Nature}\ }\textbf {\bibinfo {volume} {566}},\ \bibinfo {pages}
  {513} (\bibinfo {year} {2019})}\BibitemShut {NoStop}%
\bibitem [{\citenamefont {Grimm}\ \emph {et~al.}(2020)\citenamefont {Grimm},
  \citenamefont {Frattini}, \citenamefont {Puri}, \citenamefont {Mundhada},
  \citenamefont {Touzard}, \citenamefont {Mirrahimi}, \citenamefont {Girvin},
  \citenamefont {Shankar},\ and\ \citenamefont
  {Devoret}}]{grimm2020stabilization}%
  \BibitemOpen
  \bibfield  {author} {\bibinfo {author} {\bibfnamefont {A.}~\bibnamefont
  {Grimm}}, \bibinfo {author} {\bibfnamefont {N.~E.}\ \bibnamefont {Frattini}},
  \bibinfo {author} {\bibfnamefont {S.}~\bibnamefont {Puri}}, \bibinfo {author}
  {\bibfnamefont {S.~O.}\ \bibnamefont {Mundhada}}, \bibinfo {author}
  {\bibfnamefont {S.}~\bibnamefont {Touzard}}, \bibinfo {author} {\bibfnamefont
  {M.}~\bibnamefont {Mirrahimi}}, \bibinfo {author} {\bibfnamefont {S.~M.}\
  \bibnamefont {Girvin}}, \bibinfo {author} {\bibfnamefont {S.}~\bibnamefont
  {Shankar}},\ and\ \bibinfo {author} {\bibfnamefont {M.~H.}\ \bibnamefont
  {Devoret}},\ }\href {https://doi.org/10.1038/s41586-020-2587-z} {\bibfield
  {journal} {\bibinfo  {journal} {Nature}\ }\textbf {\bibinfo {volume} {584}},\
  \bibinfo {pages} {205} (\bibinfo {year} {2020})}\BibitemShut {NoStop}%
\bibitem [{\citenamefont {Egan}\ \emph {et~al.}(2021)\citenamefont {Egan},
  \citenamefont {Debroy}, \citenamefont {Noel}, \citenamefont {Risinger},
  \citenamefont {Zhu}, \citenamefont {Biswas}, \citenamefont {Newman},
  \citenamefont {Li}, \citenamefont {Brown}, \citenamefont {Cetina} \emph
  {et~al.}}]{egan2021fault}%
  \BibitemOpen
  \bibfield  {author} {\bibinfo {author} {\bibfnamefont {L.}~\bibnamefont
  {Egan}}, \bibinfo {author} {\bibfnamefont {D.~M.}\ \bibnamefont {Debroy}},
  \bibinfo {author} {\bibfnamefont {C.}~\bibnamefont {Noel}}, \bibinfo {author}
  {\bibfnamefont {A.}~\bibnamefont {Risinger}}, \bibinfo {author}
  {\bibfnamefont {D.}~\bibnamefont {Zhu}}, \bibinfo {author} {\bibfnamefont
  {D.}~\bibnamefont {Biswas}}, \bibinfo {author} {\bibfnamefont
  {M.}~\bibnamefont {Newman}}, \bibinfo {author} {\bibfnamefont
  {M.}~\bibnamefont {Li}}, \bibinfo {author} {\bibfnamefont {K.~R.}\
  \bibnamefont {Brown}}, \bibinfo {author} {\bibfnamefont {M.}~\bibnamefont
  {Cetina}}, \emph {et~al.},\ }\href
  {https://doi.org/10.1038/s41586-021-03928-y} {\bibfield  {journal} {\bibinfo
  {journal} {Nature}\ }\textbf {\bibinfo {volume} {598}},\ \bibinfo {pages}
  {281} (\bibinfo {year} {2021})}\BibitemShut {NoStop}%
\bibitem [{\citenamefont {Zhao}\ \emph {et~al.}(2022)\citenamefont {Zhao},
  \citenamefont {Ye}, \citenamefont {Huang}, \citenamefont {Zhang},
  \citenamefont {Wu}, \citenamefont {Guan}, \citenamefont {Zhu}, \citenamefont
  {Wei}, \citenamefont {He}, \citenamefont {Cao} \emph
  {et~al.}}]{zhao2022realization}%
  \BibitemOpen
  \bibfield  {author} {\bibinfo {author} {\bibfnamefont {Y.}~\bibnamefont
  {Zhao}}, \bibinfo {author} {\bibfnamefont {Y.}~\bibnamefont {Ye}}, \bibinfo
  {author} {\bibfnamefont {H.-L.}\ \bibnamefont {Huang}}, \bibinfo {author}
  {\bibfnamefont {Y.}~\bibnamefont {Zhang}}, \bibinfo {author} {\bibfnamefont
  {D.}~\bibnamefont {Wu}}, \bibinfo {author} {\bibfnamefont {H.}~\bibnamefont
  {Guan}}, \bibinfo {author} {\bibfnamefont {Q.}~\bibnamefont {Zhu}}, \bibinfo
  {author} {\bibfnamefont {Z.}~\bibnamefont {Wei}}, \bibinfo {author}
  {\bibfnamefont {T.}~\bibnamefont {He}}, \bibinfo {author} {\bibfnamefont
  {S.}~\bibnamefont {Cao}}, \emph {et~al.},\ }\href
  {https://doi.org/10.1103/PhysRevLett.129.030501} {\bibfield  {journal}
  {\bibinfo  {journal} {Physical Review Letters}\ }\textbf {\bibinfo {volume}
  {129}},\ \bibinfo {pages} {030501} (\bibinfo {year} {2022})}\BibitemShut
  {NoStop}%
\bibitem [{\citenamefont {Ryan-Anderson}\ \emph {et~al.}(2022)\citenamefont
  {Ryan-Anderson}, \citenamefont {Brown}, \citenamefont {Allman}, \citenamefont
  {Arkin}, \citenamefont {Asa-Attuah}, \citenamefont {Baldwin}, \citenamefont
  {Berg}, \citenamefont {Bohnet}, \citenamefont {Braxton}, \citenamefont
  {Burdick} \emph {et~al.}}]{ryan2022implementing}%
  \BibitemOpen
  \bibfield  {author} {\bibinfo {author} {\bibfnamefont {C.}~\bibnamefont
  {Ryan-Anderson}}, \bibinfo {author} {\bibfnamefont {N.}~\bibnamefont
  {Brown}}, \bibinfo {author} {\bibfnamefont {M.}~\bibnamefont {Allman}},
  \bibinfo {author} {\bibfnamefont {B.}~\bibnamefont {Arkin}}, \bibinfo
  {author} {\bibfnamefont {G.}~\bibnamefont {Asa-Attuah}}, \bibinfo {author}
  {\bibfnamefont {C.}~\bibnamefont {Baldwin}}, \bibinfo {author} {\bibfnamefont
  {J.}~\bibnamefont {Berg}}, \bibinfo {author} {\bibfnamefont {J.}~\bibnamefont
  {Bohnet}}, \bibinfo {author} {\bibfnamefont {S.}~\bibnamefont {Braxton}},
  \bibinfo {author} {\bibfnamefont {N.}~\bibnamefont {Burdick}}, \emph
  {et~al.},\ }\href {https://arxiv.org/abs/2208.01863} {\bibfield  {journal}
  {\bibinfo  {journal} {arXiv preprint arXiv:2208.01863}\ } (\bibinfo {year}
  {2022})}\BibitemShut {NoStop}%
\bibitem [{\citenamefont {Acharya}\ \emph {et~al.}(2022)\citenamefont
  {Acharya}, \citenamefont {Aleiner}, \citenamefont {Allen}, \citenamefont
  {Andersen}, \citenamefont {Ansmann}, \citenamefont {Arute}, \citenamefont
  {Arya}, \citenamefont {Asfaw}, \citenamefont {Atalaya}, \citenamefont
  {Babbush} \emph {et~al.}}]{acharya2022suppressing}%
  \BibitemOpen
  \bibfield  {author} {\bibinfo {author} {\bibfnamefont {R.}~\bibnamefont
  {Acharya}}, \bibinfo {author} {\bibfnamefont {I.}~\bibnamefont {Aleiner}},
  \bibinfo {author} {\bibfnamefont {R.}~\bibnamefont {Allen}}, \bibinfo
  {author} {\bibfnamefont {T.~I.}\ \bibnamefont {Andersen}}, \bibinfo {author}
  {\bibfnamefont {M.}~\bibnamefont {Ansmann}}, \bibinfo {author} {\bibfnamefont
  {F.}~\bibnamefont {Arute}}, \bibinfo {author} {\bibfnamefont
  {K.}~\bibnamefont {Arya}}, \bibinfo {author} {\bibfnamefont {A.}~\bibnamefont
  {Asfaw}}, \bibinfo {author} {\bibfnamefont {J.}~\bibnamefont {Atalaya}},
  \bibinfo {author} {\bibfnamefont {R.}~\bibnamefont {Babbush}}, \emph
  {et~al.},\ }\href {https://arxiv.org/abs/2207.06431} {\bibfield  {journal}
  {\bibinfo  {journal} {arXiv preprint arXiv:2207.06431}\ } (\bibinfo {year}
  {2022})}\BibitemShut {NoStop}%
\bibitem [{\citenamefont {Lebreuilly}\ \emph {et~al.}(2021)\citenamefont
  {Lebreuilly}, \citenamefont {Noh}, \citenamefont {Wang}, \citenamefont
  {Girvin},\ and\ \citenamefont {Jiang}}]{lebreuilly2021autonomous}%
  \BibitemOpen
  \bibfield  {author} {\bibinfo {author} {\bibfnamefont {J.}~\bibnamefont
  {Lebreuilly}}, \bibinfo {author} {\bibfnamefont {K.}~\bibnamefont {Noh}},
  \bibinfo {author} {\bibfnamefont {C.-H.}\ \bibnamefont {Wang}}, \bibinfo
  {author} {\bibfnamefont {S.~M.}\ \bibnamefont {Girvin}},\ and\ \bibinfo
  {author} {\bibfnamefont {L.}~\bibnamefont {Jiang}},\ }\href
  {https://arxiv.org/abs/2103.05007} {\bibfield  {journal} {\bibinfo  {journal}
  {arXiv preprint arXiv:2103.05007}\ } (\bibinfo {year} {2021})}\BibitemShut
  {NoStop}%
\bibitem [{\citenamefont {Berdou}\ \emph {et~al.}(2022)\citenamefont {Berdou},
  \citenamefont {Murani}, \citenamefont {Reglade}, \citenamefont {Smith},
  \citenamefont {Villiers}, \citenamefont {Palomo}, \citenamefont {Rosticher},
  \citenamefont {Denis}, \citenamefont {Morfin}, \citenamefont {Delbecq} \emph
  {et~al.}}]{berdou2022one}%
  \BibitemOpen
  \bibfield  {author} {\bibinfo {author} {\bibfnamefont {C.}~\bibnamefont
  {Berdou}}, \bibinfo {author} {\bibfnamefont {A.}~\bibnamefont {Murani}},
  \bibinfo {author} {\bibfnamefont {U.}~\bibnamefont {Reglade}}, \bibinfo
  {author} {\bibfnamefont {W.}~\bibnamefont {Smith}}, \bibinfo {author}
  {\bibfnamefont {M.}~\bibnamefont {Villiers}}, \bibinfo {author}
  {\bibfnamefont {J.}~\bibnamefont {Palomo}}, \bibinfo {author} {\bibfnamefont
  {M.}~\bibnamefont {Rosticher}}, \bibinfo {author} {\bibfnamefont
  {A.}~\bibnamefont {Denis}}, \bibinfo {author} {\bibfnamefont
  {P.}~\bibnamefont {Morfin}}, \bibinfo {author} {\bibfnamefont
  {M.}~\bibnamefont {Delbecq}}, \emph {et~al.},\ }\href
  {https://arxiv.org/abs/2204.09128} {\bibfield  {journal} {\bibinfo  {journal}
  {arXiv preprint arXiv:2204.09128}\ } (\bibinfo {year} {2022})}\BibitemShut
  {NoStop}%
\bibitem [{\citenamefont {Mirrahimi}\ \emph {et~al.}(2014)\citenamefont
  {Mirrahimi}, \citenamefont {Leghtas}, \citenamefont {Albert}, \citenamefont
  {Touzard}, \citenamefont {Schoelkopf}, \citenamefont {Jiang},\ and\
  \citenamefont {Devoret}}]{mirrahimi2014dynamically}%
  \BibitemOpen
  \bibfield  {author} {\bibinfo {author} {\bibfnamefont {M.}~\bibnamefont
  {Mirrahimi}}, \bibinfo {author} {\bibfnamefont {Z.}~\bibnamefont {Leghtas}},
  \bibinfo {author} {\bibfnamefont {V.~V.}\ \bibnamefont {Albert}}, \bibinfo
  {author} {\bibfnamefont {S.}~\bibnamefont {Touzard}}, \bibinfo {author}
  {\bibfnamefont {R.~J.}\ \bibnamefont {Schoelkopf}}, \bibinfo {author}
  {\bibfnamefont {L.}~\bibnamefont {Jiang}},\ and\ \bibinfo {author}
  {\bibfnamefont {M.~H.}\ \bibnamefont {Devoret}},\ }\href
  {https://iopscience.iop.org/article/10.1088/1367-2630/16/4/045014/meta}
  {\bibfield  {journal} {\bibinfo  {journal} {New Journal of Physics}\ }\textbf
  {\bibinfo {volume} {16}},\ \bibinfo {pages} {045014} (\bibinfo {year}
  {2014})}\BibitemShut {NoStop}%
\bibitem [{\citenamefont {Royer}\ \emph {et~al.}(2020)\citenamefont {Royer},
  \citenamefont {Singh},\ and\ \citenamefont
  {Girvin}}]{royer2020stabilization}%
  \BibitemOpen
  \bibfield  {author} {\bibinfo {author} {\bibfnamefont {B.}~\bibnamefont
  {Royer}}, \bibinfo {author} {\bibfnamefont {S.}~\bibnamefont {Singh}},\ and\
  \bibinfo {author} {\bibfnamefont {S.}~\bibnamefont {Girvin}},\ }\href
  {https://doi.org/10.1103/PhysRevLett.125.260509} {\bibfield  {journal}
  {\bibinfo  {journal} {Physical Review Letters}\ }\textbf {\bibinfo {volume}
  {125}},\ \bibinfo {pages} {260509} (\bibinfo {year} {2020})}\BibitemShut
  {NoStop}%
\bibitem [{\citenamefont {Gertler}\ \emph {et~al.}(2021)\citenamefont
  {Gertler}, \citenamefont {Baker}, \citenamefont {Li}, \citenamefont {Shirol},
  \citenamefont {Koch},\ and\ \citenamefont {Wang}}]{gertler2021protecting}%
  \BibitemOpen
  \bibfield  {author} {\bibinfo {author} {\bibfnamefont {J.~M.}\ \bibnamefont
  {Gertler}}, \bibinfo {author} {\bibfnamefont {B.}~\bibnamefont {Baker}},
  \bibinfo {author} {\bibfnamefont {J.}~\bibnamefont {Li}}, \bibinfo {author}
  {\bibfnamefont {S.}~\bibnamefont {Shirol}}, \bibinfo {author} {\bibfnamefont
  {J.}~\bibnamefont {Koch}},\ and\ \bibinfo {author} {\bibfnamefont
  {C.}~\bibnamefont {Wang}},\ }\href
  {https://doi.org/10.1038/s41586-021-03257-0} {\bibfield  {journal} {\bibinfo
  {journal} {Nature}\ }\textbf {\bibinfo {volume} {590}},\ \bibinfo {pages}
  {243} (\bibinfo {year} {2021})}\BibitemShut {NoStop}%
\bibitem [{\citenamefont {Reimpell}\ and\ \citenamefont
  {Werner}(2005)}]{reimpell2005iterative}%
  \BibitemOpen
  \bibfield  {author} {\bibinfo {author} {\bibfnamefont {M.}~\bibnamefont
  {Reimpell}}\ and\ \bibinfo {author} {\bibfnamefont {R.~F.}\ \bibnamefont
  {Werner}},\ }\href {https://doi.org/10.1103/PhysRevLett.94.080501} {\bibfield
   {journal} {\bibinfo  {journal} {Physical review letters}\ }\textbf {\bibinfo
  {volume} {94}},\ \bibinfo {pages} {080501} (\bibinfo {year}
  {2005})}\BibitemShut {NoStop}%
\bibitem [{\citenamefont {Fletcher}\ \emph {et~al.}(2007)\citenamefont
  {Fletcher}, \citenamefont {Shor},\ and\ \citenamefont
  {Win}}]{fletcher2007optimum}%
  \BibitemOpen
  \bibfield  {author} {\bibinfo {author} {\bibfnamefont {A.~S.}\ \bibnamefont
  {Fletcher}}, \bibinfo {author} {\bibfnamefont {P.~W.}\ \bibnamefont {Shor}},\
  and\ \bibinfo {author} {\bibfnamefont {M.~Z.}\ \bibnamefont {Win}},\ }\href
  {https://doi.org/10.1103/PhysRevA.75.012338} {\bibfield  {journal} {\bibinfo
  {journal} {Physical Review A}\ }\textbf {\bibinfo {volume} {75}},\ \bibinfo
  {pages} {012338} (\bibinfo {year} {2007})}\BibitemShut {NoStop}%
\bibitem [{\citenamefont {Noh}\ \emph {et~al.}(2018)\citenamefont {Noh},
  \citenamefont {Albert},\ and\ \citenamefont {Jiang}}]{noh2018quantum}%
  \BibitemOpen
  \bibfield  {author} {\bibinfo {author} {\bibfnamefont {K.}~\bibnamefont
  {Noh}}, \bibinfo {author} {\bibfnamefont {V.~V.}\ \bibnamefont {Albert}},\
  and\ \bibinfo {author} {\bibfnamefont {L.}~\bibnamefont {Jiang}},\ }\href
  {https://ieeexplore.ieee.org/abstract/document/8482307?casa_token=Ltbdp2oSEQoAAAAA:bILcLqTiJ6rEXCZBXON8AiIIOV51zTg2qu6GDJ0mYhmLJpbj9DvZYGI_QnMbskqNvjKn0pM}
  {\bibfield  {journal} {\bibinfo  {journal} {IEEE Transactions on Information
  Theory}\ }\textbf {\bibinfo {volume} {65}},\ \bibinfo {pages} {2563}
  (\bibinfo {year} {2018})}\BibitemShut {NoStop}%
\bibitem [{\citenamefont {Poyatos}\ \emph {et~al.}(1996)\citenamefont
  {Poyatos}, \citenamefont {Cirac},\ and\ \citenamefont
  {Zoller}}]{poyatos1996quantum}%
  \BibitemOpen
  \bibfield  {author} {\bibinfo {author} {\bibfnamefont {J.}~\bibnamefont
  {Poyatos}}, \bibinfo {author} {\bibfnamefont {J.~I.}\ \bibnamefont {Cirac}},\
  and\ \bibinfo {author} {\bibfnamefont {P.}~\bibnamefont {Zoller}},\ }\href
  {https://doi.org/10.1103/PhysRevLett.77.4728} {\bibfield  {journal} {\bibinfo
   {journal} {Physical review letters}\ }\textbf {\bibinfo {volume} {77}},\
  \bibinfo {pages} {4728} (\bibinfo {year} {1996})}\BibitemShut {NoStop}%
\bibitem [{\citenamefont {Tuckett}\ \emph {et~al.}(2018)\citenamefont
  {Tuckett}, \citenamefont {Bartlett},\ and\ \citenamefont
  {Flammia}}]{tuckett2018ultrahigh}%
  \BibitemOpen
  \bibfield  {author} {\bibinfo {author} {\bibfnamefont {D.~K.}\ \bibnamefont
  {Tuckett}}, \bibinfo {author} {\bibfnamefont {S.~D.}\ \bibnamefont
  {Bartlett}},\ and\ \bibinfo {author} {\bibfnamefont {S.~T.}\ \bibnamefont
  {Flammia}},\ }\href {https://doi.org/10.1103/PhysRevLett.120.050505}
  {\bibfield  {journal} {\bibinfo  {journal} {Physical review letters}\
  }\textbf {\bibinfo {volume} {120}},\ \bibinfo {pages} {050505} (\bibinfo
  {year} {2018})}\BibitemShut {NoStop}%
\bibitem [{\citenamefont {Tuckett}\ \emph {et~al.}(2019)\citenamefont
  {Tuckett}, \citenamefont {Darmawan}, \citenamefont {Chubb}, \citenamefont
  {Bravyi}, \citenamefont {Bartlett},\ and\ \citenamefont
  {Flammia}}]{tuckett2019tailoring}%
  \BibitemOpen
  \bibfield  {author} {\bibinfo {author} {\bibfnamefont {D.~K.}\ \bibnamefont
  {Tuckett}}, \bibinfo {author} {\bibfnamefont {A.~S.}\ \bibnamefont
  {Darmawan}}, \bibinfo {author} {\bibfnamefont {C.~T.}\ \bibnamefont {Chubb}},
  \bibinfo {author} {\bibfnamefont {S.}~\bibnamefont {Bravyi}}, \bibinfo
  {author} {\bibfnamefont {S.~D.}\ \bibnamefont {Bartlett}},\ and\ \bibinfo
  {author} {\bibfnamefont {S.~T.}\ \bibnamefont {Flammia}},\ }\href
  {https://doi.org/10.1103/PhysRevX.9.041031} {\bibfield  {journal} {\bibinfo
  {journal} {Physical Review X}\ }\textbf {\bibinfo {volume} {9}},\ \bibinfo
  {pages} {041031} (\bibinfo {year} {2019})}\BibitemShut {NoStop}%
\bibitem [{\citenamefont {Tuckett}\ \emph {et~al.}(2020)\citenamefont
  {Tuckett}, \citenamefont {Bartlett}, \citenamefont {Flammia},\ and\
  \citenamefont {Brown}}]{tuckett2020fault}%
  \BibitemOpen
  \bibfield  {author} {\bibinfo {author} {\bibfnamefont {D.~K.}\ \bibnamefont
  {Tuckett}}, \bibinfo {author} {\bibfnamefont {S.~D.}\ \bibnamefont
  {Bartlett}}, \bibinfo {author} {\bibfnamefont {S.~T.}\ \bibnamefont
  {Flammia}},\ and\ \bibinfo {author} {\bibfnamefont {B.~J.}\ \bibnamefont
  {Brown}},\ }\href {https://doi.org/10.1103/PhysRevLett.124.130501} {\bibfield
   {journal} {\bibinfo  {journal} {Physical review letters}\ }\textbf {\bibinfo
  {volume} {124}},\ \bibinfo {pages} {130501} (\bibinfo {year}
  {2020})}\BibitemShut {NoStop}%
\bibitem [{\citenamefont {{Bonilla Ataides}}\ \emph {et~al.}(2021)\citenamefont
  {{Bonilla Ataides}}, \citenamefont {Tuckett}, \citenamefont {Bartlett},
  \citenamefont {Flammia},\ and\ \citenamefont {Brown}}]{ataides2021xzzx}%
  \BibitemOpen
  \bibfield  {author} {\bibinfo {author} {\bibfnamefont {J.~P.}\ \bibnamefont
  {{Bonilla Ataides}}}, \bibinfo {author} {\bibfnamefont {D.~K.}\ \bibnamefont
  {Tuckett}}, \bibinfo {author} {\bibfnamefont {S.~D.}\ \bibnamefont
  {Bartlett}}, \bibinfo {author} {\bibfnamefont {S.~T.}\ \bibnamefont
  {Flammia}},\ and\ \bibinfo {author} {\bibfnamefont {B.~J.}\ \bibnamefont
  {Brown}},\ }\href {https://doi.org/10.1038/s41467-021-22274-1} {\bibfield
  {journal} {\bibinfo  {journal} {Nature communications}\ }\textbf {\bibinfo
  {volume} {12}},\ \bibinfo {pages} {1} (\bibinfo {year} {2021})}\BibitemShut
  {NoStop}%
\bibitem [{\citenamefont {Roffe}\ \emph {et~al.}(2022)\citenamefont {Roffe},
  \citenamefont {Cohen}, \citenamefont {Quintivalle}, \citenamefont {Chandra},\
  and\ \citenamefont {Campbell}}]{roffe2022bias}%
  \BibitemOpen
  \bibfield  {author} {\bibinfo {author} {\bibfnamefont {J.}~\bibnamefont
  {Roffe}}, \bibinfo {author} {\bibfnamefont {L.~Z.}\ \bibnamefont {Cohen}},
  \bibinfo {author} {\bibfnamefont {A.~O.}\ \bibnamefont {Quintivalle}},
  \bibinfo {author} {\bibfnamefont {D.}~\bibnamefont {Chandra}},\ and\ \bibinfo
  {author} {\bibfnamefont {E.~T.}\ \bibnamefont {Campbell}},\ }\href
  {https://arxiv.org/abs/2202.01702} {\bibfield  {journal} {\bibinfo  {journal}
  {arXiv preprint arXiv:2202.01702}\ } (\bibinfo {year} {2022})}\BibitemShut
  {NoStop}%
\bibitem [{\citenamefont {Xu}\ \emph {et~al.}(2022{\natexlab{a}})\citenamefont
  {Xu}, \citenamefont {Mannucci}, \citenamefont {Seif}, \citenamefont {Kubica},
  \citenamefont {Flammia},\ and\ \citenamefont {Jiang}}]{xu2022tailored}%
  \BibitemOpen
  \bibfield  {author} {\bibinfo {author} {\bibfnamefont {Q.}~\bibnamefont
  {Xu}}, \bibinfo {author} {\bibfnamefont {N.}~\bibnamefont {Mannucci}},
  \bibinfo {author} {\bibfnamefont {A.}~\bibnamefont {Seif}}, \bibinfo {author}
  {\bibfnamefont {A.}~\bibnamefont {Kubica}}, \bibinfo {author} {\bibfnamefont
  {S.~T.}\ \bibnamefont {Flammia}},\ and\ \bibinfo {author} {\bibfnamefont
  {L.}~\bibnamefont {Jiang}},\ }\href {https://arxiv.org/abs/2203.16486}
  {\bibfield  {journal} {\bibinfo  {journal} {arXiv preprint arXiv:2203.16486}\
  } (\bibinfo {year} {2022}{\natexlab{a}})}\BibitemShut {NoStop}%
\bibitem [{\citenamefont {Guillaud}\ and\ \citenamefont
  {Mirrahimi}(2019)}]{guillaud2019repetition}%
  \BibitemOpen
  \bibfield  {author} {\bibinfo {author} {\bibfnamefont {J.}~\bibnamefont
  {Guillaud}}\ and\ \bibinfo {author} {\bibfnamefont {M.}~\bibnamefont
  {Mirrahimi}},\ }\href {https://doi.org/10.1103/PhysRevX.9.041053} {\bibfield
  {journal} {\bibinfo  {journal} {Physical Review X}\ }\textbf {\bibinfo
  {volume} {9}},\ \bibinfo {pages} {041053} (\bibinfo {year}
  {2019})}\BibitemShut {NoStop}%
\bibitem [{\citenamefont {O'Gorman}\ and\ \citenamefont
  {Campbell}(2017)}]{o2017quantum}%
  \BibitemOpen
  \bibfield  {author} {\bibinfo {author} {\bibfnamefont {J.}~\bibnamefont
  {O'Gorman}}\ and\ \bibinfo {author} {\bibfnamefont {E.~T.}\ \bibnamefont
  {Campbell}},\ }\href {https://doi.org/10.1103/PhysRevA.95.032338} {\bibfield
  {journal} {\bibinfo  {journal} {Physical Review A}\ }\textbf {\bibinfo
  {volume} {95}},\ \bibinfo {pages} {032338} (\bibinfo {year}
  {2017})}\BibitemShut {NoStop}%
\bibitem [{\citenamefont {Schlegel}\ \emph {et~al.}(2022)\citenamefont
  {Schlegel}, \citenamefont {Minganti},\ and\ \citenamefont
  {Savona}}]{schlegel2022quantum}%
  \BibitemOpen
  \bibfield  {author} {\bibinfo {author} {\bibfnamefont {D.~S.}\ \bibnamefont
  {Schlegel}}, \bibinfo {author} {\bibfnamefont {F.}~\bibnamefont {Minganti}},\
  and\ \bibinfo {author} {\bibfnamefont {V.}~\bibnamefont {Savona}},\ }\href
  {https://arxiv.org/abs/2201.02570} {\bibfield  {journal} {\bibinfo  {journal}
  {arXiv preprint arXiv:2201.02570}\ } (\bibinfo {year} {2022})}\BibitemShut
  {NoStop}%
\bibitem [{\citenamefont {Teh}\ \emph {et~al.}(2020)\citenamefont {Teh},
  \citenamefont {Drummond},\ and\ \citenamefont {Reid}}]{teh2020overcoming}%
  \BibitemOpen
  \bibfield  {author} {\bibinfo {author} {\bibfnamefont {R.}~\bibnamefont
  {Teh}}, \bibinfo {author} {\bibfnamefont {P.}~\bibnamefont {Drummond}},\ and\
  \bibinfo {author} {\bibfnamefont {M.}~\bibnamefont {Reid}},\ }\href
  {https://doi.org/10.1103/PhysRevResearch.2.043387} {\bibfield  {journal}
  {\bibinfo  {journal} {Physical Review Research}\ }\textbf {\bibinfo {volume}
  {2}},\ \bibinfo {pages} {043387} (\bibinfo {year} {2020})}\BibitemShut
  {NoStop}%
\bibitem [{\citenamefont {Lo}\ \emph {et~al.}(2015)\citenamefont {Lo},
  \citenamefont {Kienzler}, \citenamefont {de~Clercq}, \citenamefont
  {Marinelli}, \citenamefont {Negnevitsky}, \citenamefont {Keitch},\ and\
  \citenamefont {Home}}]{lo2015spin}%
  \BibitemOpen
  \bibfield  {author} {\bibinfo {author} {\bibfnamefont {H.-Y.}\ \bibnamefont
  {Lo}}, \bibinfo {author} {\bibfnamefont {D.}~\bibnamefont {Kienzler}},
  \bibinfo {author} {\bibfnamefont {L.}~\bibnamefont {de~Clercq}}, \bibinfo
  {author} {\bibfnamefont {M.}~\bibnamefont {Marinelli}}, \bibinfo {author}
  {\bibfnamefont {V.}~\bibnamefont {Negnevitsky}}, \bibinfo {author}
  {\bibfnamefont {B.~C.}\ \bibnamefont {Keitch}},\ and\ \bibinfo {author}
  {\bibfnamefont {J.~P.}\ \bibnamefont {Home}},\ }\href
  {https://doi.org/10.1038/nature14458} {\bibfield  {journal} {\bibinfo
  {journal} {Nature}\ }\textbf {\bibinfo {volume} {521}},\ \bibinfo {pages}
  {336} (\bibinfo {year} {2015})}\BibitemShut {NoStop}%
\bibitem [{\citenamefont {Le~Jeannic}\ \emph {et~al.}(2018)\citenamefont
  {Le~Jeannic}, \citenamefont {Cavaill{\`e}s}, \citenamefont {Huang},
  \citenamefont {Filip},\ and\ \citenamefont {Laurat}}]{le2018slowing}%
  \BibitemOpen
  \bibfield  {author} {\bibinfo {author} {\bibfnamefont {H.}~\bibnamefont
  {Le~Jeannic}}, \bibinfo {author} {\bibfnamefont {A.}~\bibnamefont
  {Cavaill{\`e}s}}, \bibinfo {author} {\bibfnamefont {K.}~\bibnamefont
  {Huang}}, \bibinfo {author} {\bibfnamefont {R.}~\bibnamefont {Filip}},\ and\
  \bibinfo {author} {\bibfnamefont {J.}~\bibnamefont {Laurat}},\ }\href
  {https://doi.org/10.1103/PhysRevLett.120.073603} {\bibfield  {journal}
  {\bibinfo  {journal} {Physical Review Letters}\ }\textbf {\bibinfo {volume}
  {120}},\ \bibinfo {pages} {073603} (\bibinfo {year} {2018})}\BibitemShut
  {NoStop}%
\bibitem [{\citenamefont {Lau}\ and\ \citenamefont
  {Clerk}(2019)}]{lau2019high}%
  \BibitemOpen
  \bibfield  {author} {\bibinfo {author} {\bibfnamefont {H.-K.}\ \bibnamefont
  {Lau}}\ and\ \bibinfo {author} {\bibfnamefont {A.~A.}\ \bibnamefont
  {Clerk}},\ }\href {https://doi.org/10.1038/s41534-019-0143-1} {\bibfield
  {journal} {\bibinfo  {journal} {npj Quantum Information}\ }\textbf {\bibinfo
  {volume} {5}},\ \bibinfo {pages} {1} (\bibinfo {year} {2019})}\BibitemShut
  {NoStop}%
\bibitem [{\citenamefont {Pantaleoni}\ \emph {et~al.}(2020)\citenamefont
  {Pantaleoni}, \citenamefont {Baragiola},\ and\ \citenamefont
  {Menicucci}}]{pantaleoni2020modular}%
  \BibitemOpen
  \bibfield  {author} {\bibinfo {author} {\bibfnamefont {G.}~\bibnamefont
  {Pantaleoni}}, \bibinfo {author} {\bibfnamefont {B.~Q.}\ \bibnamefont
  {Baragiola}},\ and\ \bibinfo {author} {\bibfnamefont {N.~C.}\ \bibnamefont
  {Menicucci}},\ }\href {https://doi.org/10.1103/PhysRevLett.125.040501}
  {\bibfield  {journal} {\bibinfo  {journal} {Physical Review Letters}\
  }\textbf {\bibinfo {volume} {125}},\ \bibinfo {pages} {040501} (\bibinfo
  {year} {2020})}\BibitemShut {NoStop}%
\bibitem [{SM()}]{SM}%
  \BibitemOpen
  \href@noop {} {\emph {\bibinfo {title} {Supplementary Material}}}\BibitemShut
  {NoStop}%
\bibitem [{\citenamefont {Bennett}\ \emph {et~al.}(1996)\citenamefont
  {Bennett}, \citenamefont {DiVincenzo}, \citenamefont {Smolin},\ and\
  \citenamefont {Wootters}}]{bennett1996mixed}%
  \BibitemOpen
  \bibfield  {author} {\bibinfo {author} {\bibfnamefont {C.~H.}\ \bibnamefont
  {Bennett}}, \bibinfo {author} {\bibfnamefont {D.~P.}\ \bibnamefont
  {DiVincenzo}}, \bibinfo {author} {\bibfnamefont {J.~A.}\ \bibnamefont
  {Smolin}},\ and\ \bibinfo {author} {\bibfnamefont {W.~K.}\ \bibnamefont
  {Wootters}},\ }\href {https://doi.org/10.1103/PhysRevA.54.3824} {\bibfield
  {journal} {\bibinfo  {journal} {Physical Review A}\ }\textbf {\bibinfo
  {volume} {54}},\ \bibinfo {pages} {3824} (\bibinfo {year}
  {1996})}\BibitemShut {NoStop}%
\bibitem [{\citenamefont {Knill}\ and\ \citenamefont
  {Laflamme}(1997)}]{knill1997theory}%
  \BibitemOpen
  \bibfield  {author} {\bibinfo {author} {\bibfnamefont {E.}~\bibnamefont
  {Knill}}\ and\ \bibinfo {author} {\bibfnamefont {R.}~\bibnamefont
  {Laflamme}},\ }\href {https://doi.org/10.1103/PhysRevA.55.900} {\bibfield
  {journal} {\bibinfo  {journal} {Physical Review A}\ }\textbf {\bibinfo
  {volume} {55}},\ \bibinfo {pages} {900} (\bibinfo {year} {1997})}\BibitemShut
  {NoStop}%
\bibitem [{\citenamefont {Gross}\ \emph {et~al.}(2018)\citenamefont {Gross},
  \citenamefont {Caves}, \citenamefont {Milburn},\ and\ \citenamefont
  {Combes}}]{gross2018qubit}%
  \BibitemOpen
  \bibfield  {author} {\bibinfo {author} {\bibfnamefont {J.~A.}\ \bibnamefont
  {Gross}}, \bibinfo {author} {\bibfnamefont {C.~M.}\ \bibnamefont {Caves}},
  \bibinfo {author} {\bibfnamefont {G.~J.}\ \bibnamefont {Milburn}},\ and\
  \bibinfo {author} {\bibfnamefont {J.}~\bibnamefont {Combes}},\ }\href
  {https://iopscience.iop.org/article/10.1088/2058-9565/aaa39f/meta?casa_token=UuHxbio9iN8AAAAA:mzVdNG7MdRKIByFmMaVKJlWJdHMAAV_HomdXD-TGJVptIKtNiYg-x4Y_-HUrem_dHRyG6pxwxeCa8JpjSg}
  {\bibfield  {journal} {\bibinfo  {journal} {Quantum Science and Technology}\
  }\textbf {\bibinfo {volume} {3}},\ \bibinfo {pages} {024005} (\bibinfo {year}
  {2018})}\BibitemShut {NoStop}%
\bibitem [{\citenamefont {Chamberland}\ \emph {et~al.}(2022)\citenamefont
  {Chamberland}, \citenamefont {Noh}, \citenamefont {Arrangoiz-Arriola},
  \citenamefont {Campbell}, \citenamefont {Hann}, \citenamefont {Iverson},
  \citenamefont {Putterman}, \citenamefont {Bohdanowicz}, \citenamefont
  {Flammia}, \citenamefont {Keller} \emph {et~al.}}]{chamberland2022building}%
  \BibitemOpen
  \bibfield  {author} {\bibinfo {author} {\bibfnamefont {C.}~\bibnamefont
  {Chamberland}}, \bibinfo {author} {\bibfnamefont {K.}~\bibnamefont {Noh}},
  \bibinfo {author} {\bibfnamefont {P.}~\bibnamefont {Arrangoiz-Arriola}},
  \bibinfo {author} {\bibfnamefont {E.~T.}\ \bibnamefont {Campbell}}, \bibinfo
  {author} {\bibfnamefont {C.~T.}\ \bibnamefont {Hann}}, \bibinfo {author}
  {\bibfnamefont {J.}~\bibnamefont {Iverson}}, \bibinfo {author} {\bibfnamefont
  {H.}~\bibnamefont {Putterman}}, \bibinfo {author} {\bibfnamefont {T.~C.}\
  \bibnamefont {Bohdanowicz}}, \bibinfo {author} {\bibfnamefont {S.~T.}\
  \bibnamefont {Flammia}}, \bibinfo {author} {\bibfnamefont {A.}~\bibnamefont
  {Keller}}, \emph {et~al.},\ }\href
  {https://doi.org/10.1103/PRXQuantum.3.010329} {\bibfield  {journal} {\bibinfo
   {journal} {PRX Quantum}\ }\textbf {\bibinfo {volume} {3}},\ \bibinfo {pages}
  {010329} (\bibinfo {year} {2022})}\BibitemShut {NoStop}%
\bibitem [{Note1()}]{Note1}%
  \BibitemOpen
  \bibinfo {note} {Note that similar to the cat~\cite {puri2020bias}, the full
  error channel of the stabilized SC, which is analyzed in detail in
  supplement.~\cite {SM}, is not a Pauli error channel in general. For
  simplicity, we make the Pauli-twirling approximation only keeping the
  diagonal terms of the process matrix in the Pauli basis.}\BibitemShut {Stop}%
\bibitem [{\citenamefont {Guillaud}\ and\ \citenamefont
  {Mirrahimi}(2021)}]{guillaud2021error}%
  \BibitemOpen
  \bibfield  {author} {\bibinfo {author} {\bibfnamefont {J.}~\bibnamefont
  {Guillaud}}\ and\ \bibinfo {author} {\bibfnamefont {M.}~\bibnamefont
  {Mirrahimi}},\ }\href {https://doi.org/10.1103/PhysRevA.103.042413}
  {\bibfield  {journal} {\bibinfo  {journal} {Physical Review A}\ }\textbf
  {\bibinfo {volume} {103}},\ \bibinfo {pages} {042413} (\bibinfo {year}
  {2021})}\BibitemShut {NoStop}%
\bibitem [{Note2()}]{Note2}%
  \BibitemOpen
  \bibinfo {note} {The nonzero eigenvalue of the Lindbladian $\protect \mathcal
  {D}[\protect \hat {F}]$ with the smallest real part, which characterizes the
  rate of the population decaying into the steady-state manifold.}\BibitemShut
  {Stop}%
\bibitem [{\citenamefont {Leviant}\ \emph {et~al.}(2022)\citenamefont
  {Leviant}, \citenamefont {Xu}, \citenamefont {Jiang},\ and\ \citenamefont
  {Rosenblum}}]{leviant2022quantum}%
  \BibitemOpen
  \bibfield  {author} {\bibinfo {author} {\bibfnamefont {P.}~\bibnamefont
  {Leviant}}, \bibinfo {author} {\bibfnamefont {Q.}~\bibnamefont {Xu}},
  \bibinfo {author} {\bibfnamefont {L.}~\bibnamefont {Jiang}},\ and\ \bibinfo
  {author} {\bibfnamefont {S.}~\bibnamefont {Rosenblum}},\ }\href
  {https://arxiv.org/abs/2205.00341} {\bibfield  {journal} {\bibinfo  {journal}
  {arXiv preprint arXiv:2205.00341}\ } (\bibinfo {year} {2022})}\BibitemShut
  {NoStop}%
\bibitem [{\citenamefont {Bonilla~Ataides}\ \emph {et~al.}(2021)\citenamefont
  {Bonilla~Ataides}, \citenamefont {Tuckett}, \citenamefont {Bartlett},
  \citenamefont {Flammia},\ and\ \citenamefont {Brown}}]{bonilla2021xzzx}%
  \BibitemOpen
  \bibfield  {author} {\bibinfo {author} {\bibfnamefont {J.~P.}\ \bibnamefont
  {Bonilla~Ataides}}, \bibinfo {author} {\bibfnamefont {D.~K.}\ \bibnamefont
  {Tuckett}}, \bibinfo {author} {\bibfnamefont {S.~D.}\ \bibnamefont
  {Bartlett}}, \bibinfo {author} {\bibfnamefont {S.~T.}\ \bibnamefont
  {Flammia}},\ and\ \bibinfo {author} {\bibfnamefont {B.~J.}\ \bibnamefont
  {Brown}},\ }\href {https://doi.org/10.1038/s41467-021-22274-1} {\bibfield
  {journal} {\bibinfo  {journal} {Nature communications}\ }\textbf {\bibinfo
  {volume} {12}},\ \bibinfo {pages} {1} (\bibinfo {year} {2021})}\BibitemShut
  {NoStop}%
\bibitem [{\citenamefont {Darmawan}\ \emph {et~al.}(2021)\citenamefont
  {Darmawan}, \citenamefont {Brown}, \citenamefont {Grimsmo}, \citenamefont
  {Tuckett},\ and\ \citenamefont {Puri}}]{darmawan2021practical}%
  \BibitemOpen
  \bibfield  {author} {\bibinfo {author} {\bibfnamefont {A.~S.}\ \bibnamefont
  {Darmawan}}, \bibinfo {author} {\bibfnamefont {B.~J.}\ \bibnamefont {Brown}},
  \bibinfo {author} {\bibfnamefont {A.~L.}\ \bibnamefont {Grimsmo}}, \bibinfo
  {author} {\bibfnamefont {D.~K.}\ \bibnamefont {Tuckett}},\ and\ \bibinfo
  {author} {\bibfnamefont {S.}~\bibnamefont {Puri}},\ }\href
  {https://doi.org/10.1103/PRXQuantum.2.030345} {\bibfield  {journal} {\bibinfo
   {journal} {PRX Quantum}\ }\textbf {\bibinfo {volume} {2}},\ \bibinfo {pages}
  {030345} (\bibinfo {year} {2021})}\BibitemShut {NoStop}%
\bibitem [{\citenamefont {Puri}\ \emph {et~al.}(2020)\citenamefont {Puri},
  \citenamefont {St-Jean}, \citenamefont {Gross}, \citenamefont {Grimm},
  \citenamefont {Frattini}, \citenamefont {Iyer}, \citenamefont {Krishna},
  \citenamefont {Touzard}, \citenamefont {Jiang}, \citenamefont {Blais} \emph
  {et~al.}}]{puri2020bias}%
  \BibitemOpen
  \bibfield  {author} {\bibinfo {author} {\bibfnamefont {S.}~\bibnamefont
  {Puri}}, \bibinfo {author} {\bibfnamefont {L.}~\bibnamefont {St-Jean}},
  \bibinfo {author} {\bibfnamefont {J.~A.}\ \bibnamefont {Gross}}, \bibinfo
  {author} {\bibfnamefont {A.}~\bibnamefont {Grimm}}, \bibinfo {author}
  {\bibfnamefont {N.~E.}\ \bibnamefont {Frattini}}, \bibinfo {author}
  {\bibfnamefont {P.~S.}\ \bibnamefont {Iyer}}, \bibinfo {author}
  {\bibfnamefont {A.}~\bibnamefont {Krishna}}, \bibinfo {author} {\bibfnamefont
  {S.}~\bibnamefont {Touzard}}, \bibinfo {author} {\bibfnamefont
  {L.}~\bibnamefont {Jiang}}, \bibinfo {author} {\bibfnamefont
  {A.}~\bibnamefont {Blais}}, \emph {et~al.},\ }\href
  {https://doi.org/10.1126/sciadv.aay5901} {\bibfield  {journal} {\bibinfo
  {journal} {Science advances}\ }\textbf {\bibinfo {volume} {6}},\ \bibinfo
  {pages} {eaay5901} (\bibinfo {year} {2020})}\BibitemShut {NoStop}%
\bibitem [{\citenamefont {Xu}\ \emph {et~al.}(2022{\natexlab{b}})\citenamefont
  {Xu}, \citenamefont {Iverson}, \citenamefont {Brand{\~a}o},\ and\
  \citenamefont {Jiang}}]{xu2022engineering}%
  \BibitemOpen
  \bibfield  {author} {\bibinfo {author} {\bibfnamefont {Q.}~\bibnamefont
  {Xu}}, \bibinfo {author} {\bibfnamefont {J.~K.}\ \bibnamefont {Iverson}},
  \bibinfo {author} {\bibfnamefont {F.~G.}\ \bibnamefont {Brand{\~a}o}},\ and\
  \bibinfo {author} {\bibfnamefont {L.}~\bibnamefont {Jiang}},\ }\href
  {https://doi.org/10.1103/PhysRevResearch.4.013082} {\bibfield  {journal}
  {\bibinfo  {journal} {Physical Review Research}\ }\textbf {\bibinfo {volume}
  {4}},\ \bibinfo {pages} {013082} (\bibinfo {year}
  {2022}{\natexlab{b}})}\BibitemShut {NoStop}%
\bibitem [{\citenamefont {Yuan}\ \emph {et~al.}(2022)\citenamefont {Yuan},
  \citenamefont {Xu},\ and\ \citenamefont {Jiang}}]{yuan2022construction}%
  \BibitemOpen
  \bibfield  {author} {\bibinfo {author} {\bibfnamefont {M.}~\bibnamefont
  {Yuan}}, \bibinfo {author} {\bibfnamefont {Q.}~\bibnamefont {Xu}},\ and\
  \bibinfo {author} {\bibfnamefont {L.}~\bibnamefont {Jiang}},\ }\href
  {https://arxiv.org/abs/2208.06913} {\bibfield  {journal} {\bibinfo  {journal}
  {arXiv preprint arXiv:2208.06913}\ } (\bibinfo {year} {2022})}\BibitemShut
  {NoStop}%
\bibitem [{Note3()}]{Note3}%
  \BibitemOpen
  \bibinfo {note} {For instance, $\kappa _{c}$ is limited by filter bandwidth
  in Ref.~\cite {chamberland2022building}.}\BibitemShut {Stop}%
\bibitem [{\citenamefont {Inlek}\ \emph {et~al.}(2017)\citenamefont {Inlek},
  \citenamefont {Crocker}, \citenamefont {Lichtman}, \citenamefont {Sosnova},\
  and\ \citenamefont {Monroe}}]{inlek2017multispecies}%
  \BibitemOpen
  \bibfield  {author} {\bibinfo {author} {\bibfnamefont {I.~V.}\ \bibnamefont
  {Inlek}}, \bibinfo {author} {\bibfnamefont {C.}~\bibnamefont {Crocker}},
  \bibinfo {author} {\bibfnamefont {M.}~\bibnamefont {Lichtman}}, \bibinfo
  {author} {\bibfnamefont {K.}~\bibnamefont {Sosnova}},\ and\ \bibinfo {author}
  {\bibfnamefont {C.}~\bibnamefont {Monroe}},\ }\href
  {https://doi.org/10.1103/PhysRevLett.118.250502} {\bibfield  {journal}
  {\bibinfo  {journal} {Physical review letters}\ }\textbf {\bibinfo {volume}
  {118}},\ \bibinfo {pages} {250502} (\bibinfo {year} {2017})}\BibitemShut
  {NoStop}%
\bibitem [{\citenamefont {Bruzewicz}\ \emph {et~al.}(2019)\citenamefont
  {Bruzewicz}, \citenamefont {McConnell}, \citenamefont {Stuart}, \citenamefont
  {Sage},\ and\ \citenamefont {Chiaverini}}]{bruzewicz2019dual}%
  \BibitemOpen
  \bibfield  {author} {\bibinfo {author} {\bibfnamefont {C.}~\bibnamefont
  {Bruzewicz}}, \bibinfo {author} {\bibfnamefont {R.}~\bibnamefont
  {McConnell}}, \bibinfo {author} {\bibfnamefont {J.}~\bibnamefont {Stuart}},
  \bibinfo {author} {\bibfnamefont {J.}~\bibnamefont {Sage}},\ and\ \bibinfo
  {author} {\bibfnamefont {J.}~\bibnamefont {Chiaverini}},\ }\href
  {https://doi.org/10.1038/s41534-019-0218-z} {\bibfield  {journal} {\bibinfo
  {journal} {npj Quantum Information}\ }\textbf {\bibinfo {volume} {5}},\
  \bibinfo {pages} {1} (\bibinfo {year} {2019})}\BibitemShut {NoStop}%
\bibitem [{\citenamefont {Ringbauer}\ \emph {et~al.}(2022)\citenamefont
  {Ringbauer}, \citenamefont {Meth}, \citenamefont {Postler}, \citenamefont
  {Stricker}, \citenamefont {Blatt}, \citenamefont {Schindler},\ and\
  \citenamefont {Monz}}]{ringbauer2022universal}%
  \BibitemOpen
  \bibfield  {author} {\bibinfo {author} {\bibfnamefont {M.}~\bibnamefont
  {Ringbauer}}, \bibinfo {author} {\bibfnamefont {M.}~\bibnamefont {Meth}},
  \bibinfo {author} {\bibfnamefont {L.}~\bibnamefont {Postler}}, \bibinfo
  {author} {\bibfnamefont {R.}~\bibnamefont {Stricker}}, \bibinfo {author}
  {\bibfnamefont {R.}~\bibnamefont {Blatt}}, \bibinfo {author} {\bibfnamefont
  {P.}~\bibnamefont {Schindler}},\ and\ \bibinfo {author} {\bibfnamefont
  {T.}~\bibnamefont {Monz}},\ }\href
  {https://doi.org/10.1038/s41567-022-01658-0} {\bibfield  {journal} {\bibinfo
  {journal} {Nature Physics}\ }\textbf {\bibinfo {volume} {18}},\ \bibinfo
  {pages} {1053} (\bibinfo {year} {2022})}\BibitemShut {NoStop}%
\bibitem [{\citenamefont {Olsacher}\ \emph {et~al.}(2020)\citenamefont
  {Olsacher}, \citenamefont {Postler}, \citenamefont {Schindler}, \citenamefont
  {Monz}, \citenamefont {Zoller},\ and\ \citenamefont
  {Sieberer}}]{olsacher2020scalable}%
  \BibitemOpen
  \bibfield  {author} {\bibinfo {author} {\bibfnamefont {T.}~\bibnamefont
  {Olsacher}}, \bibinfo {author} {\bibfnamefont {L.}~\bibnamefont {Postler}},
  \bibinfo {author} {\bibfnamefont {P.}~\bibnamefont {Schindler}}, \bibinfo
  {author} {\bibfnamefont {T.}~\bibnamefont {Monz}}, \bibinfo {author}
  {\bibfnamefont {P.}~\bibnamefont {Zoller}},\ and\ \bibinfo {author}
  {\bibfnamefont {L.~M.}\ \bibnamefont {Sieberer}},\ }\href
  {https://doi.org/10.1103/PRXQuantum.1.020316} {\bibfield  {journal} {\bibinfo
   {journal} {PRX Quantum}\ }\textbf {\bibinfo {volume} {1}},\ \bibinfo {pages}
  {020316} (\bibinfo {year} {2020})}\BibitemShut {NoStop}%
\bibitem [{\citenamefont {Wang}\ \emph {et~al.}(2022)\citenamefont {Wang},
  \citenamefont {Wang},\ and\ \citenamefont {Clerk}}]{wang2022qnr}%
  \BibitemOpen
  \bibfield  {author} {\bibinfo {author} {\bibfnamefont {Y.-X.}\ \bibnamefont
  {Wang}}, \bibinfo {author} {\bibfnamefont {C.}~\bibnamefont {Wang}},\ and\
  \bibinfo {author} {\bibfnamefont {A.~A.}\ \bibnamefont {Clerk}},\ }\href
  {https://arxiv.org/abs/2203.09392} {\bibfield  {journal} {\bibinfo  {journal}
  {arXiv preprint arXiv:2203.09392}\ } (\bibinfo {year} {2022})}\BibitemShut
  {NoStop}%
\bibitem [{\citenamefont {Reiter}\ and\ \citenamefont
  {S{\o}rensen}(2012)}]{reiter2012effective}%
  \BibitemOpen
  \bibfield  {author} {\bibinfo {author} {\bibfnamefont {F.}~\bibnamefont
  {Reiter}}\ and\ \bibinfo {author} {\bibfnamefont {A.~S.}\ \bibnamefont
  {S{\o}rensen}},\ }\href {https://doi.org/10.1103/PhysRevA.85.032111}
  {\bibfield  {journal} {\bibinfo  {journal} {Physical Review A}\ }\textbf
  {\bibinfo {volume} {85}},\ \bibinfo {pages} {032111} (\bibinfo {year}
  {2012})}\BibitemShut {NoStop}%
\bibitem [{\citenamefont {Albert}(2018)}]{albert2018lindbladians}%
  \BibitemOpen
  \bibfield  {author} {\bibinfo {author} {\bibfnamefont {V.~V.}\ \bibnamefont
  {Albert}},\ }\href {https://arxiv.org/abs/1802.00010} {\bibfield  {journal}
  {\bibinfo  {journal} {arXiv preprint arXiv:1802.00010}\ } (\bibinfo {year}
  {2018})}\BibitemShut {NoStop}%
\bibitem [{\citenamefont {Gautier}\ \emph {et~al.}(2022)\citenamefont
  {Gautier}, \citenamefont {Sarlette},\ and\ \citenamefont
  {Mirrahimi}}]{gautier2022combined}%
  \BibitemOpen
  \bibfield  {author} {\bibinfo {author} {\bibfnamefont {R.}~\bibnamefont
  {Gautier}}, \bibinfo {author} {\bibfnamefont {A.}~\bibnamefont {Sarlette}},\
  and\ \bibinfo {author} {\bibfnamefont {M.}~\bibnamefont {Mirrahimi}},\ }\href
  {https://journals.aps.org/prxquantum/pdf/10.1103/PRXQuantum.3.020339}
  {\bibfield  {journal} {\bibinfo  {journal} {P R X Quantum}\ }\textbf
  {\bibinfo {volume} {3}},\ \bibinfo {pages} {020339} (\bibinfo {year}
  {2022})}\BibitemShut {NoStop}%
\end{thebibliography}


\end{document}

\title{Supplementary Material for ``Autonomous quantum error correction and fault-tolerant quantum computation with squeezed cat qubits"}

\author{Qian Xu}
\thanks{These authors contributed equally.}
\affiliation{Pritzker School of Molecular Engineering, The University of Chicago, Chicago 60637, USA}

\author{Guo Zheng}
\thanks{These authors contributed equally.}
\affiliation{Pritzker School of Molecular Engineering, The University of Chicago, Chicago 60637, USA}

\author{Yu-Xin Wang}
\affiliation{Pritzker School of Molecular Engineering, The University of Chicago, Chicago 60637, USA}

\author{Peter Zoller}
\affiliation{Institute for Theoretical Physics, University of Innsbruck, Innsbruck A-6020, Austria}
\affiliation{Institute for Quantum Optics and Quantum Information of the Austrian Academy of Sciences, Innsbruck A-6020, Austria}

\author{Aashish A. Clerk}
\affiliation{Pritzker School of Molecular Engineering, The University of Chicago, Chicago 60637, USA}

\author{Liang Jiang}
\email{liang.jiang@uchicago.edu}
\affiliation{Pritzker School of Molecular Engineering, The University of Chicago, Chicago 60637, USA}
%\affiliation{AWS Center for Quantum Computing, Pasadena, CA 91125, USA}

\maketitle

% \appendix
% \tableofcontents
\section{Subsystem decomposition of the squeezed cat}
We can divide the Hilbert space of a bosonic mode $\mathcal{H}_{\textrm{CV}} = \textrm{span}\{|n\rangle, n = 0,1,2..\}$ into two orthogonal subspaces:
$$\mathcal{H}_{CV} = \mathcal{H}_{\textrm{even}} \oplus \mathcal{H}_{\textrm{odd}},$$
where $\mathcal{H}_{\textrm{even}}$ ($\mathcal{H}_{\textrm{odd}}$) is the $+1$ ($-1$) eigenspace of the parity operator $\hat{\Pi} := e^{-i \hat{n} \pi}$. Since both $\mathcal{H}_{\textrm{even}}$ and $\mathcal{H}_{\textrm{odd}}$ are isomorphic to $\mathcal{H}_{\textrm{CV}}$, we can decompose $\mathcal{H}_{\textrm{CV}}$ into two subsystems:
$$\mathcal{H}_{\textrm{CV}}  = \mathbb{C}_L^2 \otimes \mathcal{H}_{g},$$
where $\mathbb{C}_L^2$ is a Hilbert space of dimension 2 (which we refer to as a qubit), and $\mathcal{H}_g \simeq \mathcal{H}_{\textrm{CV}}$ is a Hilbert space of infinite dimension (which we refer to as a gauge mode). Under this decomposition, $|+\rangle_L \otimes |\psi\rangle_g$ ($|-\rangle_L \otimes |\psi\rangle_g$) is an even-(odd-) parity state for any $|\psi\rangle_g \in \mathcal{H}_{g}$. The choice of the gauge mode basis is not unique. For instance, we can choose a basis for $\mathcal{H}_g$ based on the modular decomposition~\cite{pantaleoni2020modular} of the number operator: $|+\rangle_L \otimes |m\rangle_g := |0 + 2m\rangle$ ($|-\rangle_L \otimes |m\rangle_g := |1 + 2m\rangle$). For the squeezed cat (SC), it is convenient to work with another basis related to the squeezed coherent states. We first define a set of non-orthonormalized states $|\psi_{n, \pm}\rangle := \mathcal{N}_{n, \pm}  \hat{S}(r)\left[\hat{D}\left(\alpha^{\prime}\right) \pm(-1)^n \hat{D}\left(-\alpha^{\prime}\right)\right]|n\rangle$, where $\mathcal{N}_{n,\pm}$ is the normalization factor. Note that states with different parity $\pm$ are orthogonal with each other. We then apply the Gram-Schmidt orthonormalization procedure to the states within each parity branch (starting from $|\phi_{0,\pm}\rangle$ and then increasing $n$) to obtain a set of orthonomalized states $\{|\Phi_{n, \pm}\rangle\}$. Note that the states $|\Phi_{n, \pm}\rangle$ are equivalent to the shifted fock basis introduced in Ref.~\cite{chamberland2022building} by a global squeezing transformation. Finally, we define a subsystem basis as
\begin{equation}
    |\pm\rangle_L \otimes |\tilde{n} = n\rangle_g := |\Phi_{n, \pm}\rangle.
\end{equation}
The choice of this subsystem basis can describe the SC more efficiently than the fock basis since $|\pm\rangle_L \otimes |\tilde{n} = 0\rangle_g$ coincides with the SC codewords. Furthermore, the physical states of a stabilized SC usually evolve within a subspace with low excitation in the gauge mode. As such, we can apply a truncation to the gauge mode and perform the analysis within a truncated $2d$-dimensional (with $d$ being a small integer) subspace of $\mathcal{H}_{\textrm{CV}}$: $\textrm{span}\{|\pm\rangle_L \otimes |\tilde{n} = n\rangle_g, n = 0,1,..,d-1\}$.

\section{Error channel of the dissipatively stabilized squeezed cat}
\begin{figure}[h!]
    \centering
    \includegraphics[width = 0.4 \textwidth]{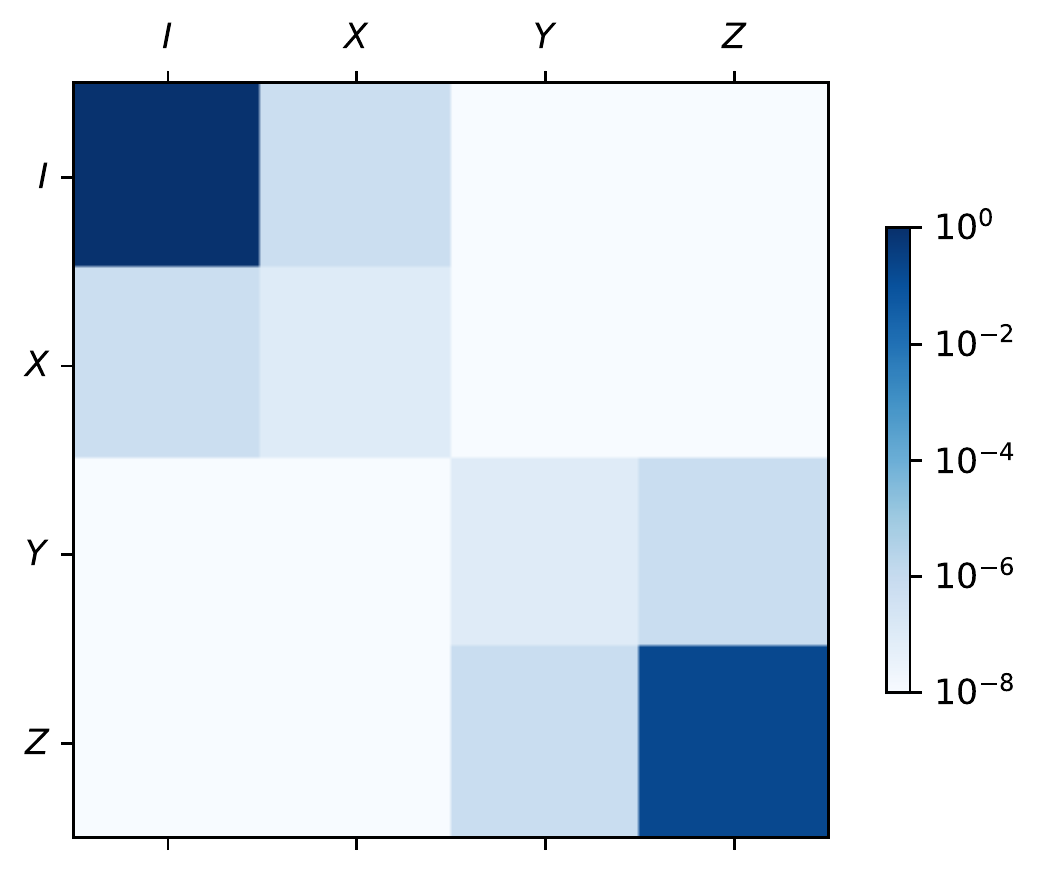}
     \caption{The absolute value of the process matrix $\chi$ for a SC stabilized by $\kappa_2 \mathcal{D}[\hat{Z}_L \hat{S}(r)(\hat{a}^2 - \alpha^{\prime 2})\hat{S}^{\dagger}(r)]$ with $\bar{n} = 4, r = 0.2$ under $\kappa_1 = 10\kappa_{\phi} = \kappa_2/100$, $n_{\textrm{th}} = 0.02$ for a time $T = 5/\kappa_2$.}
    \label{fig:channel_tomo}
\end{figure}

\hl{In this section, we provide a more detailed analysis on the error channel of the dissipatively stabilized cat, which can be represented by the process matrix $\chi$: $\mathcal{E}(\hat{\rho}) = \sum_{i,j \in \{I, X, Y, Z\}}\chi_{i,j} \hat{\sigma}_i \hat{\rho} \hat{\sigma}_j$. In Fig.~\ref{fig:channel_tomo}, we numerically obtain the matrix $\chi$ for a SC with $\bar{n} = 4, r = 0.2$ stabilized by the parity-flipping dissipator $\hat{Z}_L \hat{S}(r)(\hat{a}^2 - \alpha^{\prime 2})\hat{S}^{\dagger}(r)$ under loss, heating and dephasing. Since $\chi$ has off-diagonal elements the channel is not exactly a Pauli error channel. It is generally in the form $\mathcal{E}(\hat{\rho}) = (\lambda_I \hat{I} + \lambda_X \hat{X}) \hat{\rho} (\lambda_I \hat{I} + \lambda_X \hat{X})^{\dagger} + (\lambda_Y \hat{Y} + \lambda_Z \hat{Z})\hat{\rho}(\lambda_Y \hat{Y} + \lambda_Z \hat{Z})^{\dagger}$, with coherence between $\hat{I}, \hat{X}$ and $\hat{Y}, \hat{Z}$. Note that the cat has the same form of error channel~\cite{puri2020bias}. The Pauli error rates $\gamma_{X}, \gamma_Y, \gamma_Z$ that we use in the main text refer to the diagonal elements of $\chi$, which implicitly assumes the Pauli-twirling approximation.} 

\section{Bias-preserving operations}
In this section, we construct the remaining bias-preserving operations in $\mathcal{B} = \{\mathcal{P}_{|\pm\rangle_{c}}, \mathcal{M}_{X}, X, Z(\theta), ZZ(\theta), \text { CNOT, Toffoli} \}$, other than $Z(\theta)$ and CNOT presented in the main text.

The $ZZ$ rotation can be implemented by applying the following beam-splitter Hamiltonian to two dissipatively stabilized modes: $\hat{H}_{ZZ} = \frac{\pi}{4\alpha^{\prime 2}} e^{2r} (\hat{a}_1 \hat{a}_2^{\dagger} + \hat{a}_1^{\dagger}\hat{a}_2)$. 

The Toffoli gate can be implemented similarly to the CX gate:
$$\begin{gathered}
\frac{d}{d t} \hat{\rho}=\kappa_{2} \mathcal{D}[\hat{F}_{c_1}] + \kappa_2 \mathcal{D}[\hat{F}_{c_2}] -i[\hat{H}_{\mathrm{Tof}}, \hat{\rho}], \\
\begin{aligned}
\hat{H}_{\mathrm{Tof}} & =-\frac{\pi}{8 \alpha^{\prime 2} T}((e^{r}\hat{a}_{c_1}-\alpha^{\prime})(e^{r}\hat{a}_{c_2}^{\dagger}-\alpha^{\prime})+\text {h.c.}) \times (\hat{a}_{t}^{\dagger} \hat{a}_{t}-\alpha^{\prime 2}),\\
\end{aligned}
\end{gathered}$$
where $\hat{F}_{c_1}$ and $\hat{F}_{c_2}$ are the parity-flipping dissipator on the $c_1$ and $c_2$ mode, respectively.

The X gate can be implemented by adiabatically tuning the phases of the stabilized code states $e^{-i\frac{\pi}{T}t \hat{a}^{\dagger}\hat{a}} |SC_{r, \alpha^{\prime}}^{\pm}\rangle$ so that a $\pi$ phase rotation is implemented in time $T$. By adding a counterdiabatic drive $\hat{H}_X = \frac{\pi}{T} \hat{a}^{\dagger} \hat{a}$ the non-adiabatic effects are completely suppressed and the X gate can be implemented arbitrarily fast in principle. 

To prepare the X-basis eigenstate $|SC_{r, \alpha^{\prime}}^+\rangle$, we initialize the system into the vacuum state $|0\rangle$, and turn on the parity-preserving dissipation $\kappa_2 \mathcal{D}[\hat{S}(r)(\hat{a}^2 - \alpha^{\prime 2})\hat{S}^{\dagger}(r)]$. Since the parity is a conserved quantity, we obtain the even-parity state $|SC_{r, \alpha^{\prime}}^+\rangle$ at a time $t \gg 1/4\kappa_2 \alpha^{\prime 2}$. We note that such a process is not protected from photon losses, since the dissipation does not correct the loss-induced stochastic phase flips. We will explore the implementation of more robust state preparation against losses in future work. One possible approach is to adiabatically inflate the SC from vacuum by tuning the dissipation (with an appropriate counter-adiabatic drive~\cite{puri2017engineering}), while maintaining the phase-flip correction. 
% and adiabatically tune the dissipation so that it adiabatically inflates the cat. During such a process, we fix $\eta$ and adiabatically increase the mean photon number $\bar{n}$. The corresponding instantaneous steady state is $|SC_{r(t), \alpha^{\prime}(t)}^+\rangle$, where $r(t) = \sinh^{-1} \sqrt{(1 - \eta) \bar{n}(t)}$ and $\alpha^{\prime}(t) = \sqrt{\eta \bar{n}(t)} e^{r(t)}$. \todo{Derive the counterdiabatic drive and discuss the implementation in detail in SM.}

To perform an X-basis measurement, we turn off the engineered dissipation and apply the standard QND bosonic parity measurement using a dispersive coupling between the bosonic mode and a transmon qubit~\cite{sun2014tracking, hann2018robust, elder2020high}. A single photon loss during the dispersive coupling changes the parity and leads to stochastic measurement errors with a probability that depends on when the loss jump happens. As such, the loss-induced measurement error probability associated with such a single measurement is $p_m = \frac{1}{2}\bar{n} \kappa_1 T$. To suppress the loss-induced error, we apply the QND parity measurement three times and apply the parity-flipping dissipation to correct the loss after each measurement. Finally, we perform a majority vote to obtain the measurement outcome. Such a protocol leads to a loss-induced measurement error probability $p_m^{\prime} = \frac{3}{2}\eta \kappa_1 \bar{n} T + \mathcal{O}((\kappa_1 T)^2)$, which maintains the $\eta$ suppression factor against photon loss.

\section{\label{sec:A_map} Details of the concatenated quantum error correction}
In this section, we present the details of the concatenated QEC with the SC qubits. The physical error rates of the SC that we use for Figs. 4(a)(c) are shown in Tab.~\ref{tab:error_model1}. We fix $\bar{n} = 4$, $\eta = 0.25$ ($r = 1.32$) for all the operations. The idling time depends on the specific location in a QEC cycle (idling during the ancilla readout and initialization, idling during gate operations on other data qubits, etc.). We use the same state preparation time as in Ref.~\cite{chamberland2022building} to make sure that the leakage is below $10^{-3}$. We assume that the X-basis measurement has negligible contribution to the QEC cycle time, given that the swap between the ancilla mode and the readout mode can be implemented much faster than all the other operations~\cite{de20223, maiti2022frequency} and the time-consuming repeated parity measurements on the readout mode can be performed in parallel with the QEC cycle. Furthermore, we numerically observe that both the surface code and the repetition code have a high tolerance against ancilla measurement errors. Their logical error rates are almost unaffected for measurement errors up to $10^{-2}$. So we neglect the Z error rates from $\mathcal{M}_X$ as well. 
\begin{table}[h]
    \centering
    \begin{tabular}{|p{2cm}<{\centering}|p{2cm}<{\centering}|p{2cm}<{\centering}|p{2cm}<{\centering}|p{3cm}<{\centering}|p{3cm}<{\centering}|}
         \hline
         Operation & Idling & $\mathcal{P}_{|+\rangle}$ & $\mathcal{M}_X$ & CX (rep code) & CX (surface code)\\
         \hline 
        $\eta$ & 0.25 & 0.25 & 0.25 & 0.25 & 0.25\\
         \hline 
         Gate time & $T$ & $10*\frac{1}{\kappa_2 \alpha^{\prime 2}}$ & $\approx 0$ & $1/\kappa_2$ & $T^*$\\
         \hline
         $p_Z$ & $\kappa_1 \eta \bar{n} T$ & $\frac{10 \bar{n}}{\alpha^{\prime 2}} \frac{\kappa_1}{\kappa_2}$ & $\approx 0$ & $\approx\left(1 + \eta\right)\bar{n}\frac{\kappa_1}{\kappa_2}$ & $0.44\left(\frac{\kappa_1}{\kappa_2}\right)^{2/3}$ \\
         \hline
         $p_{X, Y}$ & $\approx 0$ &  $0$ & $0$ & $1.38\frac{e^{-2\alpha^{\prime 2}}}{\alpha^{\prime 2}}$ & $/$\\
         \hline
    \end{tabular}
    \caption{For CX gates, repetition code and surface code assumes different constraints and regimes. For repetition code, it works under the long gate time regime, and for $\kappa_1/\kappa_2 \geq 10^{-3}$, the optimal gate time is numerically found to be limited by the constraint, i.e. $T = \frac{1}{\kappa_2}$. For surface code, the optimal gate time is obtained through first principle simulations, which we denote as $T^*$. All quantities that are neglected in the simulations are labelled as $\approx$. $p_{X,Y}$ denotes the total probability of all non-dephasing errors. $p_{X, Y}$ of the CX gate in the surface-code simulation is not used.}
    \label{tab:error_model1}
\end{table}

We use the same QEC circuits and decoders (mininum-weight-perfect-matching decoder) as those in Ref.~\cite{chamberland2022building} for both the thin surface code and the repetition code. 

For surface code simulations, we presented the logical error rates varying with physical error rate ratio $\kappa_1/\kappa_2$ and threshold as a function of average photon number in $\kappa_1/\kappa_2$ in Fig.4 (a, b). We numerically obtain the optimal CX gate time for certain $\kappa_1/\kappa_2$, assuming a fixed cooling time $T_{\textrm{cool}}$. When the total Z error is optimal, we observe that the Z-error components can be described by similar relations of $p^{\textrm{opt}}_{Z}\propto \left(\frac{\kappa_1}{\kappa_2}\right)^{2/3}$. With these expressions for each Z error component, we perform circuit level simulations with Stim~\cite{gidney2021stim} to obtain logical error rates and estimations of thresholds at varying physical error ratios. In Fig.4 (b), we also includes thresholds for conventional cat encoding~\cite{chamberland2022building}, where the optimal Z error is independent of the physical error ratio. 

The repetition code simulation we performe focuses on the logical error rates and the minimal logical errors achievable, as shown in Fig.4 (c, d). For the logical error rates, the simulations are similar to surface code ones. The difference arise in the minimal logical error simulations, where the long gate time constraint is imposed. The constraint is primarily imposed such that the non-dephasing error rate of the CX gate can be expressed as $p_{X, Y} \approx 5.57 \times \frac{e^{-2 \alpha^{\prime 2}}}{\alpha^{\prime 2}}\frac{1}{\kappa_2 T}$. In this regime, the Z error is dominated by loss-induced error, which is linear in $\kappa_1/\kappa_2$. The optimal gate time for ratios larger then $\sim 10^{-3}$, is limited by this constraint. Similarly, we included minimal error rates for cat encoding, with the non-dephasing error rate expression of CX gate and Z logical error rates of repetition code shown in Table 7 and Eq. (39) in Ref.~\cite{chamberland2022building}.

\section{Physical Implementations}

As discussed in the first Methods subsection in the main text, the desired nonlinear dissipator $\mathcal{D}[\hat{F}]$ with 
$ \hat{F} \propto \hat{Z}_L \hat{S}(r) 
\left(\hat{a}^2 - \alpha^{\prime 2}\right) \hat{S}^{\dagger}(r) $ can be realized via two distinct approaches, each involving nonlinear interactions that are more compatible with circuit QED systems, or trapped ion systems, respectively. Below, we provide more details for the relevant physical implementations.

\subsection{Implementation in superconducting circuit systems: derivations of adiabatic elimination and error analysis }

Recall in the main text, we consider physical implementation via two auxiliary modes $b$ and $c$, so that the total system dynamics can be described by the following master equation 
\begin{align}
\frac{d}{dt}\hat{\rho} = 
    & -i \left [ \frac{\lambda}{2} 
\hat{Z}_L \hat{b}^{\dagger} \hat{b}  
    + (J_{ab} 
\hat{\tilde{a}}^{\dagger} \hat{b}
+ 
( J_{ac} \hat{\tilde{a}}
-i J_{bc} \hat{b} ) \hat{c}^{\dagger}
+h . c . ), \hat{\rho} \right ] + \kappa_c \mathcal{D}[\hat{c}],
\label{seq:full_three_modes_qme} 
\end{align}
where the corresponding tunnel coupling rates are given by 
\begin{equation}
J_{ab} =  {\sqrt{\Gamma_a \Gamma_b}} / {2}
, \quad 
J_{ac} =  {\sqrt{\Gamma_a \kappa_c}} / {2}
, \quad 
J_{bc} =  {\sqrt{\Gamma_b \kappa_c}} /{2}
. 
\end{equation}
We are interested in the regime where modes $b$ and $c$ effectively serve as a Markovian reservoir for the composite system consisting of logical qubit and the gauge mode degrees of freedom. In what follows, we first show how the desired system dissipator emerges following a standard adiabatic elimination procedure (e.g.~see Ref.~\cite{reiter2012effective}) and assuming the relevant parameters are chosen such that the adiabatic approximation is valid. We then exactly solve the steady state dynamics of Eq.~\eqref{seq:full_three_modes_qme} under initial state with a single excitation in the gauge mode, and use that analytical solution to quantitatively derive the conditions on the physical parameters to be in the adiabatic regime, as well as the dominating error rate introduced by Eq.~\eqref{seq:full_three_modes_qme}. 

Following the convention in Ref.~\cite{reiter2012effective}, we identify the ground and excited manifolds as the joint vacuum, and the excited state subspace of auxiliary modes $b$ and $c$, respectively. We can thus separate the total system dynamics in Eq.~\eqref{seq:full_three_modes_qme}  into contributions describing the ground (excited) state Hamiltonian $\hat{H} _{g}$ 
($\hat{H} _{e}$), the perturbative excitation (relaxation) coupling operator 
$\hat V _{+}$ ($\hat V _{-}$), as well as the decay jump operator $\hat L _{\mathrm{loss}}$ as 
\begin{align}
& \hat{H} _{g} =  0 
, \quad 
\hat{H} _{e} =  
 \frac{\lambda}{2} 
\hat{Z}_L \hat{b}^{\dagger} \hat{b}  
+ i \frac{\sqrt{\Gamma_b \kappa_c} }{2} 
(\hat{b} ^{\dagger} \hat{c}
- \hat{c}^{\dagger} \hat{b}  ) 
, \\
& \hat V _{-} = \hat V _{+} ^{\dag}
= \frac{\sqrt{\Gamma_a } }{2} 
\hat{\tilde{a}} ^{\dag} 
\left (\sqrt{\Gamma_b  }  \hat{b}
+ \sqrt{ \kappa_c}
\hat{c} \right ) 
, \quad 
\hat L _{\mathrm{loss}}
= \sqrt{ \kappa_c}
\hat{c} 
. 
\end{align}
Note that the dissipation in Eq.~\eqref{seq:full_three_modes_qme} only involves a relaxation process from the excited to the ground manifold, we can directly apply the general formulas in~\cite{reiter2012effective} to adiabatically eliminate the auxiliary modes and obtain an  effective quantum master equation involving $\hat{Z}_L $ and $ \hat{\tilde{a}} $. More specifically, the effective non-Hermitian Hamiltonian governing subspace with a single excitation in modes $b$ or $c$ can be written as \begin{align}
\hat H _{\mathrm{NH}}
& = \hat{H} _{e}   
- \frac{i }{2} 
\hat L _{\mathrm{loss}} ^{\dag} 
\hat L _{\mathrm{loss}} 
\\
& = \frac{\lambda}{2}\hat{Z}_L 
\left | 1 _{b} \rangle \langle 1 _{b} \right | 
+ i \frac{\sqrt{\Gamma_b \kappa_c} }{2} 
( \left | 1 _{b} \rangle \langle 1 _{c} \right |
- h.c.  ) 
- i \frac{\kappa_{c}}{2}
\left | 1 _{c} \rangle \langle 1 _{c} \right | 
,
\end{align}
so that we obtain the effective ground-state manifold system Hamiltonian 
$\hat H _{\mathrm{eff}}$, and the jump operator $\hat L _{\mathrm{eff}}$ as 
\begin{align}
& \hat H _{\mathrm{eff}} 
=  \hat{H} _{g} 
- \frac{1}{2} \hat V _{-} 
\left [ \hat H _{\mathrm{NH}}^{-1} 
+ (\hat H _{\mathrm{NH}}^{-1} ) ^{\dag}
\right ]  
\hat V _{+} ^{\dag}
= 0
, \\
& \hat L _{\mathrm{eff}} 
= \hat L _{\mathrm{loss}} 
\hat H _{\mathrm{NH}}^{-1} 
\hat V _{+} 
= i \sqrt{\Gamma_a } 
\hat{\tilde{a}}  
\frac{i \lambda \hat{Z}_L -\Gamma_{b}}
{i \lambda \hat{Z}_L +\Gamma_{b}} 
. 
\end{align}
We have thus derived Eq.~(18) in the main text. 

For the purpose of analyzing the dominating error due to finite decay rate 
$\kappa _{c}$ of the auxiliary reservoir mode $c$, we now view the logical qubit operator $\hat{Z}_L $ as a qubit Pauli operator, and the gauge mode operator $\hat{\tilde{a}}  $ as the lowering operator of a genuine bosonic mode. We note that this treatment is valid in the relevant parameter regime we are interested in. In such case, we can derive exactly the long-time evolution of a generic initial state with a single excitation in the gauge mode $ \tilde{a} $ and vacuum in the auxiliary modes $b$ and $c$, which can be fully captured by the coherence function 
$\hat{\Sigma} _{L-} $ (i.e.~off-diagonal density matrix element with respect to eigenbasis of $\hat{Z}_L $), as  
\begin{equation}
\label{seq:qbl.coh.bc}
\lim _{t \to \infty }
\frac{\langle \hat{\Sigma} 
_{L-} (t )  
\rangle }
{\langle \hat{\Sigma} 
_{L-} (0 )  
\rangle}
= \left (\frac{ \Gamma_{b}- i \lambda }
{\Gamma_{b} + i \lambda} \right ) ^{2} 
\frac {1 + \frac{\Gamma_a \kappa_c 
(\Gamma_{b} + i \lambda) }
{ ( \kappa_c + i \lambda )  
( \Gamma_{b}- i \lambda ) ^{2} } } 
{ 1+ \frac{\Gamma_a  
( \kappa_c ^{2} + i \Gamma_{b}\lambda )  }
{\kappa_c 
( \kappa_c + i \lambda )  
( \Gamma_{b} + i \lambda ) } }
. 
\end{equation}
To realize our desired dissipator, we set 
$\Gamma_b = \lambda $. From Eq.~\eqref{seq:qbl.coh.bc}, it is straightforward to see that the conditions for validity of adiabatic elimination can be written as  
\begin{equation}
\label{seq:bc.cond.AE}
\sqrt{\Gamma_a \Gamma_b} 
\ll  \kappa_c 
, \quad 
\Gamma_a 
\ll  \Gamma_b  
, \quad 
\Gamma_a 
\ll \kappa_c 
. 
\end{equation}
It is interesting to note that the adiabatic elimination does not require the two reservoir modes' decay rates to satisfy a hierarchy, i.e.~$\Gamma_b  \ll \kappa_c  $. This might seem surprising at the first glance, as we motivate Eq.~\eqref{seq:full_three_modes_qme} based on the intuition that mode $c$ is lossy and enables a directional coupling from $a$ to $b$. However, from the perspective of doing adiabatic approximation, we only need to ensure that the excitation exchange between system (logical qubit and gauge mode) and the auxiliary modes is much slower than the internal reservoir dynamics of $b$ and $c$, which also leads to Eq.~\eqref{seq:bc.cond.AE}.

In current superconducting circuit implementations, one of the relevant factors limiting the strength $\Gamma_a$ of effective dissipation is the finite bandwidth of the auxiliary reservoir $\kappa_c $. For convenience, we now rewrite Eq.~\eqref{seq:bc.cond.AE} as follows
\begin{equation} 
\Gamma_a 
\ll \min \{ 
\Gamma_b  
, \kappa_c 
, \kappa_c ^{2} / {\Gamma_b}
 \} 
. 
\end{equation}
One can thus show that for a fixed value of $\kappa_c $, the optimal choice of $\Gamma_b$ to maximize the effective dissipation strength $\Gamma_a$ is $ \Gamma_b = \kappa_c  $, in which case we further have 
$ \kappa_c = \Gamma_b =  \lambda $. Substituting  this expression into Eq.~\eqref{seq:qbl.coh.bc}, we can in turn derive the leading-order error rate to the logical qubit coherence function as 
\begin{equation}
\lim _{t \to \infty }
\frac{\langle \hat{\Sigma} 
_{L-} (t )  
\rangle }
{\langle \hat{\Sigma} 
_{L-} (0 )  
\rangle}
\simeq \left (\frac{ \Gamma_{b}- i \lambda }
{\Gamma_{b} + i \lambda} \right ) ^{2} 
\left (1 - \frac {\Gamma_a}{2\Gamma_{b}} 
\right )  
.   
\end{equation}
We thus recover the relation between ideal Z-error rate assuming a perfect dissipator $\mathcal{D}[\hat{F}]$, versus the simulated error rate via the discussed physical implementation, as $\eta_{\textrm{sim}} - \eta_{\textrm{pred}} = (1 - \eta_{\textrm{pred}}) ({\Gamma_a}/{2\Gamma_b})$.

\subsection{Implementation in trapped-ion system}
% In trapped-ion system, we try to implement the dissipator:
% \begin{equation}
% \begin{aligned}
%     \hat{F} & = \frac{1}{\alpha^{\prime}} \hat{S}(r) [\hat{a}(\hat{a}^2 - \alpha^{\prime 2})] \hat{S}^{\dagger}(r) \\
% \end{aligned}
% \end{equation}

\hl{Fig.~6 in the main text describes the level scheme and laser configurations
to the coupling Hamiltonians to implement the SC in trapped-ion
systems. The setup involves the motional mode of the ion with Fock
states $\ket{n}$ which is coupled to three internal states labeled
$\ket{g}$, $\ket{e}$, $\ket{f}$. Starting from the atomic ground
state $\ket{g}$, the system goes through a two-step coherent transition
$\ket{g}\rightarrow\ket{f}\rightarrow\ket{e}$ with relevant driving
lasers indicated in Fig.~6 by black and green arrows, respectively.
In the first step, a laser drives the atomic transition $\ket{g}\otimes\ket{n}\rightarrow\ket{f}\otimes\ket{n}$
and the frequency-resolved second motional sidebands $\ket{g}\otimes\ket{n}\rightarrow\ket{f}\otimes\ket{n\pm2}$,
while the second step couples coherently $\ket{f}\otimes\ket{n}\rightarrow\ket{e}\otimes\ket{n\pm1}$.
Finally, state $\ket{e}$ is assumed to decay back to the ground state
$\ket{g}$ via spontaneous emission. The effective dissipator acting
on the motional state is obtained via adiabatic elimination of states
$\ket{e},\ket{f}$. Here we present the technical details of such a protocol.}

% \todo{Q: How do we separately obtain $\hat{a}$ and $\hat{a}^2 \hat{a}^{\dagger}$?}
Neglecting the momentum kicks first, the system dynamics is described by:
\begin{equation}
    \begin{aligned}
    \frac{d}{dt}\hat{\rho} & = -i[\hat{H}, \hat{\rho}] + \Gamma \mathcal{D}[|g\rangle \langle e|]\hat{\rho} \\
    \hat{H} = &  \Omega_0 (|f\rangle \langle g| + |g\rangle \langle f|) + \sum_{i = 1}^3 \frac{1}{2}\Omega_i \cos[\eta_0 (\hat{a}e^{-i\nu t} + \hat{a}^{\dagger} e^{i\nu t})] (|f\rangle \langle g| e^{-i \delta_i t} + |g\rangle \langle f| e^{i \delta_i t}) \\ 
    & + \sum_{i =4}^5\frac{1}{2}\Omega_i \sin[\eta_0 (\hat{a}e^{-i\nu t} + \hat{a}^{\dagger} e^{i\nu t})] (|e\rangle \langle f| e^{-i\delta_i t} + |f\rangle \langle e| e^{ i\delta_i t})
    \end{aligned}
\end{equation}
By choosing $\delta_1 = -2\nu, \delta_2 = 2\nu, \delta_3 = 0, \delta_4 = - \nu, \delta_5 = \nu$ and neglecting the fast-rotating terms, we obtain:
\begin{equation}
    \hat{H} = [\epsilon_1 \hat{a}^2 + \epsilon_2 \hat{a}^{\dagger 2} + \epsilon_3 (\hat{a}^{\dagger}\hat{a} + \hat{a}\hat{a}^{\dagger}) + \epsilon_0]|f\rangle \langle g| + (\epsilon_4 \hat{a} + \epsilon_5 \hat{a}^{\dagger}) |e\rangle \langle f| + h.c.,
\end{equation}
where $\epsilon_1 = - \frac{1}{4} \eta_0^2 \Omega_1, \epsilon_2 = - \frac{1}{4}\eta_0^2 \Omega_2, \epsilon_3 = - \frac{1}{4}\eta_0^2 \Omega_3, \epsilon_0 = \frac{1}{2}(\Omega_0 + \Omega_3), \epsilon_4 = \frac{1}{2}\eta_0 \Omega_4, \epsilon_5 = \frac{1}{2}\eta_0 \Omega_5$. By setting $\epsilon_1 = \cosh^2 r \Omega^{\prime}_{gf}, \epsilon_2 = \sinh^2 r \Omega^{\prime}_{gf}, \epsilon_3 = \sinh r \cosh r \Omega^{\prime}_{gf}, \epsilon_0 = - \alpha^{\prime 2} \Omega^{\prime}_{gf}, \epsilon_4 = \frac{\cosh r}{\alpha^{\prime}} \Omega^{\prime}_{ef}, \epsilon_5 = \frac{\sinh r}{\alpha^{\prime}} \Omega^{\prime}_{ef}$, we obtain 
\begin{equation}
    \hat{H} = \Omega^{\prime}_{gf} [\hat{S}(r) (\hat{a}^2 - \alpha^{\prime 2})\hat{S}^{\dagger}(r)]|f\rangle \langle g| + \Omega^{\prime}_{ef} \frac{1}{\alpha^{\prime}} \hat{S}(r)\hat{a}\hat{S}^{\dagger}(r) |e\rangle \langle f| + h.c.
\end{equation}
In the adiabatic regime $2\alpha^{\prime} \Omega_{gf}^{\prime} \ll \Gamma$, $\Omega^{\prime}_{gf} \ll \Omega^{\prime}_{ef}$, we can obtain a reduced dynamics on the motional mode by adiabtically eliminating the $|e\rangle, |f\rangle$ states:
\begin{equation}
    \frac{d}{dt}\hat{\rho}_m = \kappa_2 \mathcal{D}[\hat{F}]\hat{\rho}_m,
\end{equation}
where $\kappa_2 = (\frac{\Omega^{\prime}_{gf}}{\Omega^{\prime}_{ef}})^2 \Gamma$. Numerically we find that we can obtain the dissipator $\hat{F}$ with the desired rate by setting $\Omega^{\prime}_{ef} = 0.5 \Gamma, \Omega^{\prime}_{gf}/\Omega^{\prime}_{ef} = 1/20$. The leading-order off-resonant term of $\hat{H}$ is $\frac{1}{2}(\Omega_1 e^{2i \nu t} + \Omega_2 e^{-2i\nu t})|f\rangle \langle g| + h.c.$, which leads to AC stark shift of the $gf$ transition frequency. One can compensate this term by adjusting the laser detuning w.r.t $\omega_f$, or add a compensation drive to the bare $gf$ transition. We can provide a rough estimate of the achievable engineered dissipation strength $\kappa_2$ based on attainable experimental parameters $\nu = 30$ MHz, $\eta_0 = 0.15$~\cite{monroe1996schrodinger}. With the leading-order off-resonant term compensated, we can obtain a good stabilization for a $\Gamma$ up to 200 KHz, which leads to $\kappa_2 = 0.5$ KHz. This estimated rate is much larger than the typical ambient noise rates of the ions, which are in the level of 10 Hz~\cite{fluhmann2019encoding}. Note also that the prominent noise in many trapped-ion platforms was identified as the dephasing~\cite{fluhmann2019encoding},  which, compared to the loss errors, can be better suppressed by the SC.

We give a brief analysis on the effect of the momentum kicks. In the Lamb-Dick regime ($\eta_0 \ll 1$), the system dynamics that include the leading-order contribution from the momentum kicks read:
\begin{equation}
    \frac{d}{dt}\hat{\rho} = -i[\hat{H}, \hat{\rho}]+\Gamma \mathcal{D}\left[|g\rangle \langle e|\right] \hat{\rho}+\left(\Gamma \eta_0^2 \int_{-1}^1 u^2 N(u) d u\right) \mathcal{D}\left[\hat{a}+\hat{a}^{\dagger}\right] |g\rangle \langle e| \hat{\rho} |e\rangle \langle g|.
\end{equation}
We can see that the third term, which results from the momentum kicks, leads to extra phase flips of the SC when the ion decays from $|e\rangle$ to $|g\rangle$. As such, the third term competes with the second term when correcting the errors on the motional mode that induce the internal-state transition cycle $|g\rangle \rightarrow |f\rangle \rightarrow |e\rangle \rightarrow |g\rangle$. Fortunately, the rate of the third term is suppressed by $\eta_0^2$. So the momentum kicks only lead to a small increase of the phase-flip suppression factor $\eta \rightarrow \eta + \mathcal{O}(\eta_0^2)$.

\hl{In the end, we present a candidate trapped-ion platform suitable for our scheme. 
Note that we require resolved sideband conditions,
i.e. the decay rate of $\ket{e}$ is assumed smaller than the trap
frequency. Such a tunable engineered effective spontaneous transition
rate can be engineered via Raman processes as optical pumping (see
below).} 

\hl{The prototypical level scheme for trapped ion is reviewed in Ref.~\cite{schindler2013quantum}
in context of implementation of trapped-ion quantum computing for
the case of $^{40}\textrm{Ca}^{+}$ ions. We now briefly describe
the mapping of our level scheme $\ket{g}$, $\ket{e}$, $\ket{f}$
and laser configuration to levels of the Ca$^{+}$ion, as available
in these experiment. $^{40}\textrm{Ca}^{+}$ has an electronic ground
state $^{2}S_{1/2}$ coupled via a quadrupole transition
to metastable excited states $^{2}D_{5/2,3/2}$, and
both the ground and metastable states can excite the short-lived $^{2}P_{3/2,1/2}$
states in dipole-allowed transitions (see Fig. 1 of Ref.~\cite{schindler2013quantum}).}

\hl{A first mapping is obtained by identifying $\ket{g}$, $\ket{e}$,
$\ket{f}$ with long-lived states $^{2}S_{1/2}$, $^{2}D_{5/2}$,
and $^{2}D_{3/2}$, respectively, where the first transition
(black arrows in Fig.~6 main text) corrsponds to the quadrupole transition,
while the second step (green arrows in Fig.~6 main text) can be implemented
as a Raman transition coupling the two metastable $D$-states. A tunable
effective decay rate back to the $S$-ground state can be obtained
as an optical pumping process via the short-lived $P$ states.} 

\hl{A second mapping identifies the two Zeeman ground states $\ket{^{2}S_{1/2}\:m=\pm1/2}$
with levels $\ket{g},\ket{f}$ and drives the second order sidebands,
as familiar from sideband cooling utilizing Raman cooling via intermediate
$P$-states (see Fig. 2d in Ref.~\cite{schindler2013quantum}). The second coherent step can
then be implemented via transitions to $D$- states, followed by engineered
decay via short-lived $P$-states.}

% \newpage
% \bibliographystyle{apsrev4-2}
% \bibliography{ref_SM}
